\documentclass[fleqn,usenatbib]{mnras}

\usepackage{newtxtext,newtxmath}

\usepackage[T1]{fontenc}

\DeclareRobustCommand{\VAN}[3]{#2}
\let\VANthebibliography\thebibliography
\def\thebibliography{\DeclareRobustCommand{\VAN}[3]{##3}\VANthebibliography}
\DeclareRobustCommand{\DE}[3]{#2}
\let\DEthebibliography\thebibliography
\def\thebibliography{\DeclareRobustCommand{\DE}[3]{##3}\DEthebibliography}

\usepackage{graphicx}	
\usepackage{amsmath}	
\usepackage{amstext}	
\usepackage{multicol}   
\usepackage{bm}	    	
\usepackage{pdflscape}	
\usepackage{color}      
\usepackage[figure,figure*]{hypcap}
\usepackage{dutchcal}   
\usepackage{ragged2e}   

\newcommand{\D}			{$^\circ$}
\newcommand{\vrot}		{V$_{\rm rot}$}
\newcommand{\vrad}		{V$_{\rm rad}$}
\newcommand{\dvsys}		{$\Delta$V$_{\rm sys}$}
\newcommand{\vsys}		{V$_{\rm sys}$}
\newcommand{\ha}		{\mbox{\rm{H}~$\alpha$}}

\newcommand{\HI}       {\mbox{\rm H\,{\small I}}}

\newcommand{\kms}		{\mbox{km\,s$^{-1}$}}
\newcommand{\msun}		{\mbox{M$_\odot$}}
\newcommand{\msmh}		{\mbox{M$_\star$/M$_{\rm halo}$}}
\newcommand{\mjybeam}		{\mbox{mJy\,beam$^{-1}$}}

\newcommand{\tclean}	{{\fontfamily{cmtt}\selectfont tclean}}
\newcommand{\emcee}	{{\sc emcee}}

\newcommand{\miriad}	{{\sc Miriad}}
\newcommand{\casa}	{{\sc casa}}
\newcommand{\nemo}	{{\sc nemo}}
\newcommand{\betastar}{$\beta^*$}
\newcommand{\betarvir}{$\beta_{1-2\%{\rm\ R_{vir}}}$}
\newcommand{\rvir}{R$_{\rm vir}$}
\newcommand{\ML}{$\Upsilon_\star$}
\newcommand{\spitzer}{{\em Spitzer}}
\newcommand{\avebetastar}{$\langle\beta^*\rangle = 0.6$}
\newcommand{\scatterbetastar}{$\pm0.1$}
\newcommand{\medbetastar}{$\tilde{\beta}^* = 0.5^{+0.4}_{-0.6}$}
\newcommand{\change}{}
\newcommand{\changes}{}

\defcitealias{relatores19a}{R19a}
\defcitealias{relatores19b}{R19b}
\newcommand{\cta}{\citetalias}
\newcommand{\cpa}{\citepalias}

\title[Cuspy dark matter profiles in massive dwarfs]{Cuspy dark matter density profiles in massive dwarf galaxies}

\author[Cooke, Levy, et al.]{Lauren H. Cooke,$^{1,2}$ Rebecca C. Levy,$^{1,\change{3}}$\thanks{\change{NSF Astronomy and Astrophysics Postdoctoral Fellow}}\thanks{Contact e-mail: \href{mailto:rebeccalevy@email.arizona.edu}{\change{rebeccalevy@email.arizona.edu}}} Alberto D. Bolatto,$^{1,\change{4,5,6}}$  Joshua D. Simon,$^{7}$ \newauthor Andrew B. Newman,$^{7}$ Peter Teuben,$^{1}$ Brandon D. Davey,$^{1,8}$ Melvyn Wright,$^{9}$ Elizabeth Tarantino,$^{1}$ \newauthor Laura Lenki\'c,$^{1,\change{10}}$ and Vicente Villanueva$^{1}$\\
$^{1}$Department of Astronomy, University of Maryland, College Park, MD 20742, USA\\
$^{2}$Holton-Arms School, Bethesda, MD 20817, USA\\
$^{3}$\change{Steward Observatory, University of Arizona, Tucson, AZ 85721, USA}\\
$^{4}$\change{Joint Space-Science Institute, University of Maryland, College Park, MD 20742, USA}\\
$^{5}$\change{Visiting Scholar at the Flatiron Institute, Center for Computational Astrophysics, NY 10010, USA}\\
$^{6}$\change{Visiting Astronomer, National Radio Astronomy Observatory, VA 22903, USA}\\
$^{7}$Observatories of the Carnegie Institution for Science, Pasadena, CA 91101, USA\\
$^{8}$Department of Physics, University of South Florida, Tampa, FL 33620, USA\\
$^{9}$Department of Astronomy, University of California, Berkeley, CA 94720, USA\\
$^{10}$\change{SOFIA Science Center, USRA, NASA Ames Research Center, M.S. N232-12, Moffett Field, CA 94035, USA}\\
}

\date{Accepted 2022 February 28. Received 2022 February 28; in original form 2021 June 21}

\pubyear{2022}

\begin{document}
\label{firstpage}
\pagerange{\pageref{firstpage}--\pageref{lastpage}}
\maketitle

\begin{abstract}
Rotation curves of galaxies probe their total mass distributions, including dark matter. Dwarf galaxies are excellent systems to investigate the dark matter density distribution, as they tend to have larger fractions of dark matter compared to higher mass systems. The core-cusp problem describes the discrepancy found in the slope of the dark matter density profile in the centres of galaxies ($\beta^*$) between observations of dwarf galaxies (shallower cores) and dark matter-only simulations (steeper cusps). We investigate $\beta^*$ in six nearby spiral dwarf galaxies for which high-resolution CO $J=1-0$ data were obtained with ALMA. We derive rotation curves and decompose the mass profile of the dark matter using our CO rotation curves as a tracer of the total potential and 4.5~$\mu$m photometry to define the stellar mass distribution. We find $\langle\beta^*\rangle = 0.6$ with a standard deviation of $\pm0.1$ among the galaxies in this sample, in agreement with previous measurements in this mass range. The galaxies studied are on the high stellar mass end of dwarf galaxies and have cuspier profiles than lower mass dwarfs, in agreement with other observations. When the same definition of the slope is used, we observe steeper slopes than predicted by the FIRE and NIHAO simulations. This may signal that these relatively massive dwarfs underwent stronger gas inflows toward their centres than predicted by these simulations, that these simulations over-predict the frequency of accretion or feedback events, or that a combination of these or other effects are at work.
\end{abstract}




\section{Introduction}

\label{sec:intro}
Dwarf galaxies have proven to be important laboratories to test principles of current dark energy and cold dark matter cosmology ($\Lambda$CDM), despite their difficulty to observe \changes{\citep[e.g.,][]{bullock17,lelli22}}. 
Cosmological simulations of cold dark matter predict a cusp-like dark matter density distribution near \changes{galaxy centres}, meaning the density has a power-law slope ($\beta$) between $1$ and $1.5$ \citep[e.g.,][]{navarro96b,moore99}\footnote{In some of the literature, the power-law slope is denoted as $\alpha$, where $\alpha\equiv -\beta$. Here we use $\beta$ throughout for consistency\changes{, such that $\rho(r) \propto r^{-\beta}$.}}. The most well-known form of this density distribution is the Navarro-Frenk-White (NFW) profile \citep{navarro96b}. Observational studies of dwarf galaxies, however, show that central density profiles range from flat core-like distributions ($\beta\approx0$) to the cusp-like profiles expected from dark matter-only simulations \citep[see e.g.,][]{deblok01b,deblok01a,deblok02,simon05,oh11,oh15,adams14,relatores19a,relatores19b}. \changes{This discrepancy in the slope of the dark matter density profile near the centres of dwarf galaxies measured from observations and simulations is called the core-cusp problem \citep[e.g.,][]{flores94,moore94,bullock17,lelli22}}. 

\change{There are at least three solutions to this problem that have been explored in the literature.} First, dark matter could behave differently than assumed under the $\Lambda$CDM model. For example, if dark matter could interact with itself (so-called self-interacting dark matter or SIDM), the transfer of energy between dark matter particles would soften the density profiles, alleviating the core-cusp problem \citep[e.g.,][]{spergel00,vogelsberger12,peter13,fry15,elbert15,cyr-racine16,vogelsberger16,kaplinghat20,leung21}. An important test for SIDM comes from galaxy clusters, where the upper-limit on the SIDM cross section is marginally consistent with the lower limit needed to alleviate the core-cusp problem, suggesting that \change{such} SIDM models are unlikely to be a solution to the core-cusp problem \citep{bullock17}. \change{There are, however, a class of well-motivated, velocity-dependent SIDM scattering cross sections, which can reproduce the dark matter distributions and properties in both galaxy clusters and dwarf galaxies \citep[e.g.,][]{yoshida00,read18,valli18,kaplinghat20,sagunski21,correa21}.} As pointed out by \citet{bullock17}, warm dark matter models produce the same density distributions as CDM and are therefore not a solution to the core-cusp problem.

\change{Another possibility is that systematic and/or measurement uncertainties in the rotation curve analysis may give rise to this tension. Rotation curves measured for some kinematic tracer\changes{s} are used to infer the circular velocity curve and hence the mass or density profiles\footnote{\change{In this paper, we use \changes{`rotation curve'} to refer to the measured rotation speed of some kinematic tracer. We use \changes{`(circular) velocity curve'} to refer to the velocity corresponding the mass distribution of some component of the galaxy \changes{(i.e., $V^2=R\frac{\partial\Phi}{\partial R}$)}.}}. In the rotation curve analysis, a failure to properly account for non-circular motions can lead to measurements of cored profiles. It has been argued in the literature that triaxiality in the dark matter halo (as opposed to the often assumed spherical geometry) can induce non-circular motions in the gas disc \citep[e.g.,][]{simon05,hayashi06,read16,genina18,marasco18,oman19,santos-santos20,jahn21}. If these non-circular motions are unaccounted for in the rotation curve analysis, they may lead to an underestimate of the true circular velocity and hence produce an apparent core-like dark matter profile. However, for these effects to completely explain the core-cusp problem, a clear explanation of why these non-circular motions only \changes{give the appearance of} cores in low-mass galaxies is necessary. Moreover, non-circular motions can also lead to an over-prediction of the central circular velocity and instead produce an apparently {\em cuspier} dark matter density profiles. These non-circular motions can induce spurious diversity in rotation curve shapes (i.e. both steeper and shallower slopes) and must be accounted for to understand any underlying physical rotation curve diversity \citep[e.g.,][]{oman15,oman19,santos-santos18,santos-santos20,jahn21}. Therefore, while care should be taken to properly account for and measure non-circular motions in rotation curve analyses, such errors are unlikely to fully alleviate the core-cusp problem \changes{\citep[e.g.,][]{kaplinghat20,lelli22}}.}

A \changes{possible} solution within the $\Lambda$CDM paradigm is that baryonic physics and feedback impact the dark matter distribution in a galaxy. This feedback, especially from supernovae, can redistribute mass in a galaxy, reshaping the potential and hence the dark matter density distribution. This feedback will be most efficient in the centres of dwarf galaxies, where the dark matter density and star formation rate surface densities are highest. Numerical simulations, including feedback, find that the peak core-formation occurs in galaxies with \change{stellar masses (M$_\star$)} $\simeq10^{8-9}$~\msun\ \citep[e.g.,][]{navarro96c,governato12,pontzen14,dicintio14,chan15,tollet16,fitts17,hopkins18,lazar20,maccio20}. At smaller masses, the star formation rates are too low for the resulting stellar feedback to significantly alter the density distribution. At larger masses, the gravitational potential is too deep for stellar feedback to significantly redistribute the dark matter into a core-like profile.

In this study, we build upon the set of observed inner dark matter density slopes using new observations of a sample of six galaxies from the Dwarf Galaxy Dark Matter (DGDM) survey \citep{truong17}. Using new $^{12}$CO $J=1-0$ observations from ALMA and \textit{Spitzer} 4.5 $\mu$m data as tracers of the total potential and stellar component, we kinematically decompose the dark matter density profiles to measure the inner dark matter density slopes (\betastar).

This work closely follows that of \citet[][hereafter \cta{relatores19a} and \cta{relatores19b}]{relatores19a,relatores19b}, who measured \betastar\ from another DGDM survey sub-sample using \ha\ observations to trace the total potential. \change{\betastar\ is defined as the slope of the dark matter density profile from 0.3--0.8~kpc, describing the shape of the dark matter density profile at small galactocentric radii. In contrast, $\beta$ is the power-law exponent of the parametrized dark density profile, whether that be a pure power-law, a NFW profile, or another form.} The \betastar\ values derived by \cta{relatores19b} agree with other CO data points from CARMA \citep{truong17}. Specifically, \cta{relatores19b} find that one third of the inner slopes were consistent with the NFW profile, while the rest were more core-like than the NFW profile predicts. 

From our sample of six dwarf galaxies, we find \change{shallow} cusp-like inner dark matter density distributions (\avebetastar\ with a standard deviation of \scatterbetastar\ among the measurements). \change{At first glance, our results broadly agree with predictions from \change{the FIRE and NIHAO} simulations, insofar as the dark matter profiles are shallower than predicted by a pure NFW profile in this stellar mass range of $10^{9.3-9.7}$~\msun\ \citep[e.g.,][]{tollet16,maccio20,lazar20}.} \change{However, when we calculate the inner dark matter density slope in the same way as these simulations (based on the virial radius, \rvir), we find that the slopes we measure are steeper than these simulations predict.} These simulations find that cores can only be maintained through accretion, mergers, and/or outflows from stellar feedback which disturb the potential. \change{We posit} that these dwarfs are likely too massive for stellar feedback to effectively alter the dark matter distributions \change{to be more core-like.}

This paper is organized as follows. We describe the CO data used for this study and the data reduction in {Section~\ref{sec:data}}. {Section~\ref{sec:rcfitting}} describes the method used to derive the CO rotation curves. The determination of the stellar components, decomposition of the dark matter density profiles, and measurement of \betastar\ are presented in {Section~\ref{sec:DMdensity}}. Our results and robustness tests are described in {Section~\ref{sec:results}}. Implications of our results are discussed in {Section~\ref{sec:discussion}}. We summarize our conclusions in {Section~\ref{sec:summary}}. 

\section{Observations and Data Reduction}
\label{sec:data}

\begin{table}
\centering
\caption{CO Data Reduction and Imaging Parameters}
\label{tab:obs}
\begin{tabular}{lcccc}
\hline
Name & Beam & \multicolumn{2}{c}{rms Noise} & \casa\ Cal. Ver.  \\
& (arcsec) & (\mjybeam) & (mK) & \\
\hline
NGC1035$^*$ & 2.90 & 4.2 & 46 & 4.5.1 \\
NGC4310 & 2.40 & 4.1 & 65 & 4.5.3 \\
NGC4451$^*$ & 2.80 & 4.2 & 49 & 4.5.1 \\
NGC4701 & 2.80 & 4.3 & 50 & 4.5.1 \\
NGC5692$^*$ & 2.85 & 4.7 & 53 & 4.5.1 \\
NGC6106$^*$ & 2.40 & 3.4 & 54 & 4.5.2 \\
\hline
\end{tabular}

\justifying\noindent{Galaxies marked with $^*$ overlap with \cta{relatores19a} and \cta{relatores19b}. The FWHM of the synthesized, circularized Gaussian beam is given. The rms noise per 2~\kms\ channel is reported for the cleaned cube for regions with \change{no signal (SNR $<3$)} based on the peak intensity in \mjybeam\ and \changes{mK}. \casa\ Cal. \changes{Ver.} lists the version of \casa\ used for calibration\changes{; all of the visibilities} were imaged with \casa\ version 5.7.2-4.}
\end{table}

The initial galaxy selection for the DGDM survey is described by \citet{truong17}, who observed these galaxies in $^{12}$CO $J=1-0$ (CO) with the Combined Array For Millimeter-wave Astronomy (CARMA). Of the initially selected 26 galaxies, 14 were robustly detected in CO with CARMA. Of these, 13 were followed up with new and more sensitive CO observations using the Atacama Large Millimeter/submillimeter Array (ALMA) as part of project number 2015.1.00820.S (PI L. Blitz). These 13 galaxies were chosen to be bright and extended in the WISE 22~$\mu$m band (W4), to have substantial IRAS 100~$\mu$m flux, and to be far enough south to be visible to ALMA. These galaxies were observed in Band 3 in the C36-1 configuration with baselines ranging from $15-640$~m and a maximum recoverable scale of $\approx 21\arcsec$ ($\sim 2$~kpc \changes{for the mean distance to these galaxies of 21~Mpc}). Mosaics of a few pointings were used for galaxies with large angular extents on the sky (e.g., NGC\,1035 and NGC\,6106).  This configuration results in $\sim 2\arcsec$ ($\sim 200$~pc \changes{for the mean distance to these galaxies}) resolution that is well matched to the \ha\ data from the Palomar Cosmic Web Imager presented by \cta{relatores19a}.

The visibilities were calibrated and flagged by the observatory using the Common Astronomy Software Application \citep[\casa;][]{casa} version listed in {Table~\ref{tab:obs}}. The data were imaged using \tclean\ in \casa\ version 5.7.2-4 with \texttt{\change{deconvolver}=`hogbom'}, \texttt{specmode=`cube'}, Briggs weighting with \texttt{robust=0.5}, and no $uv$-taper. \change{The pixel (\texttt{cell}) sizes are listed in {Table~\ref{tab:obs}}.} The baseline was fit with a linear function (\texttt{nterms=2}) to account for any change in slope over the band. The imaging was performed on the three-dimensional RA-Dec-velocity images, where each spatial pixel (spaxel) contains a spectrum. These images were cleaned until the residuals were consistent with the root-mean-square (rms) noise levels given in {Table~\ref{tab:obs}}. The final images all have a velocity resolution of 2~\kms, and we convolved the images to a circular beam, which is listed in {Table~\ref{tab:obs}}. The rms noise per channel reported in {Table~\ref{tab:obs}} is calculated where the \change{signal-to-noise ratio (SNR) $<3$} based on the peak intensity maps (described in the following paragraph).  

\begin{figure*}
\label{fig:maps}
   \centering
    \includegraphics[width=\textwidth]{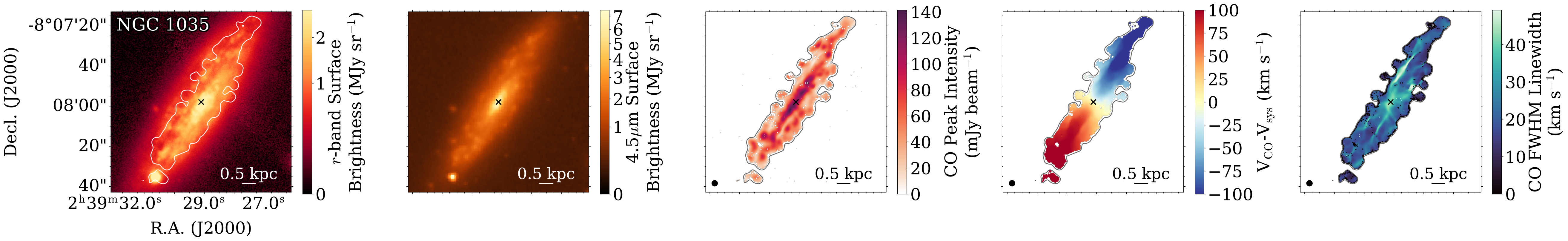}
    \includegraphics[width=\textwidth]{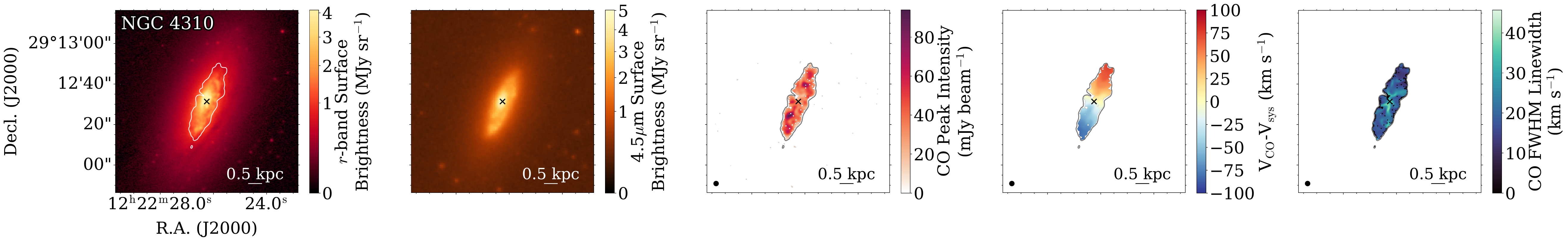}
    \includegraphics[width=\textwidth]{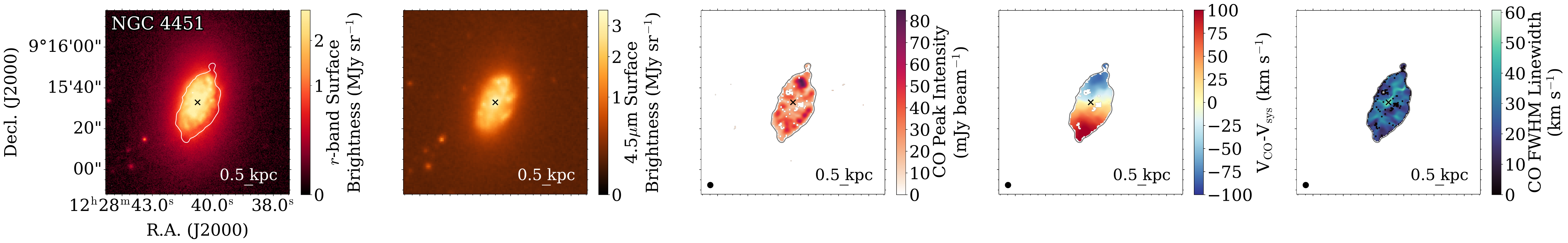}
    \includegraphics[width=\textwidth]{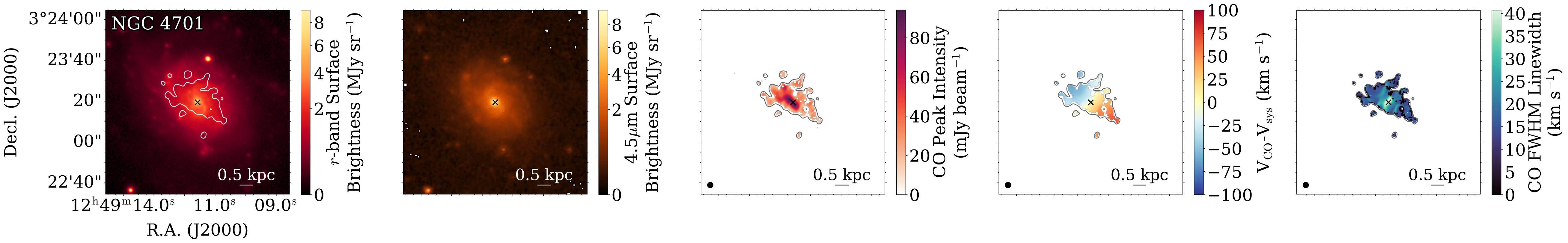}
    \includegraphics[width=\textwidth]{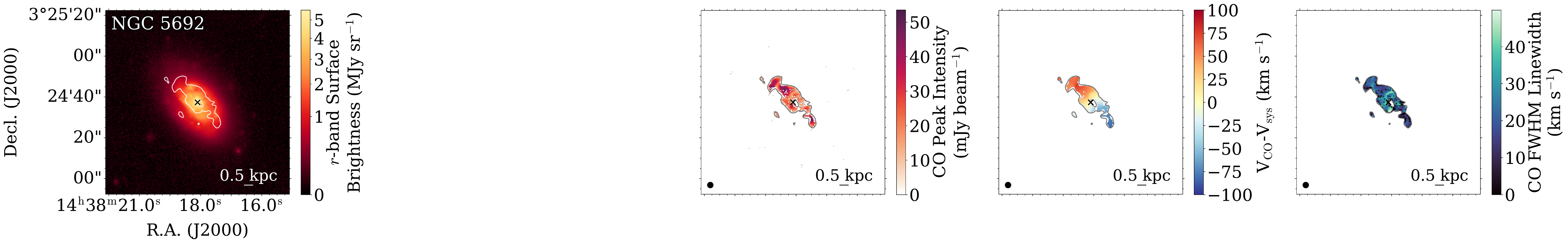}
    \includegraphics[width=\textwidth]{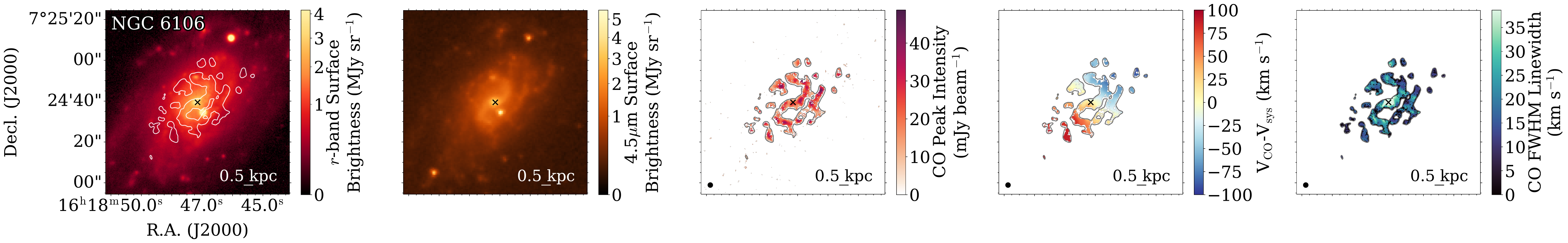}
 \caption{The six galaxies studied here, showing (from left to right) their PS1 $r$-band images, \spitzer\ 4.5~$\mu$m  images, the CO peak intensity, the CO velocity, and the CO FWHM linewidth. There is no \spitzer\ image for NGC\,5692. Each image is cropped to 1.5\arcmin\ on a side. The crosses show the location of the kinematic centre {(Table~\ref{tab:geomparams}).} The white or grey contours show SNR = 3 based on the CO peak intensity. The CO maps are masked to this level. The black ellipses in the lower left corners of the CO images show the FWHM beam sizes.}
\end{figure*}

Moment maps (peak intensity, velocity, linewidth) were produced by fitting each pixel of the CO images with a Gaussian. Error maps are also produced based on the statistical uncertainties in the Gaussian fitting. Fits that do not converge are blanked. Isolated blanked pixels (i.e., where none of the neighbouring values is blanked) are replaced with the median value of the neighbours. Pixels in the CO intensity, velocity, and velocity dispersion maps with SNR $<3$ based on the peak intensity and associated error maps are masked out. \change{We impose a minimum velocity uncertainty of 6~\kms\ to account for the random ISM motions \cpa[e.g.,][]{relatores19a,relatores19b}. This level also corresponds to $\sim$1/3 of the typical CO linewidth.} The CO peak intensity, velocity, and full-width-half-maximum (FWHM) linewidth maps are shown in {Figure~\ref{fig:maps}}. All velocities presented in this work have been converted to the relativistic velocity frame (see e.g., Appendix~A of \citealt{levy18} for details and conversions.).

Of the 13 galaxies observed with ALMA, three galaxies (NGC\,853, NGC\,1012, NGC\,4376) have insufficient signal to produce a velocity field even before our SNR mask is applied. Our SNR mask excludes three more galaxies (NGC\,4150, NGC\,4396, UGC\,8516). Finally, we exclude NGC\,4632. This galaxy is in a group with NGC\,4666 (a large starburst galaxy), NGC\,4668, and two other smaller dwarf galaxies \citep{garcia93,walter04}, though \change{these} ALMA observations only cover NGC\,4632. The CO in NGC\,4632 follows the spiral arms, with the northern spiral arm being much more prominent. A single prominent spiral arm can be a sign of a past interaction \citep[e.g.,][]{oh08_tidal}, which is possible for this galaxy given its known group membership. \change{We therefore exclude} this galaxy from the sample as the presence of other close, large galaxies and the hints of tidal interaction may affect the kinematics and dark matter halo of this system. Our final sample consists of six galaxies (NGC\,1035, NGC\,4310, NGC\,4451, NGC\,4701, NGC\,5692, NGC\,6106), shown in {Figure~\ref{fig:maps}} and listed in {Table~\ref{tab:obs}.}

These ALMA observations cover antenna spacings from $15-640$~m and \changes{filters out} some of the flux on scales larger than $\approx 21\arcsec$. Lower-sensitivity CARMA data of these galaxies taken in the C, D, and E configurations (antenna spacings from $7.5-373$~m) recover scales up to $\sim$ 40$\arcsec$ \citep{truong17}.  Importantly for our study, we are interested in the CO kinematics, as opposed to robust CO fluxes or molecular gas masses. As a test of the effects of the missing short spacings on the CO kinematics, we combined the ALMA and CARMA images using the \texttt{immerge} task in \miriad\ \citep{miriad}. We derived the velocity fields from these combined images in the same way as described above. The resulting combined velocity fields show no significant differences with the ALMA-only velocity fields. We therefore proceed using only the ALMA data in our analysis.

In addition to the CO data, \changes{we use} \spitzer\ IRAC 4.5$\mu$m (Channel 2) maps from the Spitzer Survey of Stellar Structure in Galaxies \citep[S4G;][]{sheth10,munoz-mareos13,querejeta15} to trace the stars\footnote{These images were downloaded directly from the NASA/IPAC Infrared Science Archive: \url{https://irsa.ipac.caltech.edu}.}. \spitzer\ IRAC data are not available for NGC\,5692. We instead use $r$-band images from the 3$\pi$ Steradian Survey on the PAN-STARRS1 (PS1) telescope \citep{chambers16}\footnote{\change{These images were} downloaded directly from the PS1 Image Cutout Server: \url{https://ps1images.stsci.edu/cgi-bin/ps1cutouts}.}. PS1 $r$-band and \spitzer\ 4.5~$\mu$m images are shown for these galaxies in {Figure~\ref{fig:maps}.} 

Distances to these galaxies are taken from two sources. Three of our galaxies have Tully-Fisher distances reported by \citet{tully13}. For the other three, we use the distances from the \texttt{modbest} column in the HyperLEDA catalogue \citep{makarov14}\footnote{\url{http://leda.univ-lyon1.fr/}}. These distances \changes{(and their uncertainties)} are reported in {Table~\ref{tab:geomparams}.} 

\begin{table*}
\centering
\caption{Parameters from Rotation Curve Fitting}
\label{tab:geomparams}
\begin{tabular}{cccccccccc}
\hline
Name & R.A. & Decl. & Distance & \vsys & PA & Inc  & R$_{\rm max}$ & log~M$_{\rm dyn}$(R$_{\rm max}$) & $\sigma_{\rm V}$ \\
& (J2000 hours) & (J2000 deg) & (Mpc) & (\kms) & (deg) & (deg) &  (kpc) & (log~M$_\odot$) & (\kms)\\
\hline
NGC1035 & $2^\mathrm{h}39^\mathrm{m}29.1^\mathrm{s}$ & $-8^\circ07{}^\prime58.2{}^{\prime\prime}$ & 15.9 $\pm$ 3.2 (a) & 1224 & 144 & 79 & 2.9 & 10.0 $\pm$ 0.1 & 20.6 $\pm$ 7.8 \\
NGC4310 & $12^\mathrm{h}22^\mathrm{m}26.3^\mathrm{s}$ & $+29^\circ12{}^\prime31.1{}^{\prime\prime}$ & 16.1 $\pm$ 7.4 (b) & 927 & 335 & 71 & 1.6 & 9.5 $\pm$ 0.1 & 19.0 $\pm$ 5.8 \\
NGC4451 & $12^\mathrm{h}28^\mathrm{m}40.5^\mathrm{s}$ & $+9^\circ15{}^\prime32.5{}^{\prime\prime}$ & 25.9 $\pm$ 5.2 (a) & 849 & 178 & 49 & 2.2 & 10.0 $\pm$ 0.1 & 25.3 $\pm$ 7.8 \\
NGC4701 & $12^\mathrm{h}49^\mathrm{m}11.6^\mathrm{s}$ & $+3^\circ23{}^\prime19.0{}^{\prime\prime}$ & 16.7 $\pm$ 2.7 (b) & 725 & 230 & 49 & 1.5 & 9.4 $\pm$ 0.1 & 17.1 $\pm$ 6.4 \\
NGC5692 & $14^\mathrm{h}38^\mathrm{m}18.1^\mathrm{s}$ & $+3^\circ24{}^\prime37.1{}^{\prime\prime}$ & 25.4 $\pm$ 7.4 (b) & 1605 & 36 & 53 & 1.7 & 9.5 $\pm$ 0.1 & 21.6 $\pm$ 8.4 \\
NGC6106 & $16^\mathrm{h}18^\mathrm{m}47.2^\mathrm{s}$ & $+7^\circ24{}^\prime39.3{}^{\prime\prime}$ & 24.3 $\pm$ 4.9 (a) & 1465 & 143 & 59 & 3.3 & 10.0 $\pm$ 0.1 & 18.1 $\pm$ 8.7 \\
\hline
\end{tabular}

\justifying\noindent{R.A. and Decl. list the best fit center. The distances \changes{and uncertainties are} from (a) \citet{tully13} where available or (b) \changes{the \texttt{modbest} column in} HyperLEDA otherwise (see Section \ref{sec:data}). \vsys\ is in the relativistic velocity frame. The PA is measured E of N to the receding side of the major axis. M$_{\rm dyn}$(R$_{\rm max}$) is the dynamical mass measured from the CO rotation curve at radius R$_{\rm max}$. $\sigma_{\rm V}$ is the median FWHM velocity dispersion, uncorrected for beam smearing; the uncertainty is the standard deviation.}
\end{table*}

\section{Derivation of CO Rotation Curves}
\label{sec:rcfitting}

 We derive the CO rotation curves using a first order harmonic decomposition:
\begin{equation}
\label{eq:harmdecomp}
V(r) = {\rm V_{rot}}(r)\cos\phi\sin i+{\rm V_{rad}}(r)\sin\phi\sin i+ \Delta {\rm V_{sys}}(r)
\end{equation}
which fits for the rotation (\vrot), radial (\vrad), and systemic (\dvsys) components, where $r$ is the galactocentric radius, $\phi$ is the azimuthal angle in the plane of the disk, and $i$ is the inclination and is assumed the same for all rings \citep[e.g.,][]{begeman89}. Before fitting, the central systemic velocity (\vsys) is subtracted from the map, such that the fitted systemic component (\dvsys) describes the deviation from this value. We extract the rotation velocities in concentric rings using a new Python implementation\footnote{The Python implementation of the rotation curve fitting code is publicly available at \url{https://github.com/rclevy/RotationCurveTiltedRings}.} of the original MATLAB code developed by \citet{bolatto02} and used by \citet{simon03,simon05}. \citet{levy18} modified the ring spacing algorithm so that rings are spaced in half-beam-width increments or contain at least 30 pixels per ring, whichever is larger. 

\begin{figure*}
\label{fig:CORCs}
    \centering
        \includegraphics[width=0.45\textwidth]{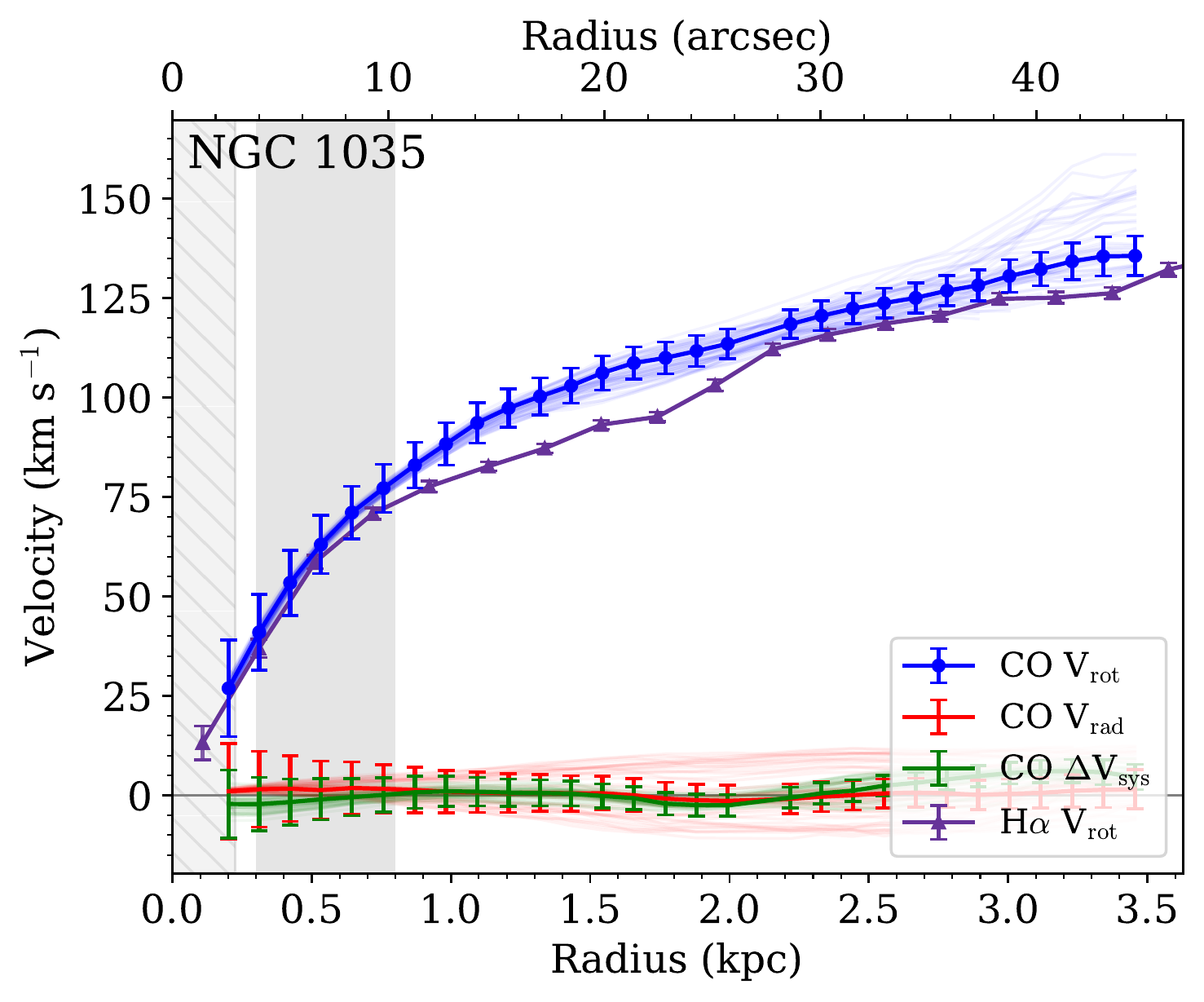}
        \includegraphics[width=0.45\textwidth]{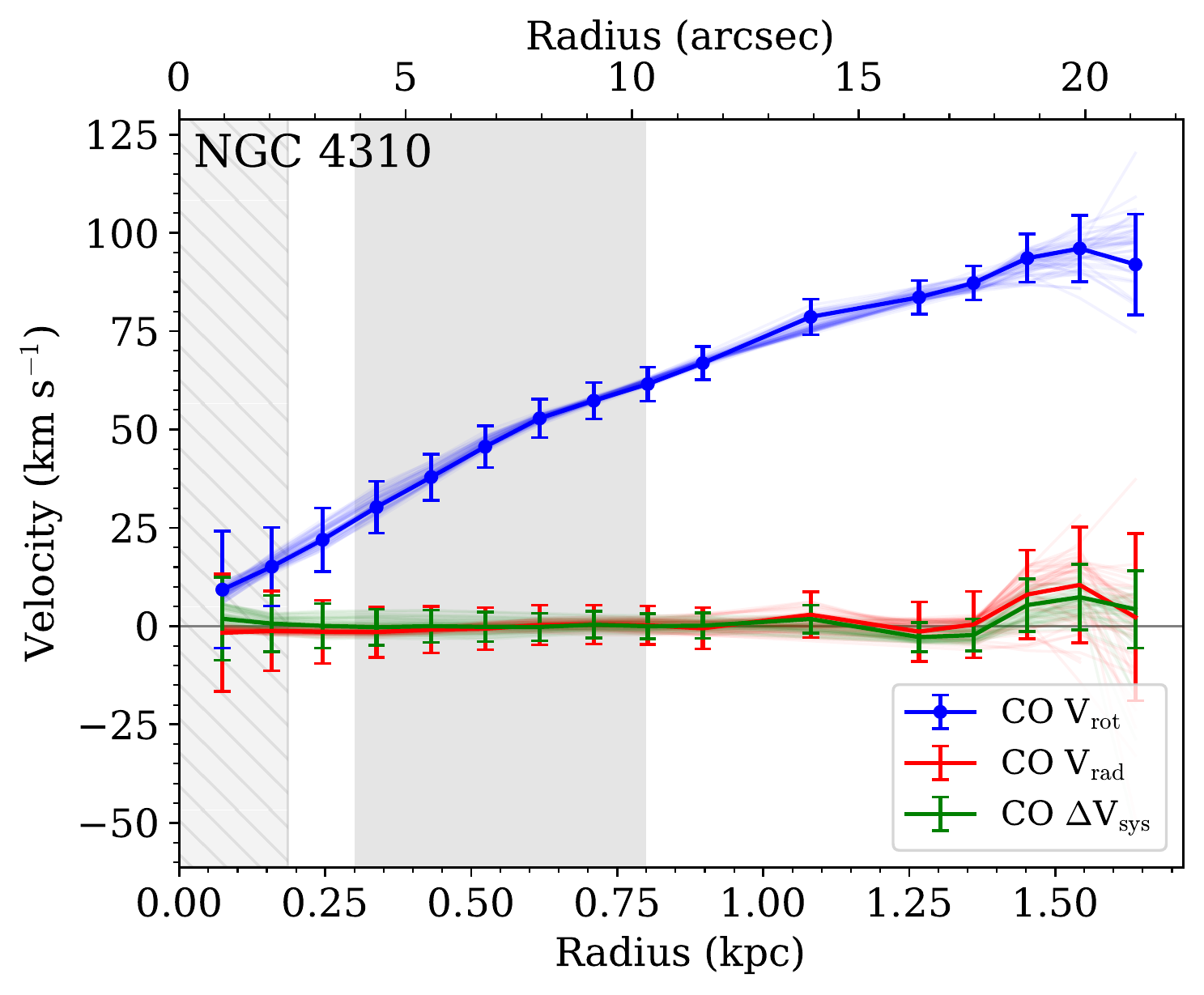}
        
        \includegraphics[width=0.45\textwidth]{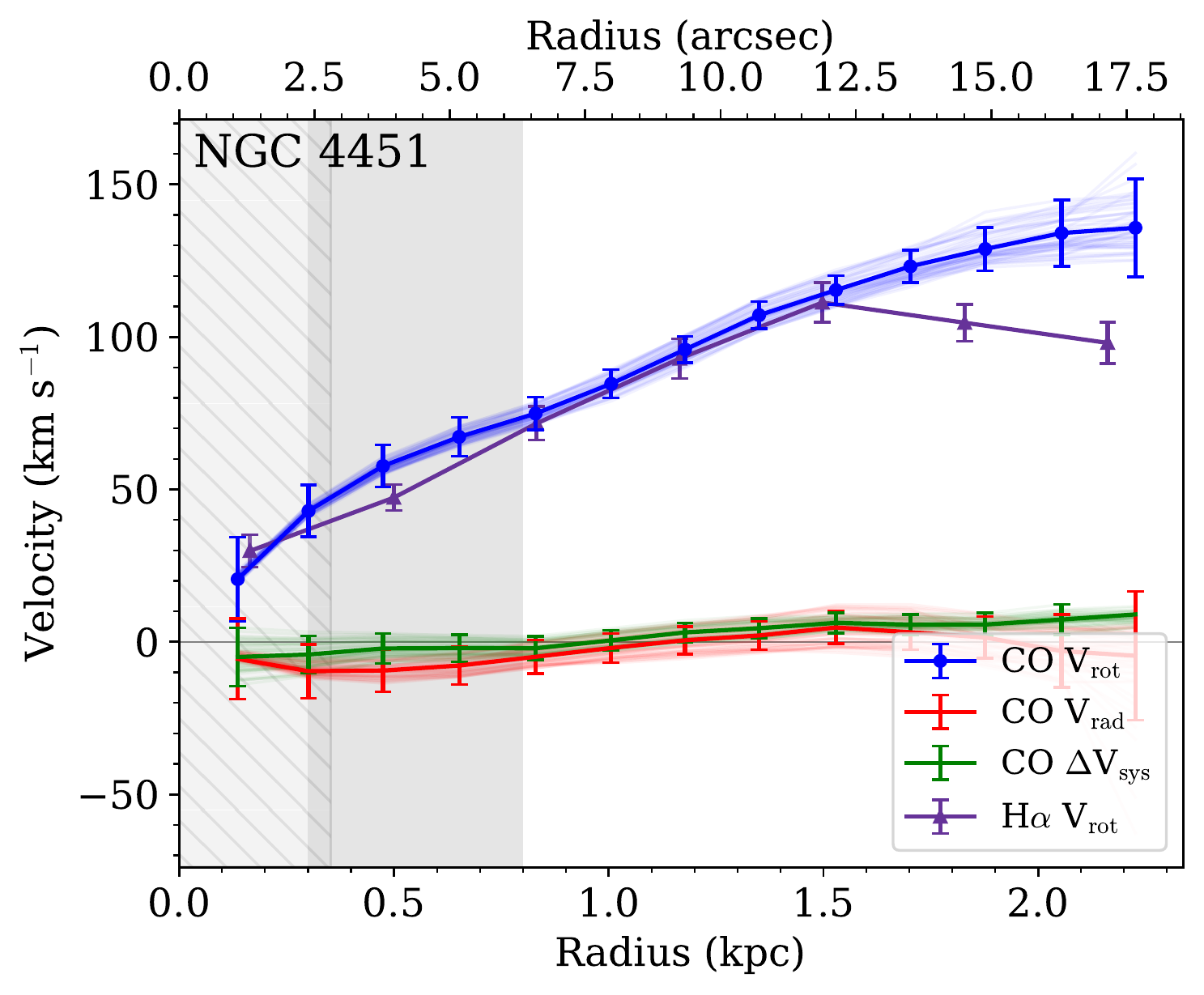}
        \includegraphics[width=0.45\textwidth]{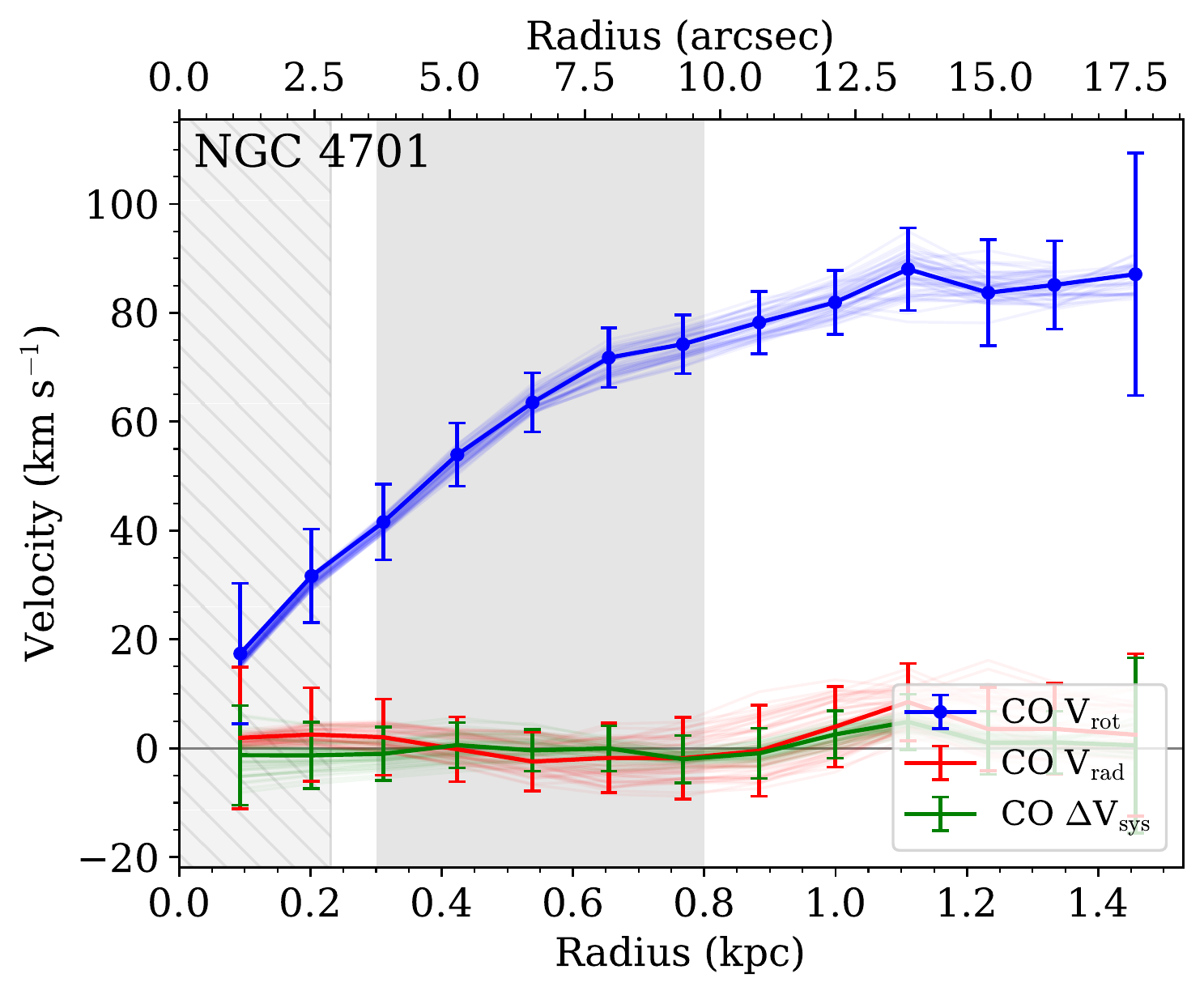}
        
        \includegraphics[width=0.45\textwidth]{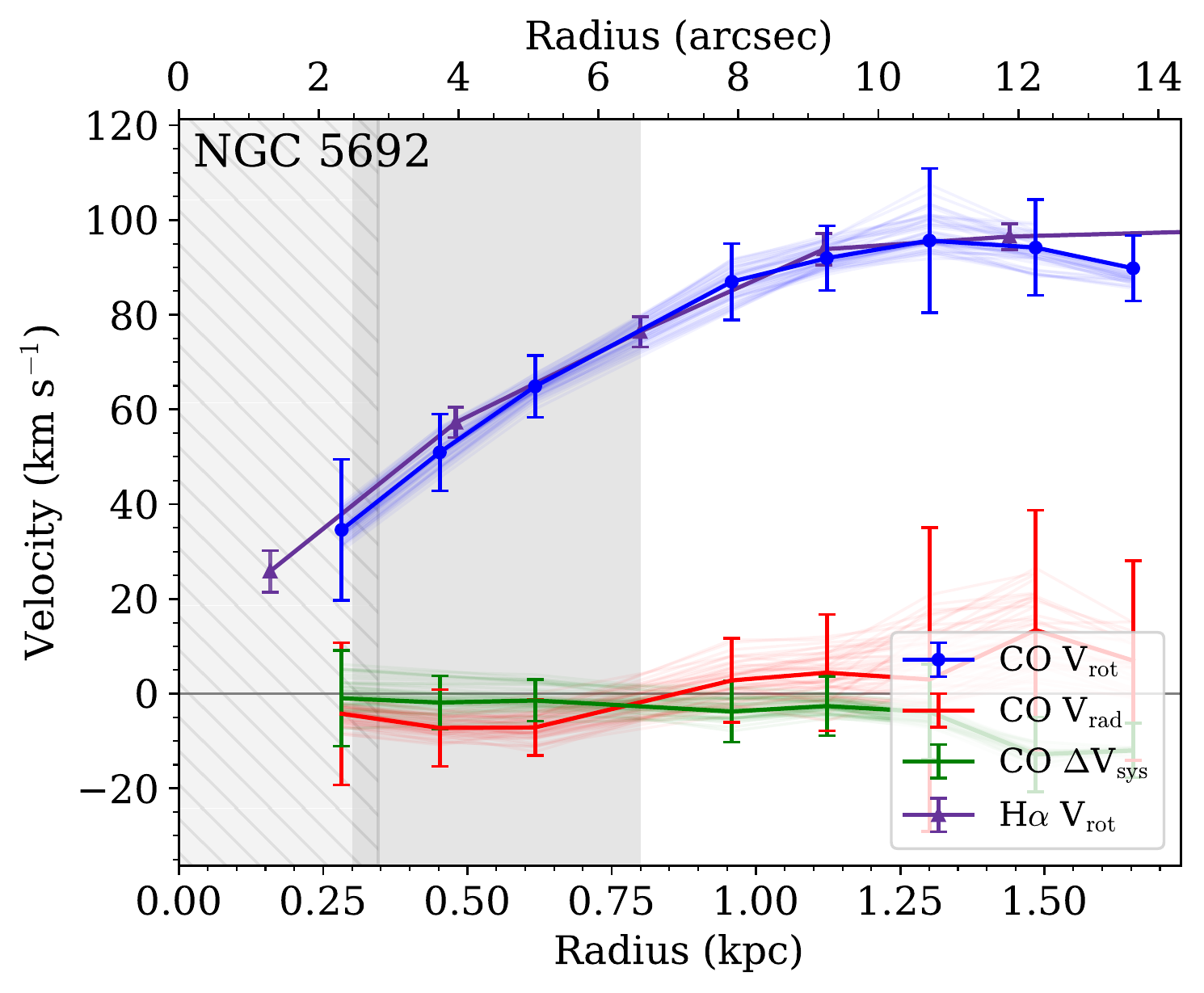}
         \includegraphics[width=0.45\textwidth]{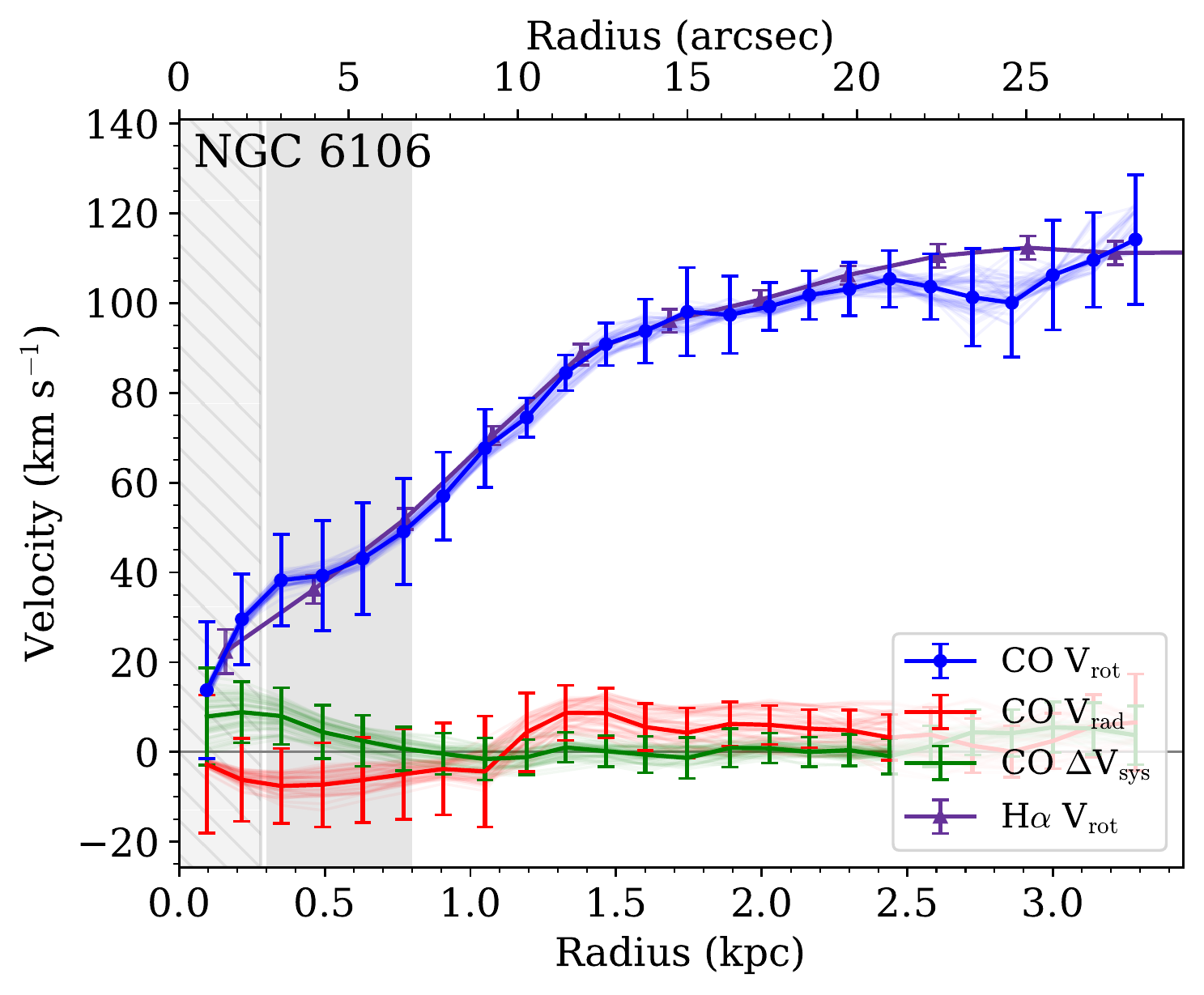}
 \caption{CO rotation curves for the six galaxies studied here using the parameters in {Table~\ref{tab:geomparams}}. The fits use a first order harmonic decomposition {(Equation~\ref{eq:harmdecomp}),} where \vrot\ is shown in blue, \vrad\ in red, and \dvsys\ in green. \change{The thick lines show the best fitting values for each component. The uncertainties shown by the error bars reflect the statistical fitting uncertainties in each ring. The thin translucent lines show the Monte Carlo trials, varying the centre position, PA, and inclination. See {Section~\ref{sec:rcfitting}} for details.} The grey hatched region shows twice the beam HWHM, indicating radii over which the CO measurements may be most affected by beam smearing. The solid grey region shows the range of radii over which \betastar\ \changes{is} calculated. The purple curves show the \ha\ rotation curve derived by \cta{relatores19b}, where available. \change{The error bars on the \ha\ rotation curves reflect the statistical fitting errors.} In general, the CO and \ha\ rotation curves agree well; see {Section~\ref{ssec:galnotes}} for notes on individual galaxies.}
\end{figure*}

The initial values for the kinematic parameters (centre, position angle, inclination, systemic velocity) were taken from \citet{truong17}, but were adjusted \change{by hand} as needed. We modify the centre position of each galaxy to minimize the \dvsys\ component at small radii, as offsets from the correct centre induce sharp deviations from zero in this component very near the centre. We then adjust the systemic velocity value (\vsys) such that \dvsys~$\approx 0$. \change{The \changes{position angle (PA)} takes values between 0$^\circ$ and 360$^\circ$, where PA = 0$^\circ$ indicates that the receding side is oriented north, and increases counterclockwise (east of north).} The PA is modified such that \vrad~$\approx0$ throughout the galaxy, as an incorrect PA \changes{induces} a constant offset in \vrad. \change{For galaxies where there are radial trends in \vrad, the PA is adjusted to minimize \vrad\ in the outer regions of the galaxy. In galaxy centres, there may be non-circular motions due to bars and/or inflows, and the \vrad\ component is sensitive to these non-circular motions. Moreover, as briefly mentioned in \changes{Section~\ref{sec:intro}}, improperly accounting for non-circular motions can affect the derived circular velocity curve and hence the inner slope of the dark matter density profile \citep[e.g.,][]{marasco18,oman19,santos-santos20}.} The inclination was adjusted based on the harmonic fits in each ring, where an incorrect inclination can result in flat-topped or peaked harmonic models whose shapes do not match the data. \change{The PA and inclination are each fixed to a single value at all radii.} Our CO rotation curves are shown in {Figure~\ref{fig:CORCs},} and the final values of the kinematic parameters are listed in {Table~\ref{tab:geomparams}.} \change{We show the residual velocity maps (the difference between the CO velocity field and our best-fitting harmonic decomposition) in {Figure~\ref{fig:velresid}} in { Appendix~\ref{app:rotcurve}}, which were used as additional checks on our geometric parameters \citep[e.g.,][]{vanderkruit78}.}

We determine the uncertainties on the rotation curves as follows. \change{First, we calculate the statistical fitting uncertainty in each annulus. \changes{This includes} the effect of correlations between pixels due to the Gaussian beam. These uncertainties are shown as the error bars in Figure~\ref{fig:CORCs}.} Next, we account for the systematic uncertainties related to the input geometric parameters using a Monte Carlo method. In each realisation, the centre position is allowed to vary uniformly by $\pm0.4\arcsec$, the position angle by $\pm5^\circ$, and inclination by $\pm4^\circ$ from the best-fitting kinematic parameters in {Table~\ref{tab:geomparams}.} These values come from the median differences in the centre position, PA, and inclination used here compared to those kinematically derived by \cta{relatores19a}. We use 50 trials. \change{The fits from each of these trials are shown as the faint thin lines in {Figure~\ref{fig:CORCs}} and their scatter reflects the systematic uncertainties due to the input geometric parameters. At most radii, uncertainties from the fitting in each ring dominates the total uncertainty of the rotation curves. However, systematic uncertainties from changing the geometric parameters can dominate at large radii.} 

The central regions of the rotation curve \changes{are, to some degree, } affected by beam smearing \citep[e.g.,][]{warner73,bosma78,begeman87,teuben02,leung18}. Beam smearing has the effect of artificially lowering the rotation velocity and increasing the velocity dispersion. This effect is worst in the central regions where the rotation curve rises sharply and for highly inclined systems \citep[e.g.,][]{leung18}. For our sample, the effect of beam smearing can be most clearly seen in the linewidth map of NGC\,1035 (upper right-most panel of {Figure~\ref{fig:maps}),} which shows the characteristic $\times$ pattern induced by this effect \citep[e.g.,][]{teuben02}. For this analysis, we do not correct for beam smearing. In {Figure~\ref{fig:CORCs},} the grey hatched regions show twice the beam half-width-half-maximum (HWHM), indicating the radii most affected by beam smearing. As discussed in the following section, the inner slope of the dark matter density profile is calculated from $0.3-0.8$~kpc. There is only minimal overlap between this radius range and where the effects of beam smearing will be most apparent, so we conclude that beam smearing \changes{does} not significantly affect the results presented here.

In order to use CO as the tracer of the total galaxy potential \change{(i.e., to use the CO rotation curve as the circular velocity curve of the total galaxy potential), the CO} must be dynamically cold. To verify this, we find the median CO velocity dispersion for each galaxy (uncorrected for beam smearing), which are listed in {Table~\ref{tab:geomparams}.} The velocity dispersions are quite low and the galaxies have $\frac{\rm V_{rot}}{\sigma_{\rm V}} > 4$, which both indicate that the CO is dynamically cold in these galaxies. We calculate the dynamical mass (M$_{\rm dyn}$) at the measured maximum radius of our CO rotation curve (R$_{\rm max}$). Our measurement of M$_{\rm dyn}$ is a lower limit on the true dynamical mass of these galaxies\changes{, especially }since the rotation curves generally continue to rise as they approach R$_{\rm max}$.

\subsection{Correlations between points and uncertainties in the CO rotation curves}
\label{ssec:co_uncert}

\change{As we will describe in {Section~\ref{ssec:dmdecompmethod}}, the method we use to decompose the dark matter density profiles assumes that the points in the CO rotation curves are independent and that the uncertainties are Gaussian and independent as well.}

\change{The beam (i.e. the point-spread function, PSF) of the CO observations is a Gaussian, and the chosen pixel size of the images oversamples the beam by a factor of $\sim$6. This means that no pixel in the CO map is perfectly independent from any other pixel. Therefore, even though the rings used to derive the rotation curves do not overlap, pixels in those rings are not perfectly independent. The choice to space the rings at half-beam FWHM intervals (i.e., Nyquist sampling) mitigates correlations between rings. As a more direct test, we increased the pixel size of the CO velocity fields to decrease this oversampling and repeated the derivation of the CO rotation curves; the resulting CO rotation curves and uncertainties were essentially unchanged from the original versions. Therefore, correlation between points in the CO rotation curve due to the beam \changes{is} relatively minor, and \changes{we proceed} with the decomposition as if they were perfectly independent ({Section~\ref{ssec:dmdecompmethod}}).}

\change{The Monte Carlo \changes{sampling} we performed over the geometric parameters means that the systematic uncertainties from this method are highly correlated from ring to ring. The PA and inclination drive these correlations: as the PA and/or inclination are changed, the entire rotation curve shifts (primarily) vertically, as shown by the thin faint lines in {Figure~\ref{fig:CORCs}}. These uncertainties from the Monte Carlo \changes{sampling} are, in general, smaller than the uncertainties from the fitting itself, which reflect the statistical fitting uncertainty in each ring. \changes{We use} the statistical fitting uncertainty as the uncertainty on the measured circular velocity curves (of the total galaxy potential) in the decomposition as described in {Section~\ref{ssec:dmdecompmethod}}. \changes{We proceed assuming that the uncertainties in the rotation curves are Gaussian and independent between points.} \changes{We include} the effects of uncertainties in the PA and inclination in the decomposition directly, as we will describe in {Section~\ref{ssec:dmdecompmethod}}.}

\subsection{Notes on individual galaxies}
\label{ssec:galnotes}
Here, we briefly provide notes on each galaxy in this sample. All morphological information is taken from the HyperLEDA catalogue. 

\change{NGC\,1035 ---} This spiral galaxy is the most highly inclined of the subsample. Its CO \vrot\ lies systematically above the \ha\ \vrot\ derived by \cta{relatores19a}, but the shapes of the rotation curves agree well. The shape of the rotation curve is most important for determining the slope of the dark matter \changes{density profile}. \change{The amplitude depends on the inclination. For this galaxy, we use a somewhat higher inclination than \cta{relatores19a} (79\D compared to their 72\D), but this does not completely explain the offsets between the CO and \ha\ rotation curves.}

\change{NGC\,4310 ---} This galaxy is classified as intermediate between a lenticular and spiral galaxy. From the optical and 4.5~$\mu$m data {(Figure~\ref{fig:maps}),} there appears to be a dust lane. The CO traces the extent of the bright optical and NIR emission. Kinematically, this is a very well-behaved galaxy, with only a small \changes{\dvsys} component in the centre. 

\change{NGC\,4451 ---} This spiral galaxy was also observed in \ha, though its mass model was graded as low-quality \cpa{relatores19b}. There is a significant offset between the CO and \ha\ rotation velocities at large radii and the \ha\ rotation velocity is slightly lower than that of the CO overall, which may be due to irregular ionized gas kinematics in the outer regions of this galaxy. In the inner regions, the slopes of the rotation curves agree well. We use a slightly larger inclination compared to \cta{relatores19a} (49\D\ compared to their 45.1\D), but this does not completely explain the small offset between the CO and \ha\ rotation curves at radii less than 1.5~kpc.

\change{NGC\,4701 ---} This spiral galaxy shows no evidence for a bar, though the spiral arms are prominent and strong towards the centre. CO is mainly detected on the blueshifted side of the galaxy. This is not due to a tuning error, but rather due to less signal on the redshifted side of the galaxy. 

\change{NGC\,5692 ---} This spiral galaxy has no evidence for a bar. The agreement between the rotation curves derived from the CO and \ha\ is better than 3~\kms\ on average. We use nearly the same inclination as \cta{relatores19a} (53\D\ compared to their 52.4\D). There is no \spitzer\ IRAC data available for this galaxy, so $r$-band data are used to measure the stellar components instead.

\change{NGC\,6106 ---} \changes{This} spiral galaxy has strong spiral arms, and the CO detections are concentrated in the spiral arms. There is no reported bar, though the spiral arms in the centre may contribute to the non-zero radial velocity component seen in the CO rotation curve fits. This is the largest galaxy in the sample, in terms of angular and physical CO extent. We use a slightly higher inclination of 59\D\ compared to \cta{relatores19a} who found 53.4\D. The agreement between the CO and \ha\ rotation curves is excellent, though there appears to be a slight increase in the CO velocity at very small radii that is not seen in the \ha, though this may be explained by the finer radial sampling of the CO data. 

\section{Determining the Dark Matter Density Profiles}
\label{sec:DMdensity}
Our method for decomposing the contributions of the various tracers to the total \change{circular velocity} curve follows that of \cta{relatores19b}. To summarize, we can determine the dark matter \change{velocity} curve, $V_{\rm DM}(r)$, knowing the \change{circular velocity} curve of the total potential, $V_{\rm tot}(r)$, and the contributions from the stars, $V_{\star}(r)$, atomic gas, $V_{\rm atomic}(r)$, and molecular gas, $V_{\rm mol}(r)$:
\begin{equation}
    \label{eq:modelvcomp}
    V_{\rm tot}(r)^2 = V_{\rm DM}(r)^2 + V_{\star}(r)^2 + V_{\rm atomic}(r)^2 + V_{\rm mol}(r)^2.
\end{equation}
$V_\star(r)$ is calculated from \spitzer\ 4.5$\mu$m or $r$-band photometric data {(Section~\ref{ssec:stellarrotcurves}).} The contributions of the atomic and molecular gas components are negligible compared to the stars and dark matter {(Section~\ref{ssec:atomicandmolecular}).} We include these components when available from \cta{relatores19b}, otherwise their contributions are set to zero. We use the CO rotation curves as our measurement of V$_{\rm tot}(r)$. As described in {Section~\ref{sec:rcfitting},} the measured CO velocity dispersions are low {(Table~\ref{tab:geomparams}),} even without a correction for beam smearing, meaning that the CO is a robust, dynamically cold tracer of the total potential. 

Therefore, having measured the total potential, the stellar component, and the contributions from the atomic and molecular components where available, we can solve for the contribution from the dark matter. As we describe in {Section~\ref{ssec:dmdecompmethod},} we use a forward-modelling \change{Markov Chain Monte Carlo (MCMC)} approach to measure the dark matter component. From the best-fitting dark matter density profile, we compute \betastar, the slope of the dark matter density profile from $0.3-0.8$~kpc.

\subsection{Stellar mass and velocity profiles}
\label{ssec:stellarrotcurves}

\begin{figure*}
\label{fig:stellarRCcomp}
    \centering
        \includegraphics[width=0.45\textwidth]{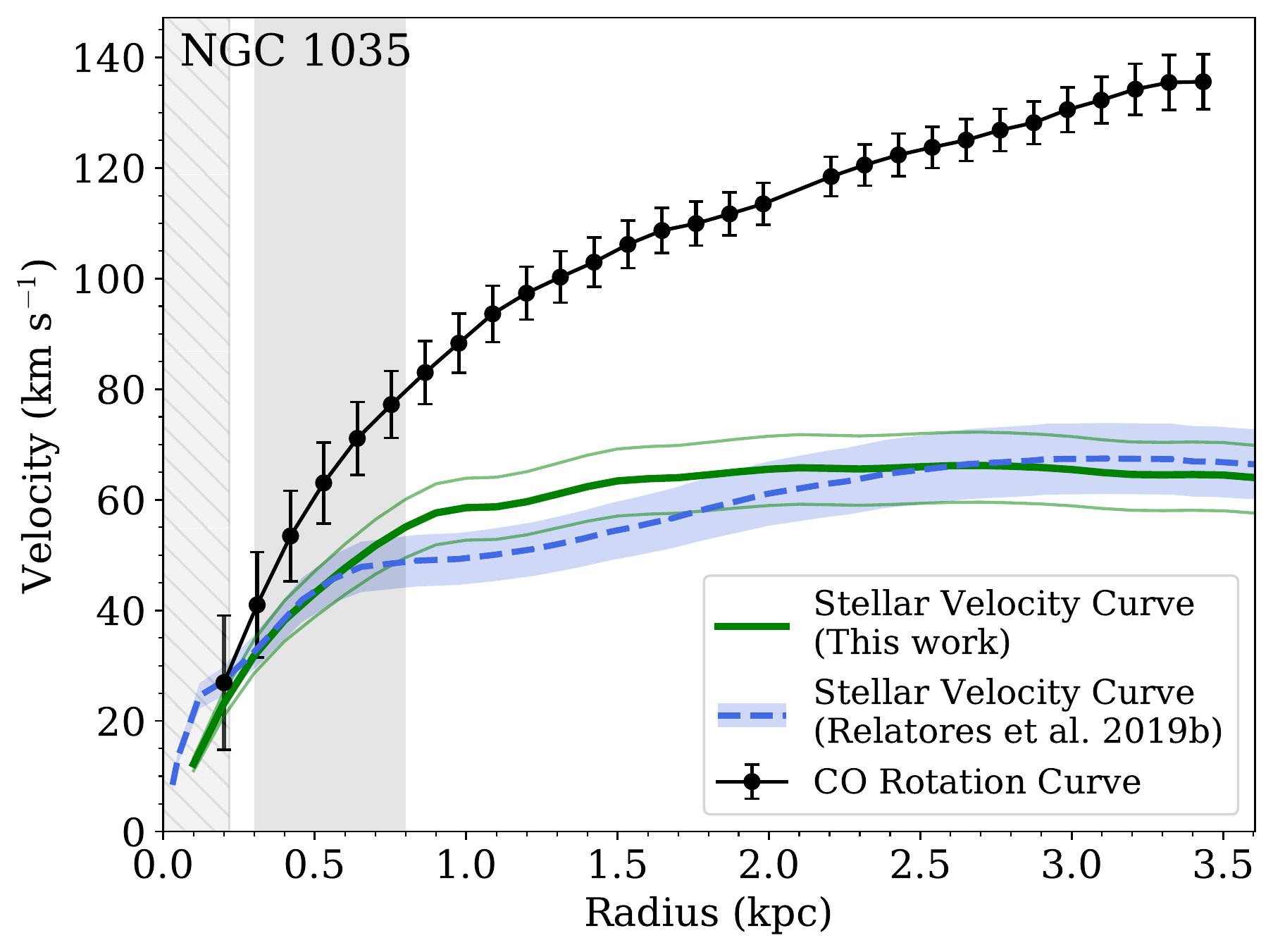}
        \includegraphics[width=0.45\textwidth]{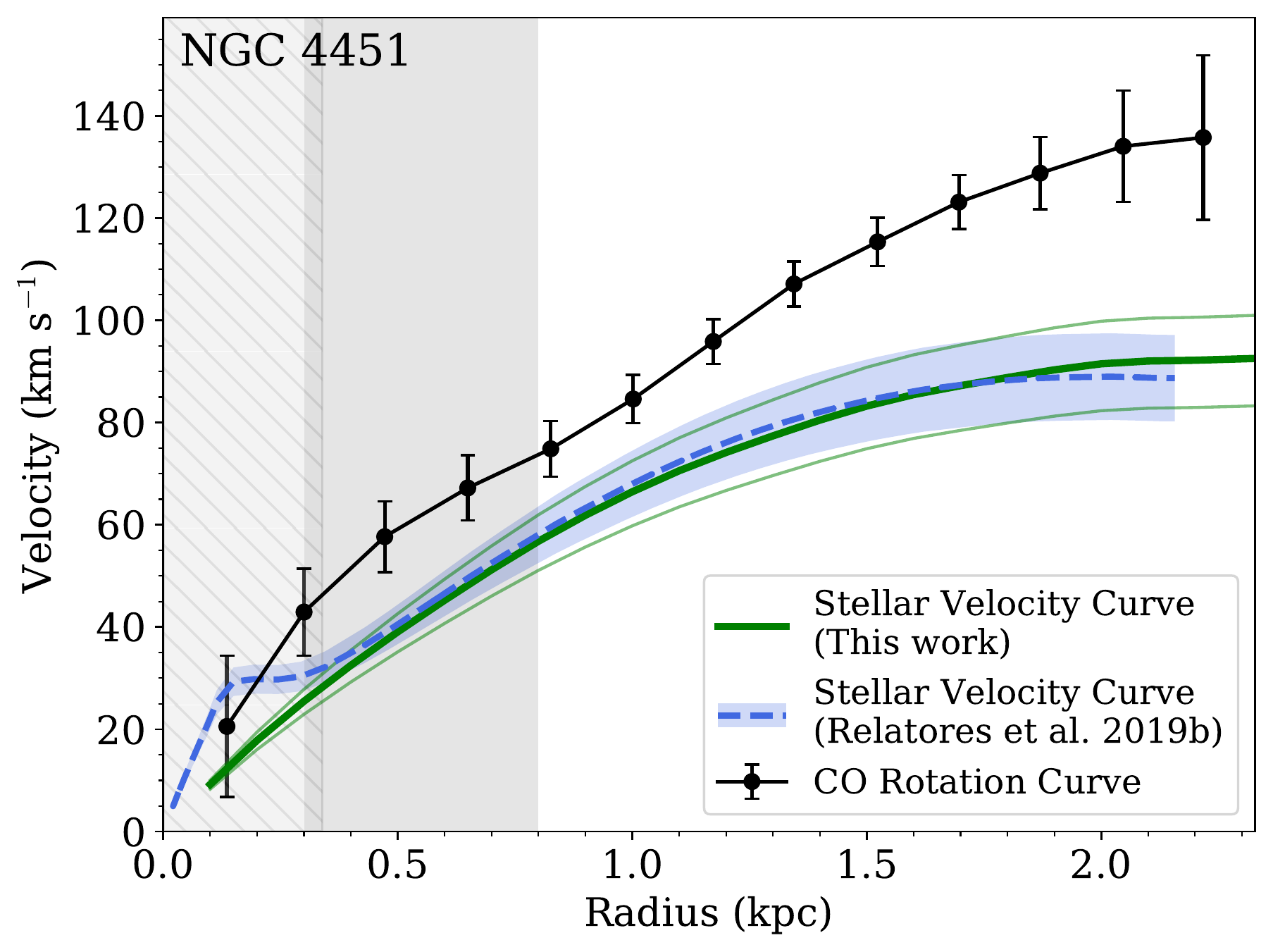}

        \includegraphics[width=0.45\textwidth]{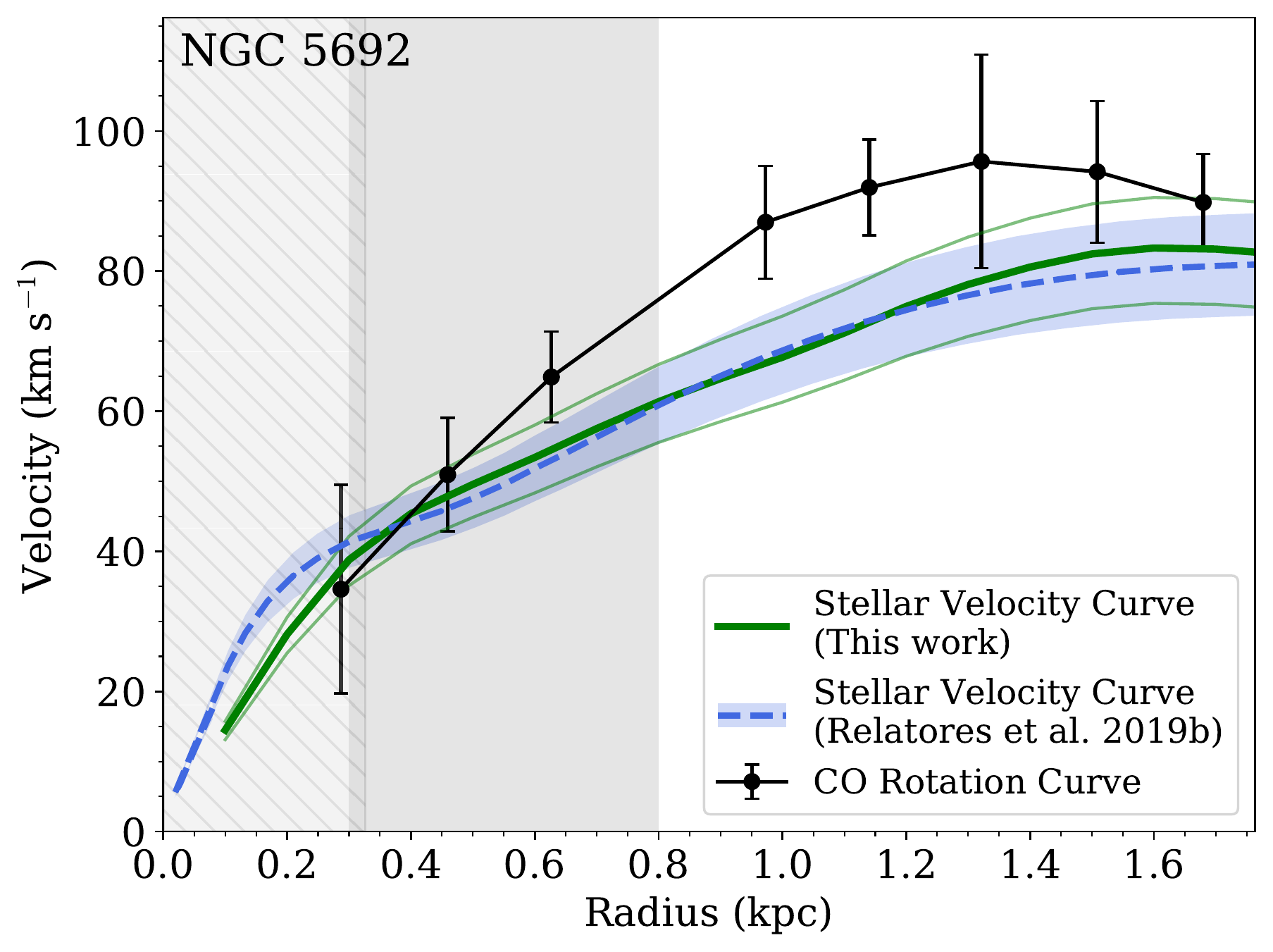}
         \includegraphics[width=0.45\textwidth]{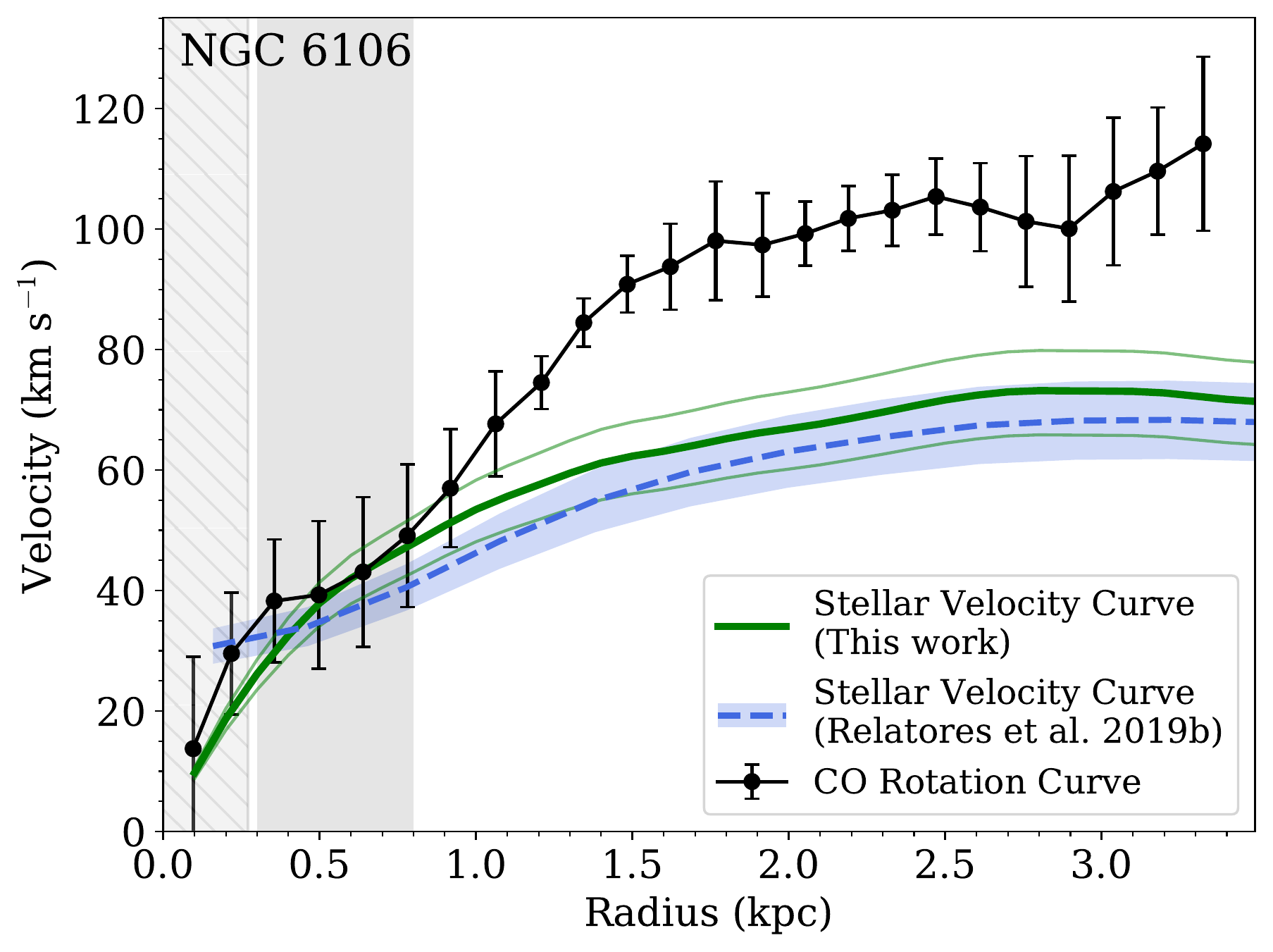}
 \caption{A comparison of the stellar \changes{velocity} curves derived as described in {Section~\ref{ssec:stellarrotcurves}} (green) compared to those derived by \cta{relatores19b} (blue dashed). \change{The thin green curves show the effect of changing the \ML\ within the range listed in {Table~\ref{tab:decompparams}}, whereas the thick green curve shows the fiducial value of \ML.} We also show the CO rotation curves (black) which trace the total potential. In general, the agreement between these methods is good, especially in the radial regime where $\beta^*$ \changes{is} calculated (grey shaded region). The grey hatched region shows twice the CO beam HWHM, indicating the central regions in which the CO rotation velocities may be affected by beam smearing.}
\end{figure*}

We derive the stellar \change{circular velocity} curves starting from the \textit{Spitzer} IRAC 4.5$\mu$m maps {(Figure~\ref{fig:maps}).} As discussed by \cta{relatores19a} and \cta{relatores19b}, the 4.5$\mu$m data are preferred since they more closely trace the stellar mass distribution than the $r$-band data. Using the geometric parameters listed in {Table~\ref{tab:geomparams},} the 4.5$\mu$m maps were divided into annuli centred on the centre of the galaxy. For radii that overlap with the CO rotation curves, the radial binning matches that of the CO. For radii beyond the CO extent, the annuli are 5 pixels wide and extend out to a radius ${\rm R_{max,\star}}$ (listed in {Table~\ref{tab:decompparams}).} The cumulative stellar mass profiles do not continue to rise substantially at ${\rm R_{max,\star}}$, so stellar masses calculated at this radius encompass nearly all of the stellar mass of the galaxies. The surface brightness was summed in each annulus and converted to a luminosity. To convert to solar luminosities, we find the luminosity of the Sun at 4.5 $\mu$m using the IRAC Channel 2 zero-point \citep[0 Vega mag = 179.7 Jy;][]{reach05} and the absolute magnitude of the Sun at 4.5 $\mu$m \citep[$M_{\odot}^{4.5} = 3.27$ Vega mag;][]{oh08}. This yields $L_{\nu,\odot}^{4.5} = 1.06\times10^{18} {\rm\ erg\ s^{-1}\ Hz^{-1}}$ or $\nu L_{\nu,\odot}^{4.5} = 7.05\times10^{31} {\rm\ erg\ s^{-1}} = 0.018\ L_{\odot}$. To convert to mass, we assume a constant mass-to-light ratio (\ML) from stellar population synthesis models reported by \cta{relatores19b} (see their Table 1). For those galaxies that do not overlap with \cta{relatores19b}'s sample, we take their average \ML~$= 0.21 \pm 0.04$. We report the \ML\ value\changes{s} used in this work in {Table~\ref{tab:decompparams}.} 

\textit{Spitzer} IRAC 4.5 $\mu$m data are not available for NGC\,5692. Like \cta{relatores19b}, we instead use $r$-band images from PS1 {(Figure~\ref{fig:maps})} and employ the same method as described above to calculate the surface brightness profiles. The filters on PS1 are quite similar to the corresponding Sloan Digital Sky Survey (SDSS) filters \citep{tonry12}. The absolute magnitude of the Sun is 4.65 AB magnitudes in the SDSS $r$-band \citep{willmer18}, and the effective central wavelength of the PS1 $r$-band filter is 617 nm \citep{tonry12}. This yields $L_{\nu,\odot}^{r} = 5.99\times10^{18} {\rm\ erg\ s^{-1}\ Hz^{-1}}$ or $\nu L_{\nu,\odot}^{r} = 2.91\times10^{33} {\rm\ erg\ s^{-1}} = 0.746\ L_\odot$. To convert to mass, we use the $r$-band \ML\ from stellar population synthesis models derived by \cta{relatores19b} (\ML~$= 0.83\pm0.15$; see their {Table~1} and our {Table~\ref{tab:decompparams}).} 

We use \nemo\ \citep{teuben95} to determine the \change{stellar velocity component} from the extracted \change{stellar} surface brightness profiles. We assume that the stars are in a thin disk. In reality, the stars are likely in an oblate spheroid distribution. However, the axis ratios of stellar distributions tend to be small \citep[e.g.,][]{kregel02}, and all of the galaxies used here are classified as spirals. For their modelling, \cta{relatores19b} assume an oblate spheroid with $c/a = 0.14$ motivated by observations. Where possible, we compare our derived stellar \change{circular velocity} curves (assuming a thin disk) to those derived by \cta{relatores19b} {(Figure~\ref{fig:stellarRCcomp}).} In general, the agreement (both in terms of shape and amplitude) are good, especially over the range of radii where \betastar\ \changes{is} calculated. The stellar \change{circular velocity} curves derived by \cta{relatores19b} tend to be higher than ours in the centres, which may be due to our coarser binning. The total stellar masses of the galaxies in our sample (listed in {Table~\ref{tab:decompparams})} agree with those measured by \cta{relatores19b} within the uncertainties, where available.  

\change{If the stellar velocity component is higher than the circular velocity based on the CO rotation curve}, we fix $V_\star$ at those radii to a maximal disk (i.e. where the stellar component accounts for the total potential). We increase the uncertainty on the stellar \change{component} in these bins by adding the velocity difference to the existing uncertainty in quadrature. The first five bins in NGC\,4701 and the first bin in NGC\,5692 are affected by this change. 

\subsubsection{Correlations between points in the stellar rotation curves}
\label{sssec:stellar_uncert}

\change{As we will describe in {Section~\ref{ssec:dmdecompmethod}}, the points in the stellar rotation curve are assumed to be independent, and the uncertainties are assumed to be Gaussian and independent as well. Similar to the CO maps (Section~\ref{ssec:co_uncert}), the PSF of the {\em Spitzer} 4.5$\mu$m observations \changes{induces some} correlation between pixels in the maps. The {\em Spitzer} PSF at 4.5$\mu$m has a FWHM of $1.6$\arcsec\ \citep{sheth10}, about half that of the Gaussian CO beams (Table~\ref{tab:obs}). Where the decomposition can be performed, the stellar rotation curve rings have the same spacing as the CO. Therefore, \changes{points in the stellar rotation curves are} even less correlated than in the case of the CO rotation curves. \changes{We proceed} with the decomposition as if they were perfectly independent (Section~\ref{ssec:dmdecompmethod}). }

\subsection{Atomic and molecular gas components}
\label{ssec:atomicandmolecular}

For the four galaxies that overlap with \cta{relatores19b}, we also include contributions from the atomic and molecular gas in the decomposition. We refer the reader to that paper for a detailed discussion of how these components were measured. We do not include uncertainties on the atomic and molecular components for this analysis (nor are they given by \cta{relatores19b}). For the other two galaxies, however, we assume that these components are a negligible contribution to the mass. As we will discuss in {Section~\ref{ssec:checks},} including or excluding these components does not affect the final results.

\subsection{Decomposing the dark matter component}
\label{ssec:dmdecompmethod}

\begin{figure*}
\label{fig:decomp}
    \centering
        \includegraphics[width=0.45\textwidth]{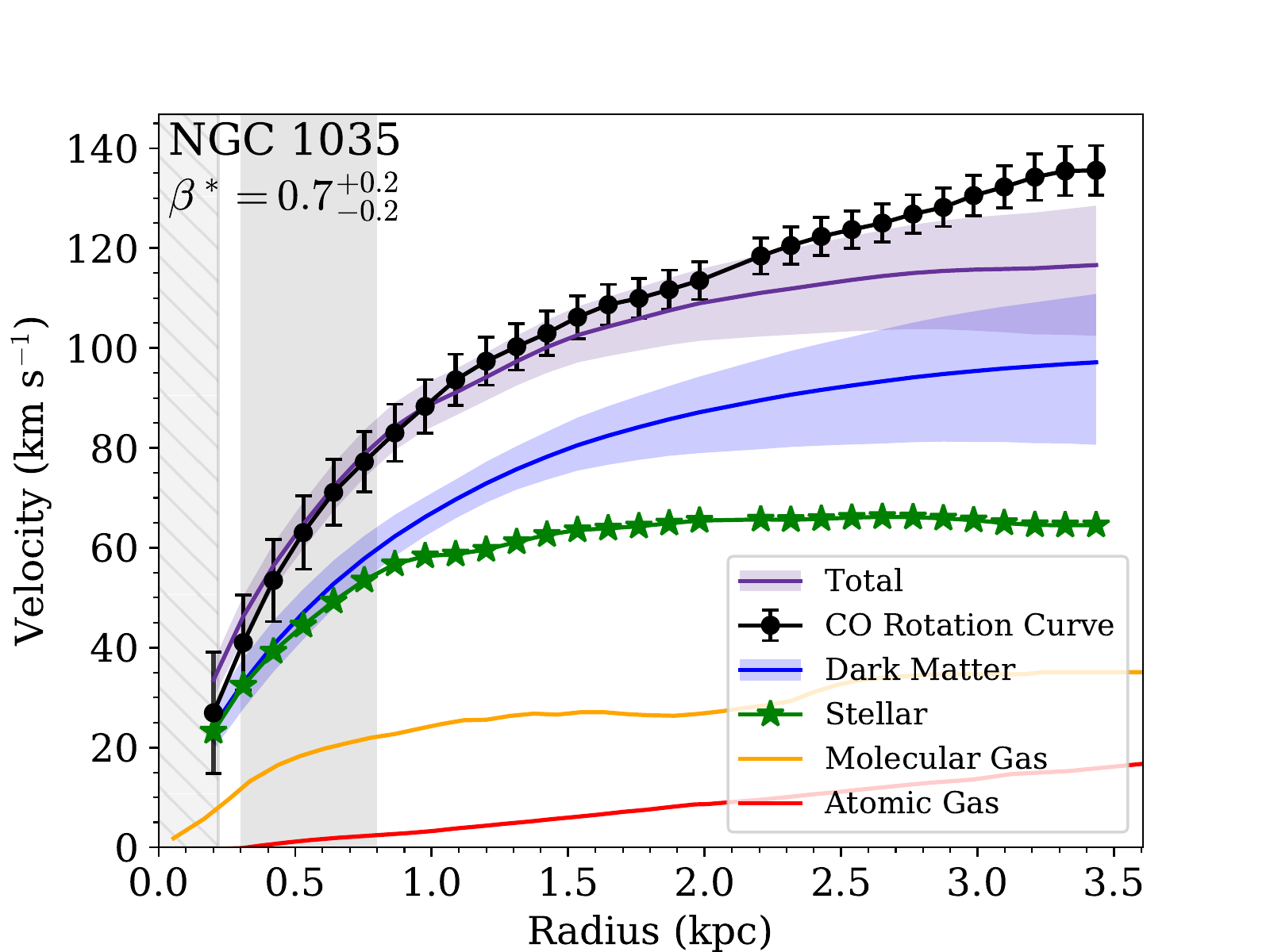}
        \includegraphics[width=0.45\textwidth]{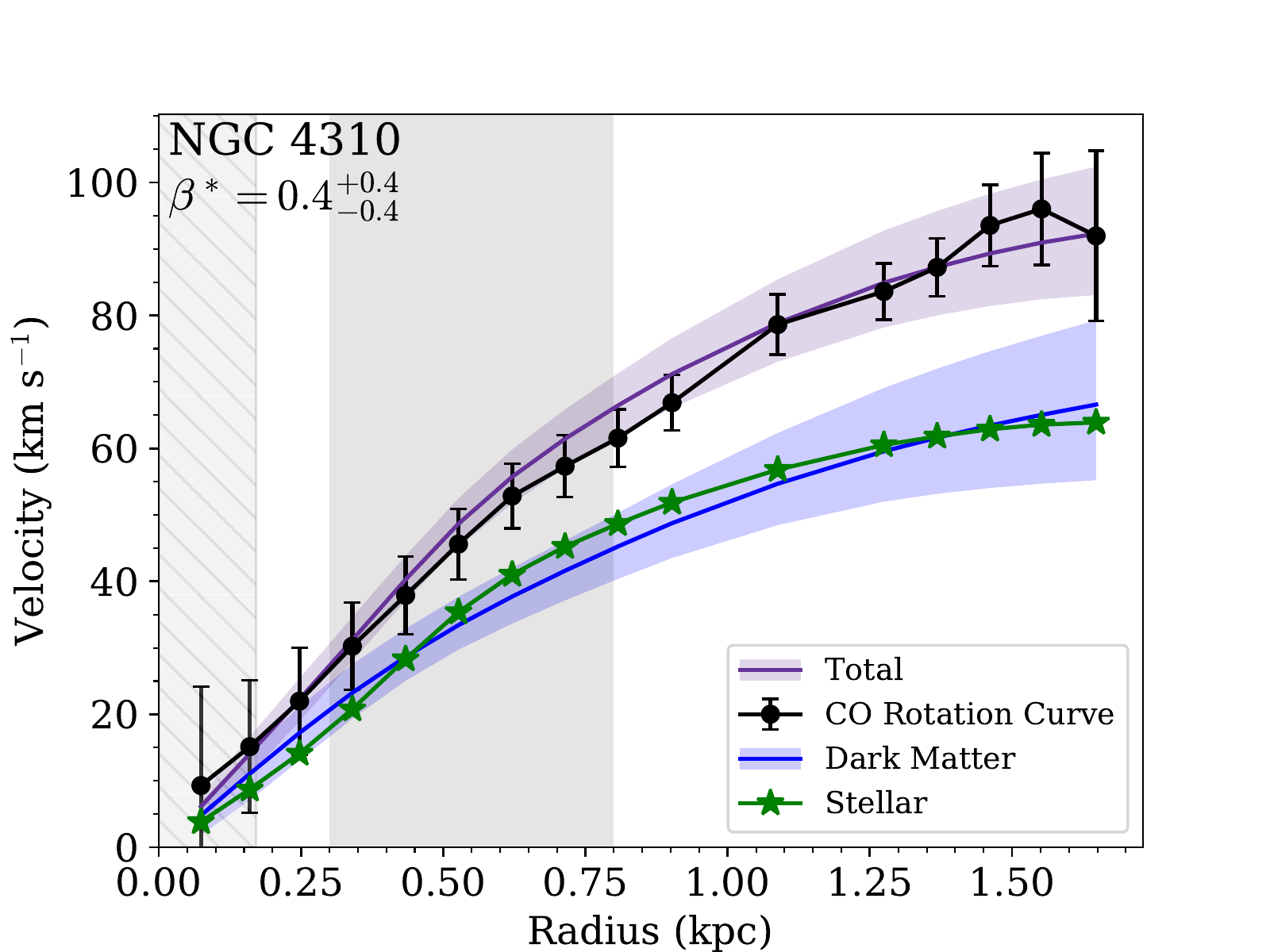}
       
        \includegraphics[width=0.45\textwidth]{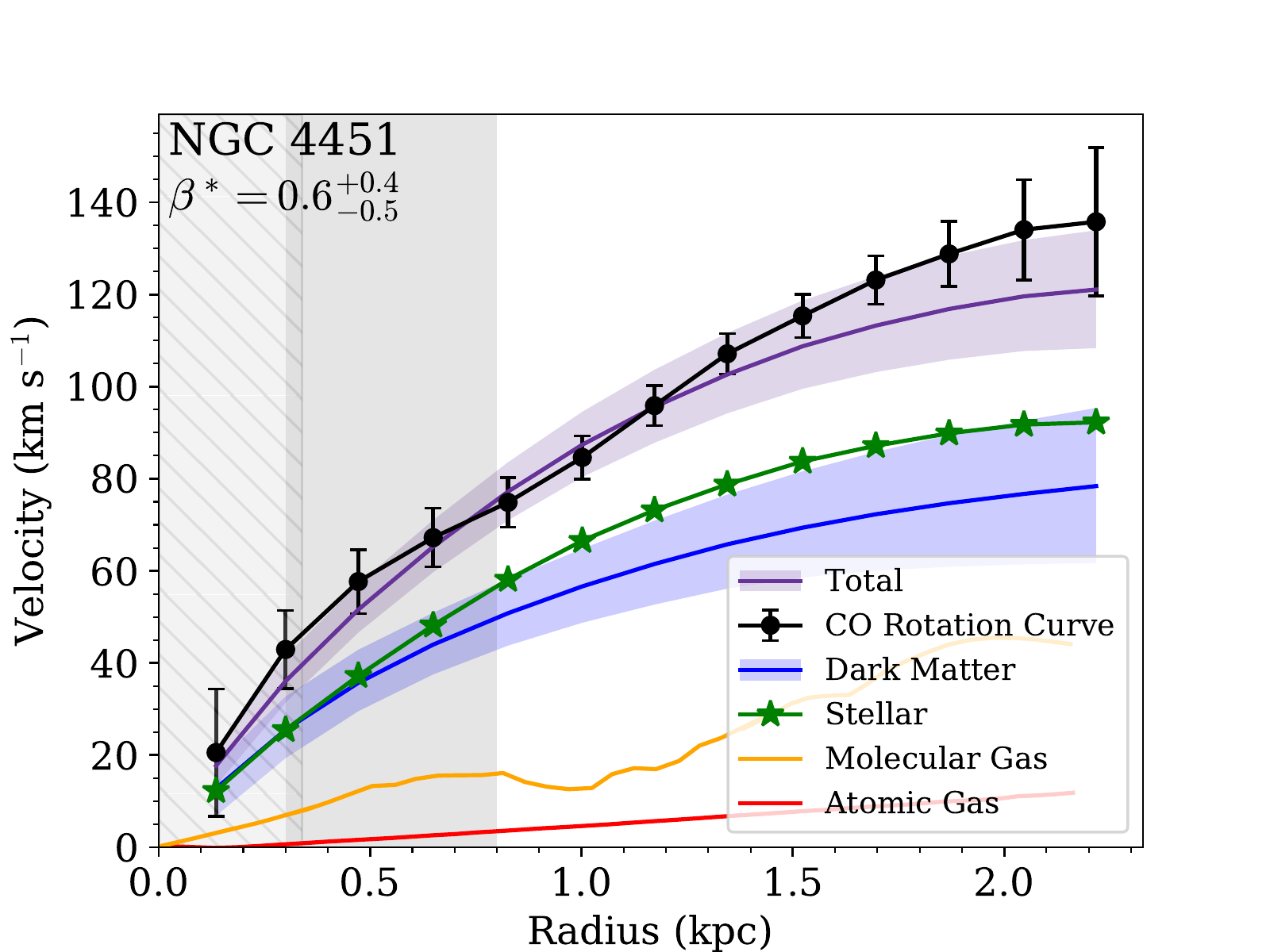}
        \includegraphics[width=0.45\textwidth]{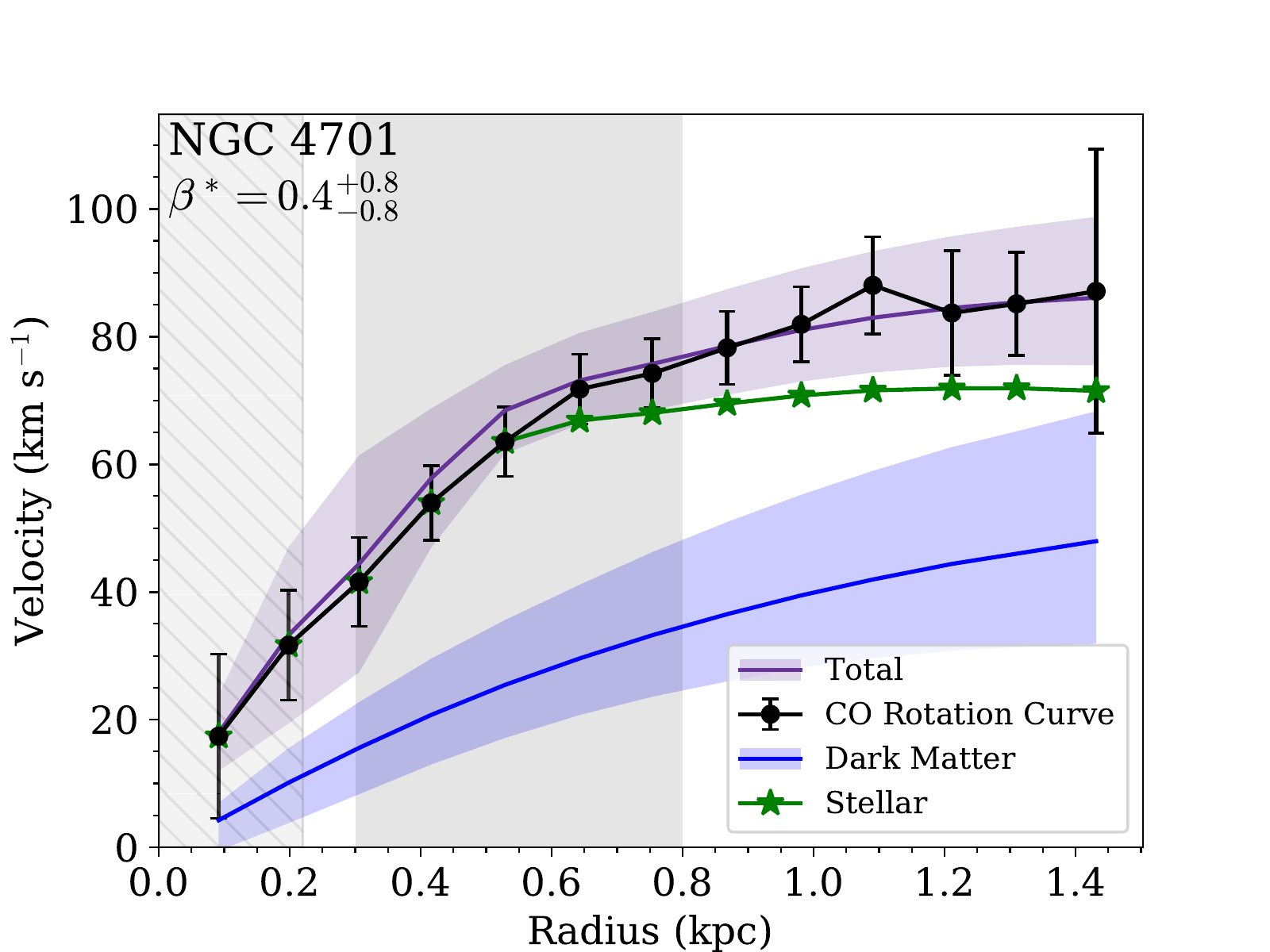}
        
        \includegraphics[width=0.45\textwidth]{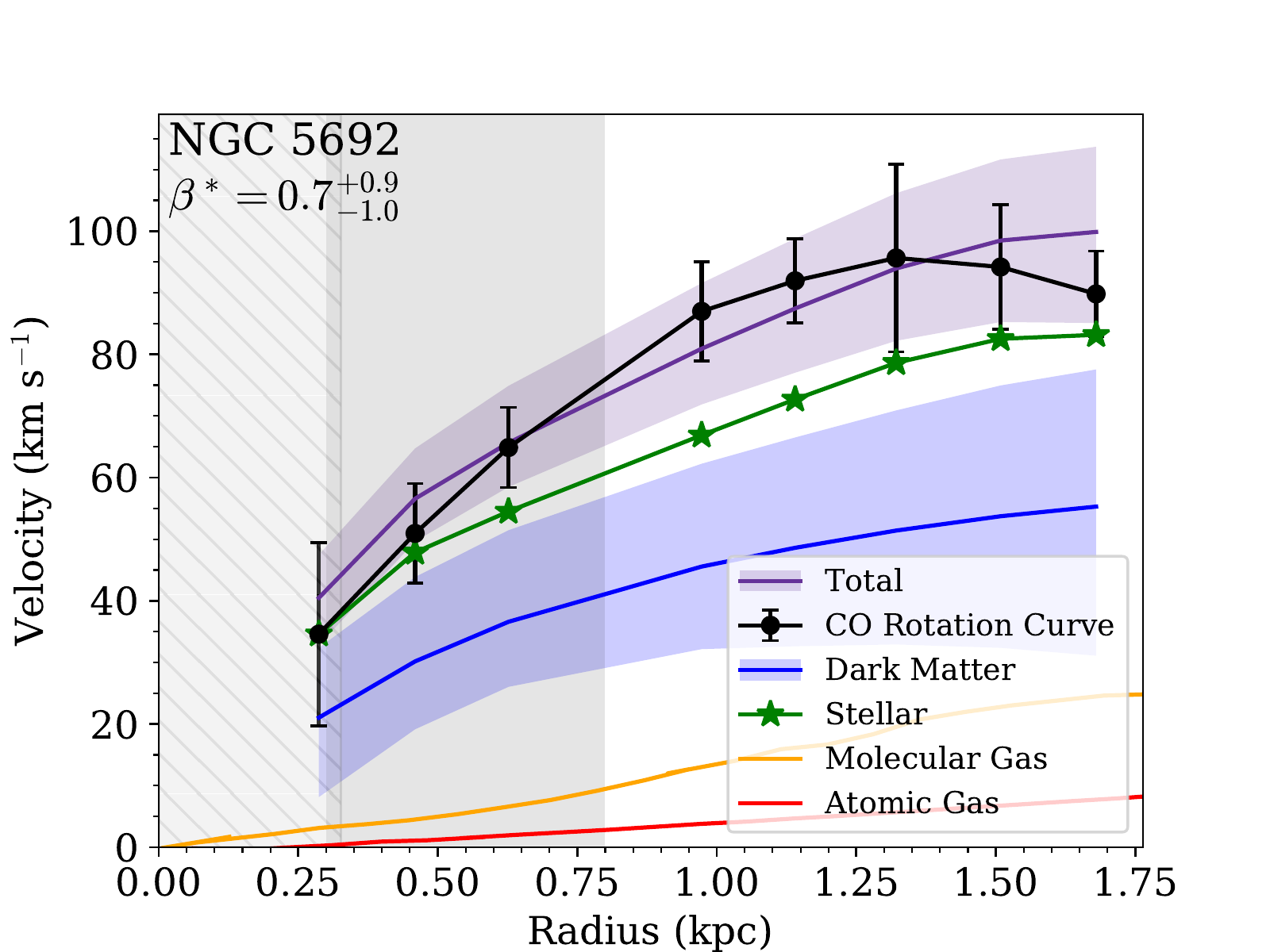}
        \includegraphics[width=0.45\textwidth]{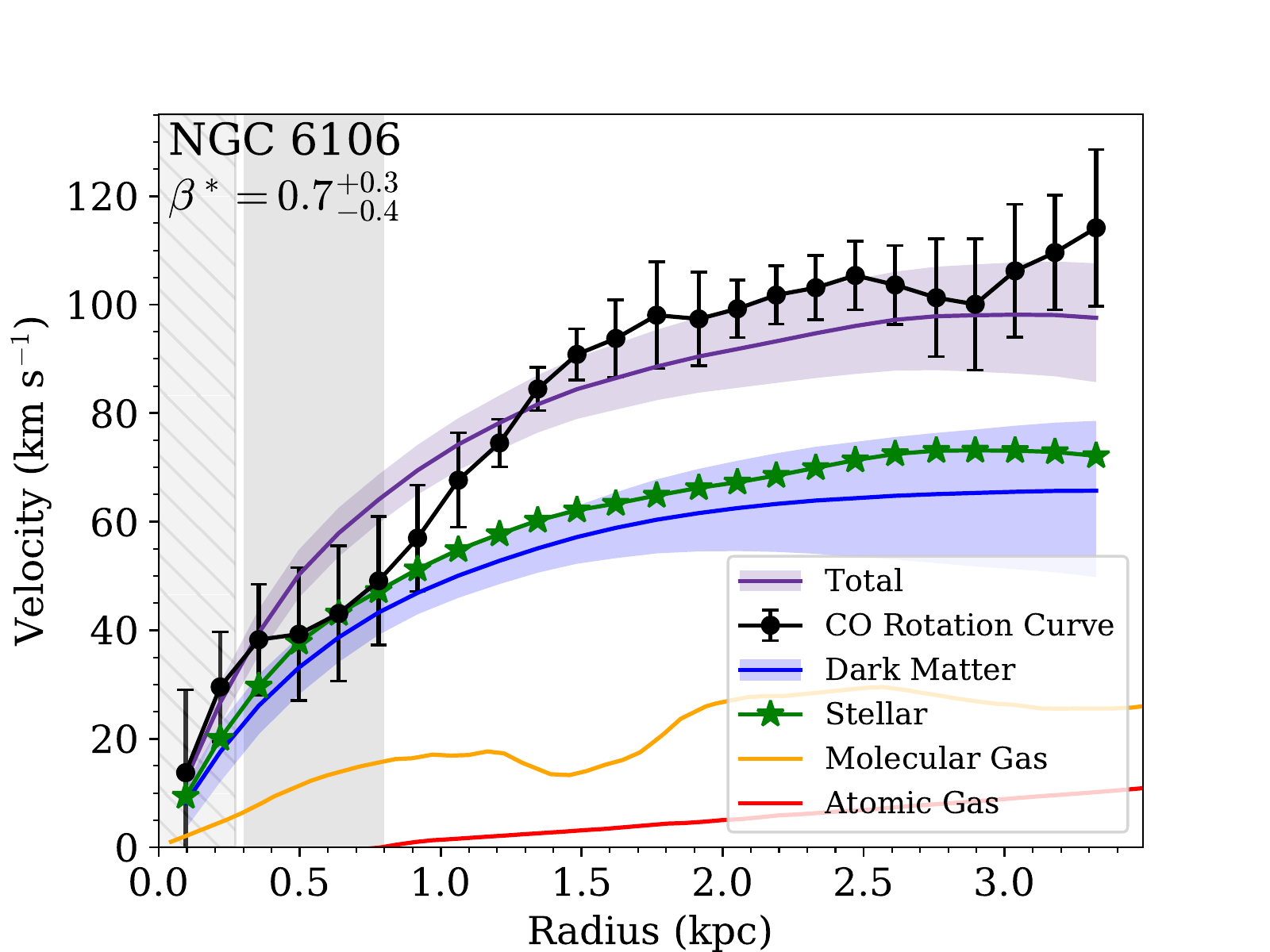}
 \caption{\change{The results of the density profile decomposition, plotted in terms of the tangential velocities that correspond to the mass distribution of each component.} The CO rotation curve (tracing the total potential) is shown as the black dots with error bars {(Section~\ref{sec:rcfitting})}. \change{The green stars show the measured stellar velocity curve for the fiducial value of \ML\ listed in {Table~\ref{tab:decompparams}} {(Section~\ref{ssec:stellarrotcurves}}). The mass contributions from the atomic and molecular components (if available) are shown in red and orange {(Section~\ref{ssec:atomicandmolecular}).}} The blue curves and shaded regions show the dark matter component, and the purple curves and shaded regions show the total modelled rotation curves of all the components {(Equation~\ref{eq:modelvcomp}),} which should closely match the CO. The grey hatched region shows twice the CO beam HWHM, indicating the central regions which may be affected by beam smearing. The grey shaded region shows the radii over which \betastar\ is calculated {(Equation~\ref{eq:betastar}),} and the best-fitting \betastar\ values are listed in the upper left corners.}
\end{figure*}

\begin{table*}
\begin{center}
\caption{Fitted Stellar and Dark Matter Density Parameters}
\label{tab:decompparams}
\begin{tabular}{cccccccccc}
\hline
Name & $\Upsilon_*$ & R$_{\mathrm{max},*}$  & logM$_*(<{\mathrm{R_{max,*}}})$ & R$_{\mathrm{vir}}$ & 1$-$2\% R$_{\mathrm{vir}}$& logM$_{200}$ & $c_{200}$ & $\beta^*$ & \betarvir\\
 & & (kpc) & (logM$_{\odot}$) & (kpc) & (kpc) & (logM$_{\odot}$) & & & \\
\hline
NGC\,1035 & 0.21$\pm$0.04 & 6.0 & 9.5 & 79{\raisebox{0.5ex}{\tiny$^{+49}_{-21}$}} & 0.8{\raisebox{0.5ex}{\tiny$^{+0.5}_{-0.2}$}}$-$1.6{\raisebox{0.5ex}{\tiny$^{+1.0}_{-0.4}$}} & 11.4{\raisebox{0.5ex}{\tiny$^{+2.0}_{-0.3}$}} & 27{\raisebox{0.5ex}{\tiny$^{+22}_{-13}$}} & 0.7{\raisebox{0.5ex}{\tiny$^{+0.2}_{-0.2}$}} & 1.1{\raisebox{0.5ex}{\tiny$^{+0.6}_{-0.4}$}} \\
NGC\,4310 & 0.21$\pm$0.04 & 6.2 & 9.3 & 64{\raisebox{0.5ex}{\tiny$^{+30}_{-15}$}} & 0.6{\raisebox{0.5ex}{\tiny$^{+0.3}_{-0.1}$}}$-$1.3{\raisebox{0.5ex}{\tiny$^{+0.6}_{-0.3}$}} & 11.2{\raisebox{0.5ex}{\tiny$^{+1.4}_{-0.3}$}} & 36{\raisebox{0.5ex}{\tiny$^{+23}_{-13}$}} & 0.4{\raisebox{0.5ex}{\tiny$^{+0.4}_{-0.4}$}} & 0.8{\raisebox{0.5ex}{\tiny$^{+0.7}_{-0.6}$}} \\
NGC\,4451 & 0.21$\pm$0.04 & 9.7 & 9.7 & 90{\raisebox{0.5ex}{\tiny$^{+29}_{-14}$}} & 0.9{\raisebox{0.5ex}{\tiny$^{+0.3}_{-0.1}$}}$-$1.8{\raisebox{0.5ex}{\tiny$^{+0.6}_{-0.3}$}} & 11.5{\raisebox{0.5ex}{\tiny$^{+0.9}_{-0.2}$}} & 40{\raisebox{0.5ex}{\tiny$^{+20}_{-14}$}} & 0.6{\raisebox{0.5ex}{\tiny$^{+0.4}_{-0.5}$}} & 1.1{\raisebox{0.5ex}{\tiny$^{+0.5}_{-0.6}$}} \\
NGC\,4701 & 0.21$\pm$0.04 & 6.5 & 9.4 & 65{\raisebox{0.5ex}{\tiny$^{+40}_{-14}$}} & 0.7{\raisebox{0.5ex}{\tiny$^{+0.3}_{-0.2}$}}$-$1.3{\raisebox{0.5ex}{\tiny$^{+0.8}_{-0.3}$}} & 11.1{\raisebox{0.5ex}{\tiny$^{+2.3}_{-0.3}$}} & 45{\raisebox{0.5ex}{\tiny$^{+43}_{-24}$}} & 0.4{\raisebox{0.5ex}{\tiny$^{+0.8}_{-0.8}$}} & 0.9{\raisebox{0.5ex}{\tiny$^{+1.0}_{-1.0}$}} \\
NGC\,5692 & 0.83$\pm$0.15 & 4.8 & 9.3 & 75{\raisebox{0.5ex}{\tiny$^{+43}_{-12}$}} & 0.8{\raisebox{0.5ex}{\tiny$^{+0.4}_{-0.2}$}}$-$1.5{\raisebox{0.5ex}{\tiny$^{+0.9}_{-0.2}$}} & 11.3{\raisebox{0.5ex}{\tiny$^{+2.3}_{-0.2}$}} & 46{\raisebox{0.5ex}{\tiny$^{+47}_{-25}$}} & 0.7{\raisebox{0.5ex}{\tiny$^{+0.9}_{-1.0}$}} & 1.2{\raisebox{0.5ex}{\tiny$^{+1.0}_{-1.1}$}} \\
NGC\,6106 & 0.21$\pm$0.04 & 8.9 & 9.7 & 68{\raisebox{0.5ex}{\tiny$^{+115}_{-12}$}} & 0.7{\raisebox{0.5ex}{\tiny$^{+1.1}_{-0.1}$}}$-$1.4{\raisebox{0.5ex}{\tiny$^{+2.3}_{-0.3}$}} & 11.2{\raisebox{0.5ex}{\tiny$^{+15.5}_{-0.2}$}} & 47{\raisebox{0.5ex}{\tiny$^{+140}_{-20}$}} & 0.7{\raisebox{0.5ex}{\tiny$^{+0.3}_{-0.4}$}} & 1.1{\raisebox{0.5ex}{\tiny$^{+1.1}_{-0.5}$}} \\
\hline
\end{tabular}
\end{center}

\justifying\noindent{$\Upsilon_*$ is the mass-to-light ratio assumed here reproduced from the stellar population synthesis model results presented by \cta{relatores19b}. M$_*$ is the total stellar mass within a radius R$_{\mathrm{max},*}$ from the fitting described in \mbox{Section \ref{ssec:stellarrotcurves}}; uncertainties are $\pm$0.1 dex. R$_{\mathrm{vir}}$ is the virial radius derived from the best-fitting total density profile, at which the density is 200$\times$ the critical density. Uncertainties are propagated from the uncertainties on the total density profile. M$_{200}$ is the mass contained within R$_{200}\equiv{\mathrm{R_{vir}}}$. Uncertainties are propagated from ${\mathrm{R_{vir}}}$. $c_{200}$ is the concentration parameter, defined as R$_{200}$/$r_s$. Uncertainties are propagated from ${\mathrm{R_{vir}}}$ and $r_s$. \betastar\ is the inner slope of the dark matter density profile, as defined in Equation \ref{eq:betastar}. \change{\betarvir\ is the inner slope of the dark matter density profile measured from 1$-$2\% of R$_{\mathrm{vir}}$.}}
\end{table*}

Knowing the contributions from the stars (and atomic and molecular gas, if available) and the total potential, we can derive the dark matter density profile, following {Equation~\ref{eq:modelvcomp}.} We take a forward-modelling approach, generating trials of the dark matter density profile and minimizing the difference between the model total circular velocity curve and that observed from the CO as described below.

\changes{We assume that the dark matter density profile is described by a} generalised Navarro-Frenk-White (gNFW; \citealt{zhao96}) profile of the form
\begin{equation}
    \label{eq:NFW}
    \rho_{\rm DM}(r) = \frac{\rho_o}{(r/r_s)^\beta(1+r/r_s)^{3-\beta}}
\end{equation}
where $r_s$ is the scale radius, $\rho_o$ is the characteristic density, and $\beta$ sets the slope as $r\rightarrow0$. Setting $\beta=1$ recovers the original NFW profile.

We transform this density profile into its corresponding \change{velocity} curve, $V_{\rm DM}(r)$, where
\begin{equation}
    \label{eq:densityvelocity}
    V(r)^2 = \frac{4\pi G}{r}\int_0^r \mathcal{r}^{2}\rho(\mathcal{r})d\mathcal{r}.
\end{equation}

We use an affine invariant MCMC method \citep[\emcee;][]{emcee} to draw samples of $\rho_o$, $r_s$, and $\beta$ \change{(as well as \ML, PA, and inclination as described below)} to generate trials of the dark matter density and associated \change{velocity} curve {(Equations~\ref{eq:NFW}} and {\ref{eq:densityvelocity}).} The uniform priors on \change{$\rho_o$, $r_s$, and $\beta$} are set such that $0 < \frac{\rho_o}{{\rm M_\odot~pc^{-3}}} < 1.0 $, $0 < \frac{r_s}{{\rm kpc}} < 10$, and $\change{-2.5} < \beta < 2.5$.

\change{As described in Section~\ref{sssec:stellar_uncert}, \ML\ is included as a parameter in the model which scales the stellar velocity curve as $\Upsilon_\star^{\frac{1}{2}}$. Therefore, $V_\star(r|\Upsilon_\star)$ explicitly. The uniform priors on \ML\ are given by the ranges in Table~\ref{tab:decompparams}.}

\change{There are known degeneracies between the geometric parameters and properties of the dark matter halo (for example, between the halo mass and inclination; e.g., \citealt{maller97,oguri05,read16}). We therefore include the PA and inclination as (nuisance) parameters to be fit in the MCMC. Rather than re-fitting the rotation curve at each step in the MCMC, which would be computationally expensive, we first generate a grid of rotation curves by varying the PA and inclination. We allow the \changes{priors on the} PA (inclination) to vary uniformly by $\pm5^\circ$ ($\pm4^\circ$) from the best-fitting value in steps of $1^\circ$, which is based on the range from the Monte Carlo analysis discussed in {Section~\ref{sec:rcfitting}}. We use the grid of rotation curves as a lookup table, choosing the CO rotation curve with the nearest PA and inclination to each MCMC trial, acknowledging that there \changes{is} some (small) error introduced by discretising the PA and inclination in this way. }

\change{Our likelihood function is given by
\begin{equation}
    \label{eq:likelihood}
    \ln\mathcal{L} = \frac{-1}{2}\sum_i\frac{ \left(V_{{\rm CO},i}-V_{{\rm tot},i} \right)^2}{\sigma_{{\rm CO},i}^2}.
\end{equation}
where $V_{\rm tot}$ is our model of the total circular velocity curve of the galaxy given by {Equation~\ref{eq:modelvcomp}}. As described in {Section~\ref{ssec:co_uncert}}, $\sigma_{\rm CO}$ are the uncertainties directly from the harmonic decomposition fitting and reflect the statistical fitting uncertainty in each ring (see {Section~\ref{ssec:co_uncert}} for more details).}

We initialize the MCMC with $\rho_o = 0.5$ \msun\ pc$^{-3}$, $r_s$ equivalent to the maximum radius in the CO rotation curve, $\beta = 0.5$, \change{the PA and inclination given in {Table~\ref{tab:geomparams}}, and \ML\ given in {Table~\ref{tab:decompparams}}. We use 300 walkers for 300 steps.} By examining the chains, the MCMC has forgotten the initial conditions after 50 steps. To be conservative, we discard the initial 100 steps.

The best-fitting parameters are given by the median of the marginalized posterior parameter distributions. The uncertainties are given by the 16$^{\rm th}$ and 84$^{\rm\ th}$ percentiles, which is equivalent to 1-$\sigma$ for a Gaussian distribution. \change{The best-fitting values of $\rho_o$, $r_s$, $\beta$, \ML, PA, and inclination are given in {Table~\ref{tab:fittedparams}}, and the posterior distributions of $\rho_o$, $r_s$, and $\beta$ are shown in {Figure~\ref{fig:mcmcposteriors}} in {Appendix~\ref{app:mcmc}}. The fitted values of \ML, PA, and inclination are consistent well within the uncertainties with the fiducial values in {Tables~\ref{tab:geomparams}} and {\ref{tab:decompparams}}. The best-fitting velocity decompositions are shown in {Figure~\ref{fig:decomp}.}}

\subsection{Measuring the inner slope of the dark matter density profile}
\label{ssec:betastar}
We are interested in the inner slope of the dark matter density profile. Following \cta{relatores19b}, we define \changes{`inner'} as radii from $0.3-0.8$~kpc and denote the slope in this radial range as \betastar:
\begin{equation}
    \label{eq:betastar}
    \beta^* \equiv -\frac{\log\left(\frac{\rho_{\rm DM}(0.8 {\rm~kpc})}{\rho_{\rm DM}(0.3 {\rm~kpc})}\right)}{\log\left(\frac{0.8 {\rm~kpc} }{0.3 {\rm~kpc}}\right)}.
\end{equation}
We calculate \betastar\ at each step in the MCMC chain. The best-fitting value of \betastar\ is given by the median of the distribution and the quoted uncertainties are the 16$^{\rm th}$ and 84$^{\rm th}$ percentiles (upper right corner panels of {Figure~\ref{fig:mcmcposteriors}).} Our final values of \betastar\ and uncertainties are reported in {Table~\ref{tab:decompparams}.}

\changes{As shown in Table~\ref{tab:geomparams}, the uncertainties on the distances to these galaxies are $>20$~per~cent. These uncertainties affect the conversion from angular to physical distance and hence the radial range over which \betastar\ is measured. As a test of how the distance uncertainties affect our measurement of \betastar, we re-scale the velocity curve radii and re-calculate \betastar\ assuming the minimum and maximum distances to the galaxy. We find these values of \betastar\ are consistent with the fiducial values within the uncertainties. Therefore, though the uncertainties in the galaxy distances has some effect, it is not the dominate source of uncertainty on our measurement of \betastar. }

\changes{For NGC\,4701 (and the innermost point of NGC\,5692), we fix the stellar velocity curve to a maximal disk where the observed stellar velocity is larger than the CO rotation curve (see Section~\ref{ssec:stellarrotcurves}).  As a test, we relax the maximal disk assumption for NGC\,4701 and re-fit \betastar. Though we find a slightly shallower slope (\betastar~$=~0.1^{+0.6}_{-0.5}$), it is consistent with the slope measured for a maximal disk (\betastar~$=~0.4^{+0.8}_{-0.8}$) well within the uncertainties.}

\changes{Rather than our definition of \betastar\ in Equation~\ref{eq:betastar}, the inner slope of the dark matter density profile is often, especially in simulation-based analyses, defined between $1-2$~per~cent of the virial radius (\rvir).} It is often convenient to define R$_{\rm vir}\equiv{\rm R}_{200}$, where ${\rm R}_{200}$ is the radius at which the density is 200 times the critical density \citep[$\rho_{\rm crit}$; e.g.,][]{bullock17}. We refer to this density as $\rho_{200}~ (\equiv 200\rho_{\rm crit})$. The mass enclosed within ${\rm R}_{200}$ is ${\rm M}_{200}$. Assuming a local Universe value of $H_0 = 73.2 {\rm\ km\ s^{-1}\ Mpc^{-1}}$ \citep{riess21,divalentino21} yields $\rho_{\rm crit}=148.7{\rm\ M_\odot~kpc^{-3}}$. We extrapolate our best-fitting density profile for the total potential to estimate R$_{\rm vir}~ (\equiv {\rm R}_{200})$ and ${\rm M_{200}}$ {(Table~\ref{tab:decompparams}).} Given the uncertainties on the total density profile, we estimate the uncertainty on \rvir\ and hence ${\rm M_{200}}$. We find that $0.3-0.8$~kpc corresponds to $0.3-1.6$~per~cent of \rvir\ for these galaxies. Said another way, $1-2$~per~cent of \rvir\ corresponds to $0.6-1.8$~kpc for these galaxies {(Table~\ref{tab:decompparams}).} Accounting for the uncertainties on \rvir, the $1-2$~per~cent of \rvir\ spans a range of $0.1-5.0$~kpc. Therefore, our definition of \betastar\ tends to probe smaller physical radii in these galaxies than the definition used in many \changes{simulation-based analyses.} This will be discussed further in {Section~\ref{sec:discussion}} when we compare our results to simulations.

We also compute the central concentration ($c_{200}$) implied by our measurements, where $c_{200} \equiv \frac{{\rm R_{200}}}{r_s} \equiv \frac{{\rm R_{\rm vir}}}{r_s}$ \citep[e.g.,][]{dutton14,newman15}. We report our values of $c_{200}$ in {Table~\ref{tab:decompparams},} where the uncertainties are substantial and propagated from the uncertainties on \rvir\ {(Table~\ref{tab:decompparams})} and $r_s$ {(Table~\ref{tab:fittedparams}).} We compare with \cta{relatores19b} where possible,\footnote{Though \cta{relatores19b} do not report $c_{200}$ directly, it is related to $\beta$ and $c_{-2}$, which they report in their Table 4, where $c_{200} = c_{-2}(2-\beta)$. $c_{-2}$ is the concentration measured at the radius at which $\rho\propto r^{-2}$ locally. See \citet{newman15} and \cta{relatores19b} for more details.} and we agree within the (large) uncertainties on both calculations of $c_{200}$.

\section{Results}
\label{sec:results}

\begin{figure}
    \centering
    \includegraphics[width=\columnwidth]{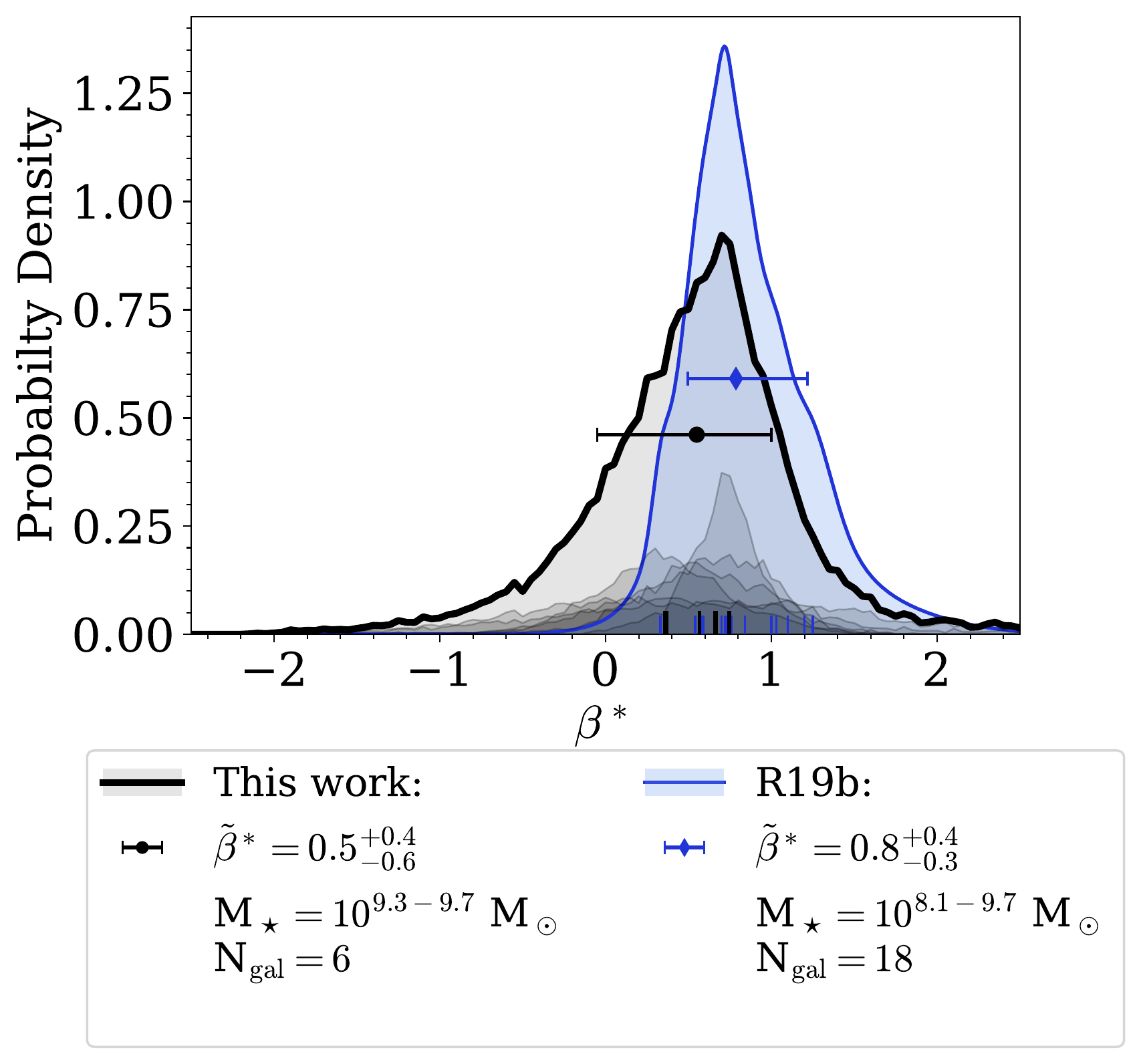}
    \caption{A KDE showing the distribution of \betastar\ values we measure in this study (\change{thick} black curve). \change{The light grey distributions show the posterior distributions of \betastar\ for each galaxy, normalised to their fractional contribution to the total KDE (thick black curve).} The black tick marks at the bottom show the individual \betastar\ values. The black dot shows the median of the distribution ($\tilde{\beta}^*$); the error bars show the 16$^{\rm th}$ and 84$^{\rm th}$ percentiles. The blue KDE, tick marks, diamond, and error bars show the same quantities but for the galaxies studied by \cta{relatores19b} with reliable mass decompositions. The legend also notes the number of galaxies in each KDE and the stellar mass range covered by those galaxies.}
    \label{fig:kde}
\end{figure}

\begin{figure*}
\centering
\includegraphics[width=\textwidth]{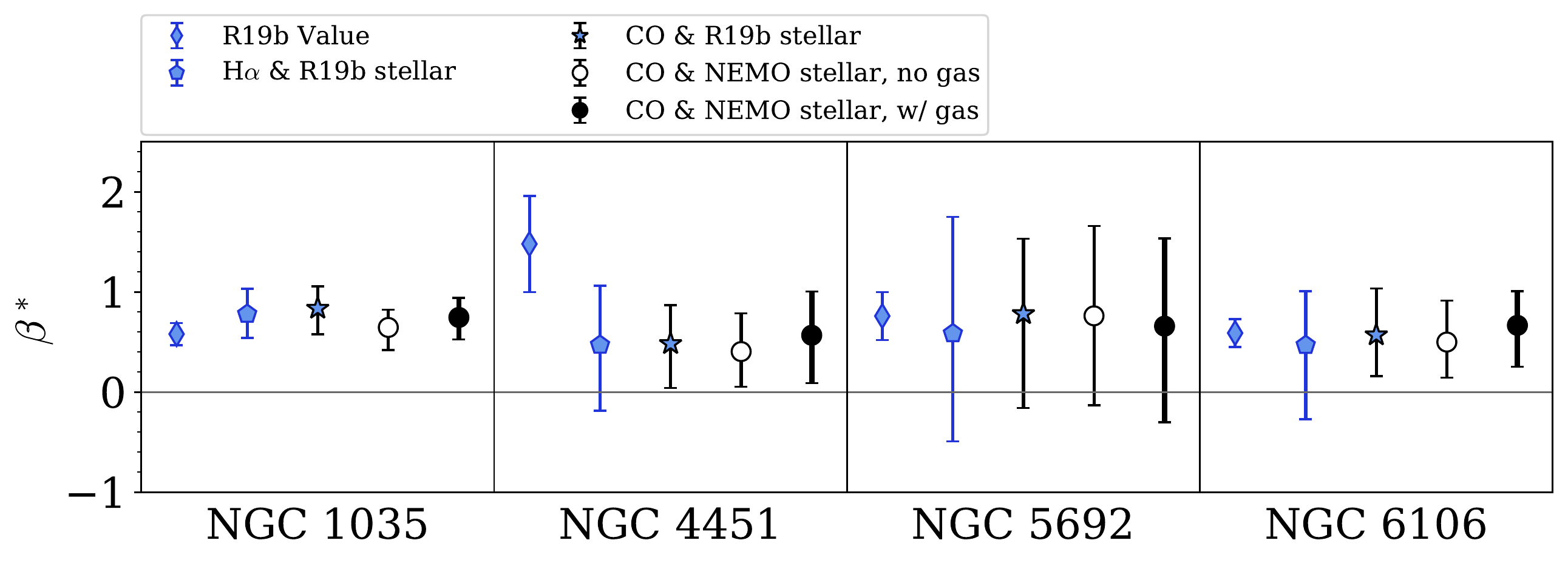}
\caption{A comparison of the \betastar\ values derived for different sets of assumptions as described in {Section~\ref{ssec:checks}} for the galaxies that overlap with \cta{relatores19b}. The blue diamonds show the values derived by \cta{relatores19b}. The blue pentagons show the \betastar\ we derive using \ha\ rotation curves to trace the total potential \cpa{relatores19a}, the stellar rotation curves from \cta{relatores19b}, and including the atomic and molecular components as a check on the accuracy of our decomposition method. The blue filled stars show the results when using our CO rotation curves to trace the total potential, the stellar rotation curves from \cta{relatores19b}, and including the atomic and molecular components. The circles show the values of \betastar\ derived using our CO rotation curves to trace the total potential, the stellar rotation curves we derive using \nemo\ {(Section~\ref{ssec:stellarrotcurves}),} and either excluding (open) or including (filled) the atomic and molecular components. Within each galaxy bin, points are artificially offset along the horizontal axis for clarity. In general, we do not find significant differences in the values of \betastar\ derived for each set of assumptions\change{,} given the uncertainties.}
\label{fig:betastarcomp}
\end{figure*}

We are interested in measuring the inner slope of the dark matter density distribution for our sample of dwarf galaxies. We define \changes{`inner'} as from $0.3-0.8$~kpc and denote the slope in this radial range as \betastar\ {(Equation~\ref{eq:betastar}).} We use the CO rotation curves that we derive here {(Figure~\ref{fig:CORCs};} {Section~\ref{sec:rcfitting})} as a robust tracer of the total potential. We use an MCMC method to construct model dark matter density profiles, assuming a gNFW form {(Equation~\ref{eq:NFW}).} By including the stellar component {(Section~\ref{ssec:stellarrotcurves}),} we minimize the difference between the CO rotation curve and our model \change{circular velocity curve} of the total potential (including the stars and dark matter; {Section~\ref{sec:DMdensity}).} We measure \betastar\ from the fitted dark matter density profiles {(Figure~\ref{fig:decomp}).} \changes{Across this sample of galaxies, we find no trends between} \betastar\ and galaxy parameters such as inclination, distance, ${\rm R_{max}}$, \rvir, ${\rm M_{dyn}}$, or M$_\star$.

For these six galaxies, we find \betastar\ values that are on average consistent with \change{shallow cusp-like} dark matter density profiles (\avebetastar\ with a standard deviation of \scatterbetastar; {Table~\ref{tab:decompparams}).} We show the distribution of our \betastar\ values in black in {Figure~\ref{fig:kde}} as a kernel density \changes{estimate} (KDE). We make this KDE by summing the posterior distributions of \betastar\ for each galaxy {(Figure~\ref{fig:mcmcposteriors})} and renormalising to unit area. The median of this distribution is \medbetastar, where the lower and upper uncertainties correspond to the 16$^{\rm th}$ and 84$^{\rm th}$ percentiles of the distribution (1-$\sigma$ for a Gaussian distribution). By our KDE construction, this quoted uncertainty reflects both the measurement uncertainties and the galaxy-to-galaxy scatter (though counted in a different way than the standard deviation of the best-fitting values). The average and scatter derived from {Table~\ref{tab:decompparams}} (\avebetastar\ with standard deviation of \scatterbetastar) reflect only the galaxy-to-galaxy scatter in the best-fitting \betastar\ values. \change{The light grey distributions in {Figure~\ref{fig:kde}} show the posterior distributions of \betastar\ for each galaxy (see also {Figure~\ref{fig:mcmcposteriors}}).}

We compare the distribution of our \betastar\ measurements to those from \cta{relatores19b}. A KDE of their values is shown in blue in {Figure~\ref{fig:kde}.} For this KDE, we represent each of their measurements as a Gaussian, where the width is set by their measurement uncertainties (see their Table 3, grades 1 and 2). The individual Gaussians are summed and normalised to unit area. \change{Our distribution of the slopes is shifted towards somewhat shallower slopes than \cta{relatores19b}, though \changes{there is substantial overlap between the distributions.}} In comparing these distributions, it is important to keep in mind that our sample only includes six galaxies, whereas \cta{relatores19b} have 18 galaxies.

\subsection{Checks on the method and assumptions}
\label{ssec:checks}

Below we describe several tests we performed to check the robustness of our modelling and results to various assumptions. We show the resulting values of \betastar\ for these different tests in {Figures~\ref{fig:betastarcomp} and \ref{fig:betastar_dmprofiles}.} For the discussion in {Section~\ref{sec:discussion},} \changes{we adopt} the values of \betastar\ we derive using our \change{CO rotation curves and stellar velocity curves}, including the atomic and molecular gas components where available, as reported in {Table~\ref{tab:decompparams}.} In general, we do not find significant differences in the values of \betastar\ derived for each set of assumptions given the uncertainties.

\subsubsection{Decomposition method}

As a check on our decomposition method, we measure \betastar\ using the \ha\ rotation curves from \cta{relatores19a} as a tracer of the total potential and the stellar rotation curves from \cta{relatores19b} to most closely reproduce their results. Any differences in these values compared to those presented by \cta{relatores19b} \changes{are} due to our fitting method, as the assumptions about the total and stellar potentials and the form of the dark matter density are the same.

As shown in {Figure~\ref{fig:betastarcomp},} the results of this test (blue pentagons) are in good agreement with the values reported by \cta{relatores19b} (blue diamonds), meaning that our fitting routine \changes{does} not introduce biases in the measurement of \betastar. The most discrepant galaxy\changes{, in terms of the absolute differences in \betastar, is NGC\,4451, though the measurements are} consistent within the uncertainties. \change{In general, the uncertainties on our points are larger than those measured by \cta{relatores19b}.}

\subsubsection{CO v. H$\alpha$ as tracers of the total potential}

A difference between this analysis and that presented by \cta{relatores19b} is that we use CO to trace the total potential while they use \ha. As discussed in {Section~\ref{sec:rcfitting}} (and by \cta{relatores19b}), there is good agreement in general between the CO and \ha\ rotation curves {(Figure~\ref{fig:CORCs}).} For the \ha, this indicates that it is indeed a dynamically cold tracer. The low velocity dispersions of the CO {(Table~\ref{tab:geomparams})} confirm that it is also dynamically cold.

We compare the values of \betastar\ we derive using the CO rotation curves (blue stars) with those \changes{we derive using the \ha\ rotation curves from \cta{relatores19a}}  (blue pentagons) in {Figure~\ref{fig:betastarcomp}}. \changes{In both measurements, we use} the stellar \changes{velocity} curves derived by \cta{relatores19b} to purely test the effect of the different tracers of the total potential. Given the uncertainties, the \betastar\ measurements from CO and \ha\ are in excellent agreement with one another. 

\subsubsection{Stellar \changes{velocity} curve derivation}

As described in {Section~\ref{ssec:stellarrotcurves},} while we use the same photometric data to measure the stellar components as \cta{relatores19b}, we derive the stellar light profiles and \change{velocity} curves using different methods and assumptions. As shown in {Figure~\ref{fig:stellarRCcomp},} the agreement between our derived stellar \change{velocity} curves is quite good in general.

In {Figure~\ref{fig:betastarcomp},} we compare the values of \betastar\ derived using our stellar \change{velocity} curves (black filled circles) and those from \cta{relatores19b} (blue stars), with all other variables the same. These values are in excellent agreement within the uncertainties. Therefore, our calculation of the stellar \change{velocity components} assuming a thin disk \changes{does} not greatly affect the results.

\subsubsection{Excluding the atomic and molecular components}

We do not have information about the atomic and molecular \change{mass} components readily available for two of the galaxies in this sample. We test whether the measured \betastar\ values are biased by not including the atomic and molecular gas in the mass decomposition. We repeat the decomposition to determine \betastar\ with and without this gas data for the four galaxies that overlap between our sample and the \cta{relatores19b} sample, using the CO data to trace the total potential and using our stellar \change{velocity} curves from \nemo. The \betastar\ values with (filled circles) and without (open circles) the gas are compared directly in {Figure~\ref{fig:betastarcomp}.} For all galaxies, the derived \betastar\ values are consistent well within the uncertainties. \change{We therefore conclude} that excluding the atomic and molecular gas data from the decomposition \changes{does} not significantly affect the results. 

\subsubsection{Assumed functional form of the dark matter density profile}

As we discuss in {Appendix~\ref{app:mcmc},} we re-run the decompositions and the aforementioned tests assuming a two parameter power-law dark matter density profile (instead of the gNFW profile given by {Equation~\ref{eq:NFW}).} The density profiles yield similar dark matter profiles and \betastar\ values as the gNFW {(Figure~\ref{fig:betastar_dmprofiles}).} This is likely because our CO observations probe small radii where the $(r/r_s)^\beta$ \change{factor} in the gNFW profile dominates, effectively reducing $\rho_{\rm DM}$ to a power-law. In some galaxies, the second \change{factor} in the gNFW profile does become important, as the power-law profiles continue to increase whereas the gNFW profiles soften and better represent the data. Therefore, we proceed with the results using the gNFW profiles.

\section{Discussion}
\label{sec:discussion}

\subsection{Comparison to previous observational studies}

\begin{figure*}
\label{fig:beta_obs_sim}
\centering
    \includegraphics[width=0.9\textwidth]{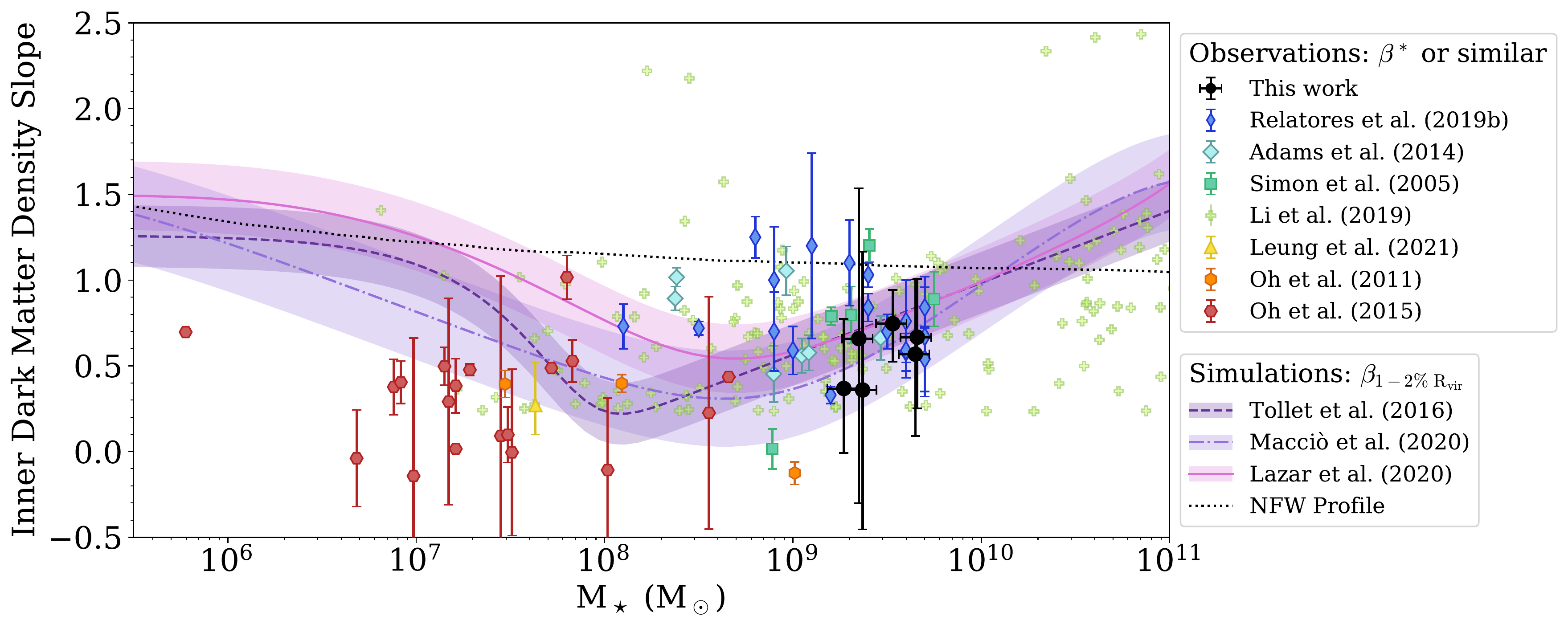}
    \includegraphics[width=0.9\textwidth]{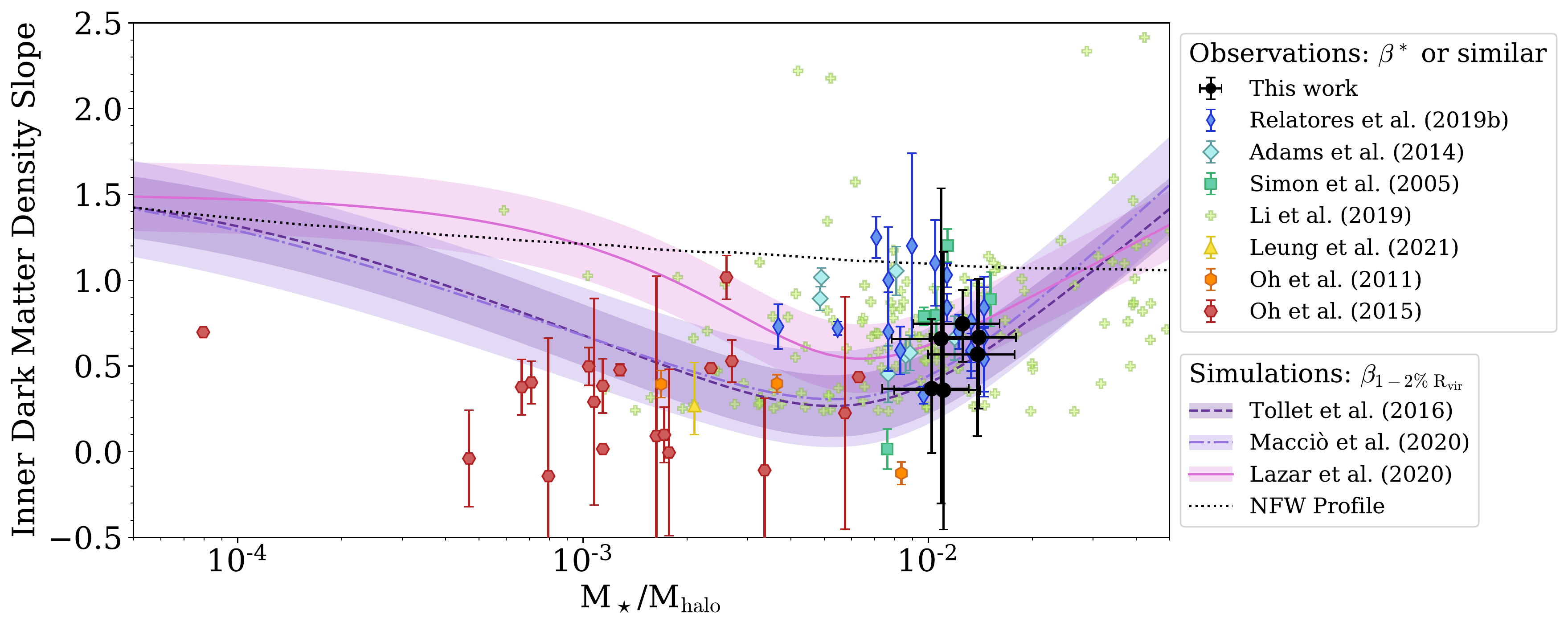}
    \includegraphics[width=0.9\textwidth]{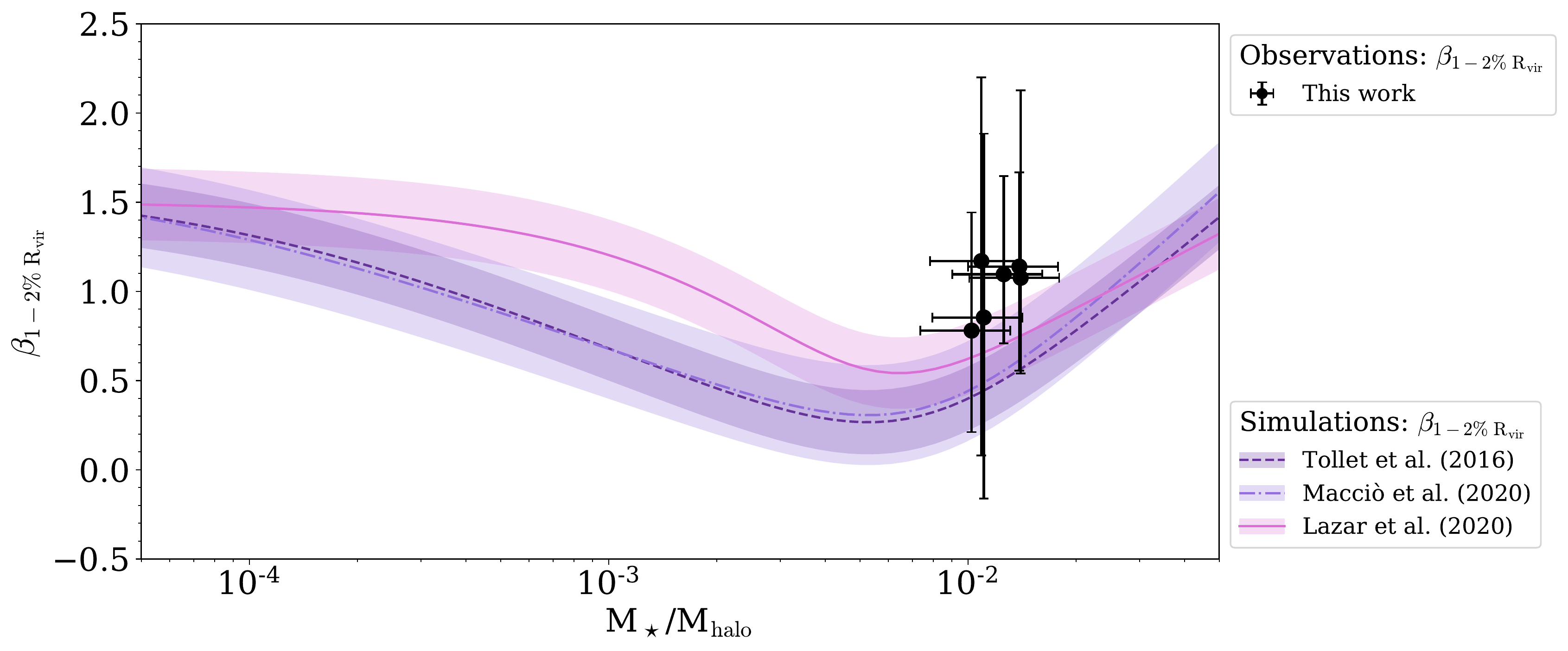}
\caption{The inner slope of the dark matter density profile as a function of stellar mass (top) and the ratio of stellar mass to halo mass (middle). \change{There are several definitions and methods of measuring the inner slope of the dark matter density profile shown in this figure.} The \change{\betastar\ values} derived here are shown in the black circles. The horizontal error bars on our points reflect the uncertainty in our stellar mass measurements (top) or the scatter in the assumed stellar mass-halo mass relation (bottom). We compare with other observational studies \change{\citep[colored symbols;] []{simon05,oh11,oh15,adams14,relatores19b,li19,leung21} and predictions from the NIHAO \citep{tollet16,maccio20} and FIRE \citep{lazar20} simulations (shaded curves).} The black dotted line shows the expectation from an NFW profile. Cores have \betastar$ \sim0$ and cusps have \betastar$ \gtrsim 0.5$. Our measurements are in good agreement with previous measurements at similar M$_\star$ and \msmh, given the scatter and uncertainties. Galaxies in this range of M$_\star$ and \msmh\ tend to have cuspy dark matter density profiles. \change{The bottom panel shows \betarvir\ for our measurements compared to the NIHAO \citep{tollet16,maccio20} and FIRE \citep{lazar20} simulations. When the inner slopes of the dark matter density profiles are measured in the same way, our measurements tend to be steeper than predicted by these simulations.}} 
\end{figure*}

We compare our measurements of the inner dark matter density slope to other observational studies of dwarf galaxies as a function of the stellar mass of the galaxy {(Figure~\ref{fig:beta_obs_sim}\change{,}} top). These observational studies span a range of tracers and methodologies, as described briefly here. There is no overlap between our sample of galaxies and those from \citet{simon05}, \citet{oh11,oh15}, \citet{adams14}, \change{\citet{li19}}, or \citet{leung21}. In {Figure~\ref{fig:beta_obs_sim},} we have tried to use similar colours for studies with similar methodologies, as described below. 

\citet{simon05} use a combination of \ha\ and CO data (where available) in five galaxies to measure \betastar. The rotation curves were extracted using the same method as described in Section~{\ref{sec:rcfitting},} and they fit their stellar rotation curves using a similar method to the one used here {(Section~\ref{ssec:stellarrotcurves}).} They measure their dark matter slopes from a power-law fit to the full dark matter \change{velocity} curves which extend out to $2-5$~kpc from the centre. Assuming a fiducial ${\rm R_{vir}\sim100}$~kpc, these data probe $\sim2-5$~per~cent of ${\rm R_{vir}}$. This is corresponds to somewhat farther out in the dark matter halo than probed by our measurements which probe $\sim0.3-1.6$~per~cent of ${\rm R_{vir}}$ (see {Section~\ref{sec:DMdensity}).}

\citet{adams14} use integral field measurements, which trace the kinematics of the stars and ionized gas, in a sample of six dwarf galaxies. They extract their circular velocity curves using dynamical modeling of both the stars and ionized gas. As described by \cta{relatores19b}, we use their gNFW fits to calculate \betastar\ following {Equation~\ref{eq:betastar}.} 

\citet{oh11,oh15} use \HI\ data from the THINGS and LITTLE THINGS surveys to trace the galaxy potentials and \spitzer\ 3.6~$\mu$m data to remove the stellar contributions. They derive their rotation curves using a tilted ring model\changes{, which allows for radially varying geometric parameters to account for features such as warps,} and apply an asymmetric drift correction to their data. Their method of deriving the stellar mass profiles is similar to ours, but using the 3.6~$\mu$m data. For their mass models, they use both a NFW profile (which produces a cusp) and a pseudo-isothermal model (which produces a core), finding that the pseudo-isothermal profiles are a better fit to their observations. They measure the inner slope of their dark matter density profiles by fitting a power-law at radii smaller than a \changes{`break'} radius, which is the radius where the slope changes most rapidly. This \changes{break} radius changes for each galaxy, but is $\lesssim1$~kpc, so we are measuring \betastar\ over a similar range of radii. 

\change{We note that \citet{iorio17} performed an independent re-analysis of the LITTLE THINGS data by fitting the kinematics of the \HI\ data in three dimensions, rather than the two dimensional approach taken by \citet{oh11,oh15}. \citet{iorio17} find that half of their rotation curves are consistent with those derived by \citet{oh15} within the uncertainties. For those galaxies where the rotation curves do differ, the slopes near the centres are not systematically steeper or shallower. In other words, if the mass modelling were repeated using the rotation curves from \citet{iorio17}, we would not expect systematically steeper (or shallower) dark matter density profiles compared to \citet{oh15}. }

As discussed previously, \cta{relatores19b} use \ha\ measurements to trace the total potential of the galaxies and \spitzer\ 4.5~$\mu$m data to trace the stellar components. They obtain their stellar profiles using a multi-Gaussian expansion method, as opposed to our method of extracting the stellar light in concentric annuli. We base our decomposition method on theirs and adopt their definition of \betastar. {Figure~\ref{fig:betastarcomp}} shows the comparison of our \betastar\ measurements to theirs for the \change{four} common galaxies.

\change{\citet{li19} derive dark matter mass models for the 175 galaxies in the {\em Spitzer} Photometry and Accurate Rotation Curves (SPARC) data base \citep{lelli16}. The SPARC galaxies span a broad range of morphologies, luminosities, and surface brightnesses. Their rotation curves tracing the total potential of the galaxies come from either \HI\ or \ha\ data. The stellar components are traced by {\em Spitzer} 3.6~$\mu$m data. The stellar mass estimates come from \citet{lelli16}, assuming \ML~$=0.5$, though \citet{lelli16} note that there are disagreements between this value and other studies. For the mass models, we use the \citet{dicintio14} fits (a version of the gNFW profile) with flat priors from \citet{li19} to reconstruct their dark matter density profiles (see their Table~A2). From these, we measure \betastar\ according to Equation~\ref{eq:betastar}. }

\citet{leung21} use \HI\ data and stellar dynamical modelling to constrain the dark matter distribution in the dwarf irregular galaxy WLM. They use a gNFW profile, but also allow for flattening of the dark matter distribution, finding a prolate geometry is favoured. Their reported best-fitting value of the dark matter density slope is $\beta$ in the gNFW profile (not \betastar; {Equation~\ref{eq:NFW}).} To best compare with our measurements, we use their value of the dark matter slope for a spherical dark matter distribution including the gas and stars ($0.27^{+0.25}_{-0.17}$) which agrees within the uncertainty with their best-fitting slope allowing the dark matter geometry to vary.

From the observations, there is a general trend of lower mass galaxies exhibiting core-like inner dark matter density profiles with slopes \betastar$\sim0$ \changes{(}{Figure~\ref{fig:beta_obs_sim}\change{,}} top), though this trend is driven by the results of \citet{oh11,oh15} and supported by \citet{leung21} \change{and \citet{li19}}. Higher mass galaxies---like those in this study---tend to have cuspy dark matter density slopes (\betastar$\gtrsim0.5$), but the scatter is large. In general, our measurements of \betastar\ agree well with other observations at similar stellar masses {(Figure~\ref{fig:beta_obs_sim}\change{,}} top).

Observations of dwarf galaxies at similar stellar masses also show a wide range of dark matter density slopes \changes{(Figure~\ref{fig:beta_obs_sim}, top).} \change{While this scatter may reflect real galaxy-to-galaxy variations, uncertainties in the rotation curves and differences in how \betastar\ is measured likely contribute to the scatter, perhaps substantially (see discussions in, for example, \cta{relatores19b} and \citealt{santos-santos20}).} We find no clear trends between \betastar\ and galaxy properties such as M$_\star$, M$_{\rm dyn}$, or \rvir. Similarly, \cta{relatores19b} find no trends between \betastar\ and the galaxy effective radius and stellar surface density in the inner kpc of the galaxy. Some recent simulations also confirm or reproduce this observed diversity in the dark matter density profiles at fixed stellar mass \citep[e.g.,][]{oman15,oman19,kamada17,santos-santos18,santos-santos20,orkney21}. 

\subsection{Comparison to simulations}

We also compare our measurements to those from simulations {(Figure~\ref{fig:beta_obs_sim}}) as a function of both the stellar mass and the ratio of the stellar mass to the halo mass (\msmh). The curves show the results from the FIRE \citep{lazar20} and NIHAO \citep{tollet16,maccio20} simulations. The simulations measure \betastar\ from $1-2$~per~cent of ${\rm R_{vir}}$\change{, as opposed to the fixed radial range of $0.3-0.8$~kpc we use (Equation~\ref{eq:betastar})}.

We convert the curve presented by \citet{lazar20} in terms of \msmh\ to M$_\star$ using their stellar mass-halo mass relation (see their Figure 1)\footnote{\change{We note that the stellar mass-halo mass relation in \citet{lazar20} quantifies the typical stellar mass at fixed halo mass. However, for our conversions, we require the halo mass at fixed stellar mass. These conversions are not necessarily the same \citep[e.g.,][]{behroozi10}. However, for the range of masses considered here, this is likely a small effect. This is substantiated by the fact that the M$_{\rm halo}$ values implied by this conversion agree well with our estimates of M$_{200}$.} }. We use this same relation to convert the observations from M$_\star$ to \msmh\footnote{\change{As part of their fitting, \citet{li19} report \msmh\ values for the SPARC galaxies. However, the definitions of the slope parameters in the \citet{dicintio14} dark matter profiles depend explicitly on \msmh, and there is a strong correlation between \betastar\ and \msmh\ induced by this relation. Therefore, for the points from \citet{li19} shown in the bottom panel of {Figure~\ref{fig:beta_obs_sim}}, we recompute \msmh\ using the independently measured stellar masses from \citet{lelli16} and the same stellar mass-halo mass relation as for the other observations.}}. The horizontal errorbars on our points in the \change{middle and bottom panels} of {Figure~\ref{fig:beta_obs_sim}} include the scatter in the stellar mass-halo mass relation in addition to the uncertainties on the stellar mass measurements. 

The above simulations predict a stellar mass range (M$_\star\sim10^{7.5-9.5}$ \msun) where feedback is efficient in softening the dark matter density profile in the centres of galaxies. In reality, this trend depends on \msmh\ and evolves with time \citep[e.g.,][]{dicintio14,tollet16}. For example, \citet{tollet16} find that all galaxies begin with cuspy profiles. For low mass galaxies, \citet{tollet16} find that the central regions evolve passively once star formation has finished, and so they retain their cuspy profiles until $z=0$. Other groups, however, find that very low mass systems can \change{develop} core-like profiles due to late-time minor mergers \citep{orkney21} or starburst phases \citep{read16}. For galaxies of intermediate \msmh, the amount of gas is rapidly changing due to inflows, accretion of satellites, mergers, and star formation-driven outflows. As a result, the potential undergoes rapid changes and the centre of mass of the gas changes with respect to the dark matter, both of which help develop and maintain the core-like density profile \citep{tollet16}. For the most massive galaxies, cores may form due to accretion and stellar feedback, resulting in rapid changes in \msmh\ in the central regions as for the intermediate-mass systems. However, \citet{tollet16} find that the net result for these higher mass galaxies is that gas flows inward towards the centre, triggering more star formation and hence increasing the stellar mass relative to the halo mass in the centre. As the potential well deepens due to the gas inflows, feedback-driven outflows become less important in terms of expelling mass. As a result of the deepened central potential, dark matter becomes concentrated at the centre and hence the cuspy profile is restored. There is some evidence from simulations that Milky Way mass galaxies (M$_\star\approx10^{10-11}$~\msun, M$_{\rm halo}\sim10^{12}$~\msun) can retain small ($0.5-2$~kpc) cores, which are maintained by stellar feedback \citep{chan15,lazar20}. Outflows from active galactic nuclei can also soften the cuspy profile in the most massive galaxies (M$_{\rm halo}>10^{12}$~\msun), though the slopes are still within the cusp-like regime \citep{maccio20}.

The galaxies in this study have M$_\star=10^{8.9-9.7}$~\msun. Assuming a stellar mass-halo mass relation like \citet{lazar20} implies M$_{\rm halo}\approx10^{11.0-11.5}$~\msun. \change{This agrees with our estimates of M$_{200}$ (\changes{as an approximation of M$_{\rm halo}$ to within $\sim90$~per~cent\footnote{\changes{\citet{lazar20} adopt the virial definition of the halo mass from \citet{bryan88}, in which the virial overdensity is $\sim100$ times the critical density. Under this definition, M$_{\rm halo}$/M$_{\rm 200}\sim1.1$. This difference is less than the scatter in Figure~\ref{fig:beta_obs_sim} and is, therefore, only a minor concern for this analysis.}}}) listed in Table~\ref{tab:decompparams} within the uncertainties. }Therefore, for these relatively massive dwarfs at $z=0$, we would expect to find more cuspy dark matter profiles, which we quantitatively confirm with these measurements.

\begin{figure}
    \centering
    \includegraphics[width=\columnwidth]{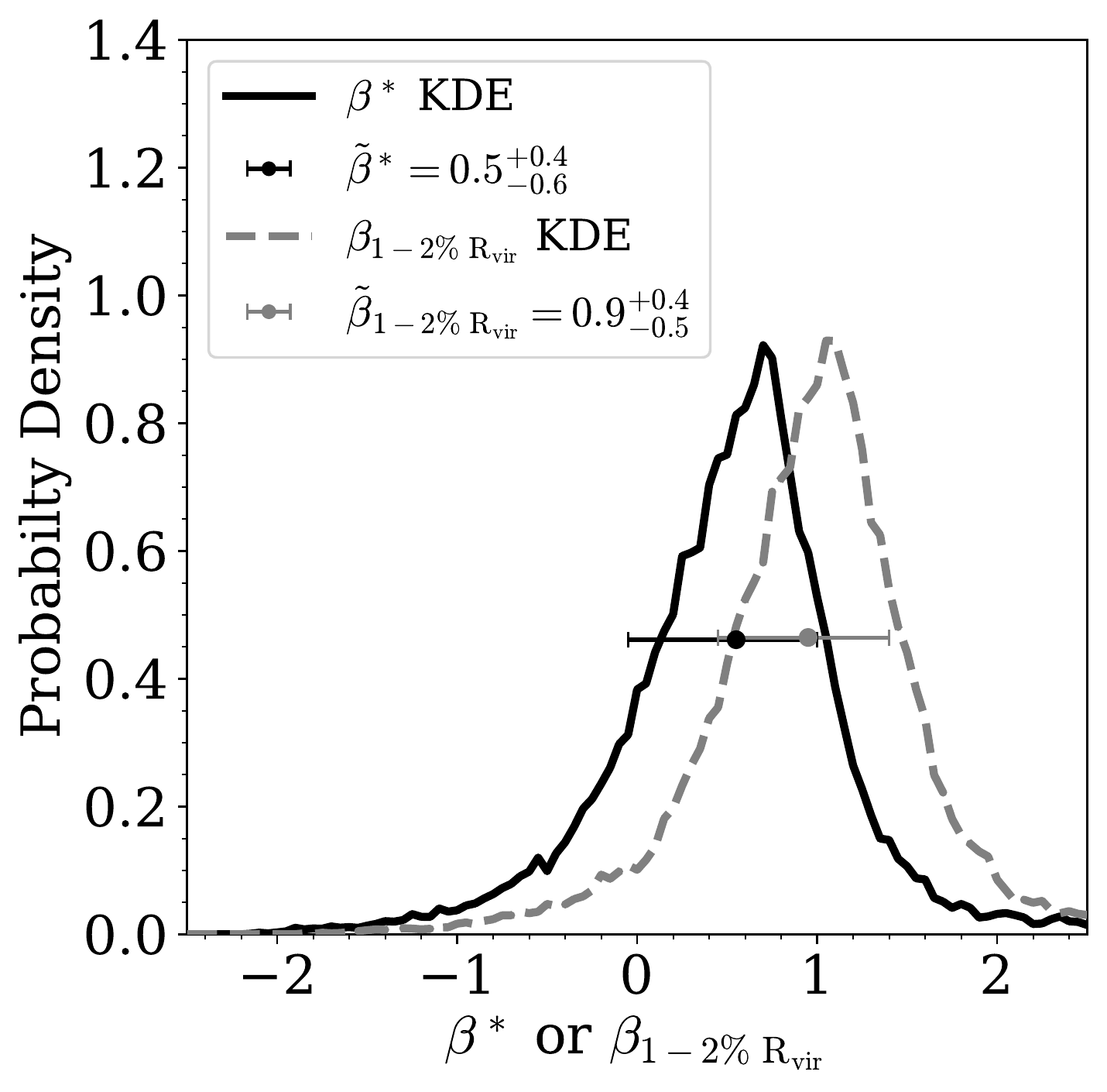}
    \caption{\change{KDEs comparing the inner slope of the dark matter density profile as parametrized by \betastar\ (black solid) or \betarvir\ (grey dashed). The black and grey dots show the median of each distribution; the error bars show the 16$^{\rm th}$ and 84$^{\rm th}$ percentiles. The \betastar\ KDE is the same as in Figure~\ref{fig:kde}. The dark matter slopes parametrized by \betarvir\ are steeper than \betastar.}}
    \label{fig:betastar_rvir_kde}
\end{figure}

As discussed in {Section~\ref{ssec:dmdecompmethod},} our definition of \betastar\ probes smaller radii than the simulations. The simulations measure the slope of the dark matter density profile from radii between $1-2$~per~cent of ${\rm R_{vir}}$, which we report in {Table~\ref{tab:decompparams}.} However, our adopted definition of \betastar\ {(Equation~\ref{eq:betastar})} covers a fixed radial range from $0.3-0.8$~kpc, which corresponds to $\sim0.3-1.6$~per~cent of ${\rm R_{vir}}$ for these galaxies. In an effort to best compare with simulations, we also compute \betarvir, given our estimates of ${\rm R_{vir}}$ and its uncertainties in {Table~\ref{tab:decompparams}.} Like \betastar, \betarvir\ is calculated at every step in the MCMC chain during the decomposition, and hence \betastar\ and \betarvir\ are calculated for the same dark matter density profile at each step and any differences between them \changes{reflects} only the differences in the definitions of \betastar\ and \betarvir. The uncertainty on \betarvir\ also incorporates the uncertainty on \rvir. \change{We report our best-fitting \betarvir\ values in {Table~\ref{tab:decompparams}}. {Figure~\ref{fig:betastar_rvir_kde}} compares the KDEs of \betastar\ and \betarvir, which are the sums of the posterior distributions over all of the galaxies normalised to unit area. The distribution of \betarvir\ is shifted towards steeper slopes than the distribution of \betastar. We compare our \betarvir\ measurements to the simulations in {Figure~\ref{fig:beta_obs_sim}} (bottom).}

When the measured dark matter density slope is parametrized by \betastar, most of our measured points agree with the predictions from \change{the FIRE and NIHAO simulations} {(Figure~\ref{fig:beta_obs_sim}}\change{, top and middle}). Observations at similar M$_\star$ and \msmh\ show a similar distribution of \betastar, though the scatter is large (see also {Figure~\ref{fig:kde}).} If we instead use \betarvir\ as our parametrization of the inner dark matter density slope---which is the value reported by the simulations---we find that most of our galaxies have steeper dark matter density profiles than predicted by \change{these} simulations {(Figure~\ref{fig:beta_obs_sim}}\change{, bottom}). Therefore, in this most equal comparison, our observations tend to have cuspier dark matter profiles than predicted by \change{the FIRE and NIHAO simulations}. 

As described above, one way for a galaxy to restore a cuspy profile is through gas inflows which trigger a burst of star formation. This both deepens the potential well, causing dark matter to concentrate in the centre of the galaxy, and increases the stellar mass of the centre relative to the halo mass. One possible explanation for the more cuspy profiles observed in these (and other) galaxies is that these galaxies had stronger gas inflows towards the centre than predicted by the simulations, resulting in more centrally concentrated dark matter at the centres. An alternate possibility is that the simulations over-predict the strength or frequency of accretion, merger, and/or outflow events for galaxies in this stellar mass range. As described by \citet{tollet16}, these feedback events stir up the potential, allowing the dark matter distribution to be shallower and maintaining a core-like central profile. If there were less frequent or intense accretion or outflows, the dip in the curves from simulations shown in \change{{Figure~\ref{fig:beta_obs_sim}}} would be both shallower and narrower. It is also possible that both effects are at work, or that there are other factors responsible for this population of galaxies with cuspy profiles given their M$_\star$ and \msmh.

\section{Summary}
\label{sec:summary}

We have investigated the inner slope of the dark matter density distribution (\betastar) using observations of six nearby spiral dwarf galaxies. We summarize our results below.

\begin{enumerate}
\itemsep0em

\item Using new CO observations from ALMA, we measure the CO rotation curves of these galaxies, which provide a dynamically cold tracer of the total potential {(Section~\ref{sec:rcfitting},} {Figure~\ref{fig:CORCs}).} We find excellent agreement between the CO and \ha\ rotation curves \change{in 3/4} galaxies where both are available.

\item We decompose the dark matter \change{velocity} curve and density profile using an MCMC method, knowing the total potential and the contribution from the stars {(Section~\ref{sec:DMdensity},} {Figure~\ref{fig:decomp}).} From these, we measure \betastar, finding \avebetastar\ with a standard deviation of \scatterbetastar\ among the best-fitting \betastar\ values for galaxies in this sample {(Table~\ref{tab:decompparams}).} Considering the full posterior distributions of \betastar, the median \betastar\ for this sample is \medbetastar, where the lower and upper uncertainties correspond to the 16$^{\rm th}$ and 84$^{\rm th}$ percentiles of the distribution {(Figure~\ref{fig:kde}).}

\item We find that the relatively massive dwarf galaxies (M$_\star = 10^{9.3-9.7}$~\msun) in this study show cuspier dark matter distributions than lower mass dwarf galaxies, in agreement with other observations at similar M$_\star$ and \msmh {(Figures~\ref{fig:kde} and \ref{fig:beta_obs_sim}).}

\item \change{To accurately compare our measurements with results from the FIRE and NIHAO simulations, we compute \betarvir\ for our galaxies. The inner dark matter density slopes parametrized by \betarvir\ are steeper than \betastar\ ({Figure~\ref{fig:betastar_rvir_kde}}). We find that our galaxies have steeper slopes on average than predicted by the FIRE and NIHAO simulations {(Figure~\ref{fig:beta_obs_sim}}, bottom). This may signal that these galaxies have undergone a stronger gas inflow than implemented in the simulations, that the simulations overpredict the frequency of accretion/merger/outflow events, or that a combination of these or other effects are at work.}
\end{enumerate}

Analyses such as these are often limited by the sensitivity needed to probe out to large radii in faint, often gas-poor dwarf galaxies. Future facilities, especially the Next Generation Very Large Array, will transform our ability to measure gas kinematics in dwarf galaxies. Owing to its increased sensitivity and angular resolution and the ability to observe both CO and \HI, we can trace the rotation curves out to larger radii more robustly. The increased spatial resolution will allow for more accurate measurements in the centres, precisely in the range of radii where \betastar\ is measured. Finally, the combination of the increased resolution and sensitivity will allow us to probe lower mass and more distant systems to better understand how the dark matter distributions change over time and as a function of environment. 

\section*{Acknowledgements}
\change{The authors sincerely thank the anonymous referee for their thoughtful and constructive feedback that greatly improved this paper. We also gratefully acknowledge the work of Leo Blitz, who wrote and led the ALMA proposal to obtain the data upon which this work is based.} L.H.C. and B.D.D. thank the Department of Astronomy at the University of Maryland for hosting them during the completion of this work. R.C.L. acknowledges partial support for this work provided by the National Science Foundation (NSF) through Student Observing Support Program (SOSP) award 7-011 from the NRAO \change{and by an NSF Astronomy and Astrophysics Postdoctoral Fellowship under award AST-2102625}. \change{A.D.B. acknowledges partial support from NSF-AST2108140.} B.D.D. acknowledges funding and support from the Graduate Resources Advancing Diversity with Maryland Astronomy and Physics (GRAD-MAP) program (\url{www.umdgradmap.org}), which is funded by the University of Maryland College of Computer, Mathematical, and Natural Sciences and the Departments of Astronomy and Physics and the NSF PIRE (Grant No. 1545949) and AAG (Grant No. AST-1615960) programs. V.V. acknowledges support from the scholarship 
ANID-FULBRIGHT BIO 2016 - 56160020 and funding from NRAO \change{through SOSP award} SOSPA7-014.

This paper makes use of the following ALMA data: ADS/JAO.ALMA\#2015.1.00820.S. ALMA is a partnership of ESO (representing its member states), NSF (USA) and NINS (Japan), together with NRC (Canada), NSC and ASIAA (Taiwan), and KASI (Republic of Korea), in cooperation with the Republic of Chile. The Joint ALMA Observatory is operated by ESO, AUI/NRAO and NAOJ. The National Radio Astronomy Observatory is a facility of the National Science Foundation operated under cooperative agreement by Associated Universities, Inc. This work is based in part on observations made with the \spitzer\ Space Telescope, which was operated by the Jet Propulsion Laboratory, California Institute of Technology under a contract with NASA. The Pan-STARRS1 Surveys (PS1) and the PS1 public science archive have been made possible through contributions by the Institute for Astronomy, the University of Hawaii, the Pan-STARRS Project Office, the Max-Planck Society and its participating institutes, the Max Planck Institute for Astronomy, Heidelberg and the Max Planck Institute for Extraterrestrial Physics, Garching, The Johns Hopkins University, Durham University, the University of Edinburgh, the Queen's University Belfast, the Harvard-Smithsonian Center for Astrophysics, the Las Cumbres Observatory Global Telescope Network Incorporated, the National Central University of Taiwan, the Space Telescope Science Institute, the National Aeronautics and Space Administration under Grant No. NNX08AR22G issued through the Planetary Science Division of the NASA Science Mission Directorate, the National Science Foundation Grant No. AST-1238877, the University of Maryland, Eotvos Lorand University (ELTE), the Los Alamos National Laboratory, and the Gordon and Betty Moore Foundation. This study makes use of numerical values from the HyperLEDA database (\url{http://leda.univ-lyon1.fr}). This research has made use of NASA’s Astrophysics Data System Bibliographic Services. This research made use of Astropy \citep{astropy}, \casa\ \citep{casa}, corner \citep{corner}, \emcee\ \citep{emcee}, MatPlotLib \citep{matplotlib}, \miriad\ \citep{miriad}, \nemo\ \citep{teuben95}, NumPy \citep{numpy}, pandas \citep{pandas}, photutils \citep{photutils}, SciPy \citep{scipy}, and seaborn \citep{seaborn}.

\section*{Data Availability}

The ALMA data used for this study are part of program ADS/JAO.ALMA\#2015.1.00820.S and are publicly available on the ALMA Science Archive (\url{https://almascience.nrao.edu/asax/}). The data products used in this analysis can be downloaded at \url{https://github.com/rclevy/RotationCurveTiltedRings}. The \spitzer\ images are publicly available on the NASA/IPAC Infrared Science Archive (\url{https://irsa.ipac.caltech.edu/}). The PS1 data are publicly available on the PS1 Image Cutout Server (\url{https://ps1images.stsci.edu/cgi-bin/ps1cutouts}).

\bibliographystyle{mnras}

\appendix
\section{Details of the Rotation Curve Fitting}
\label{app:rotcurve}

\begin{figure*}
    \centering
        \includegraphics[width=0.33\textwidth]{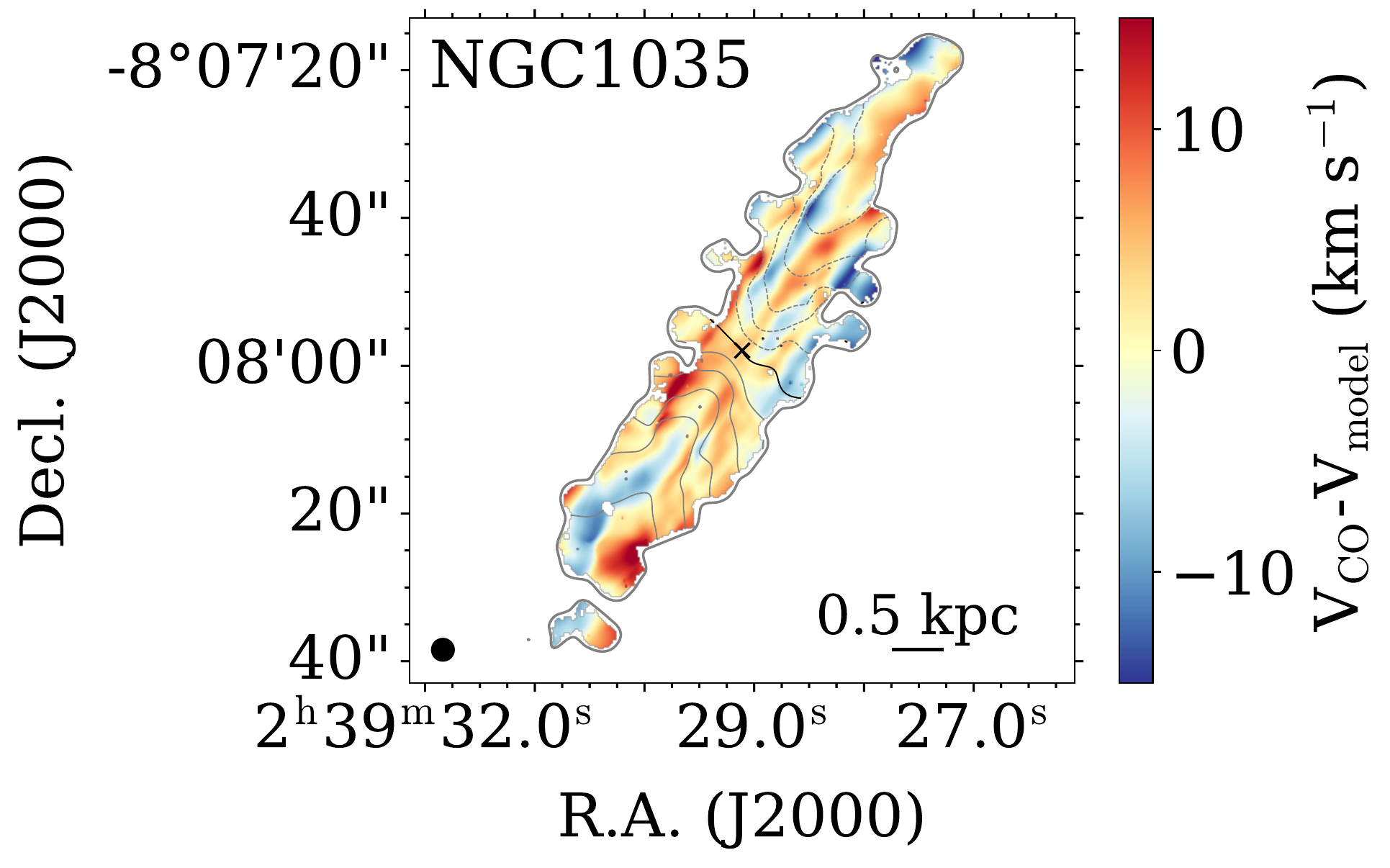}
        \includegraphics[width=0.33\textwidth]{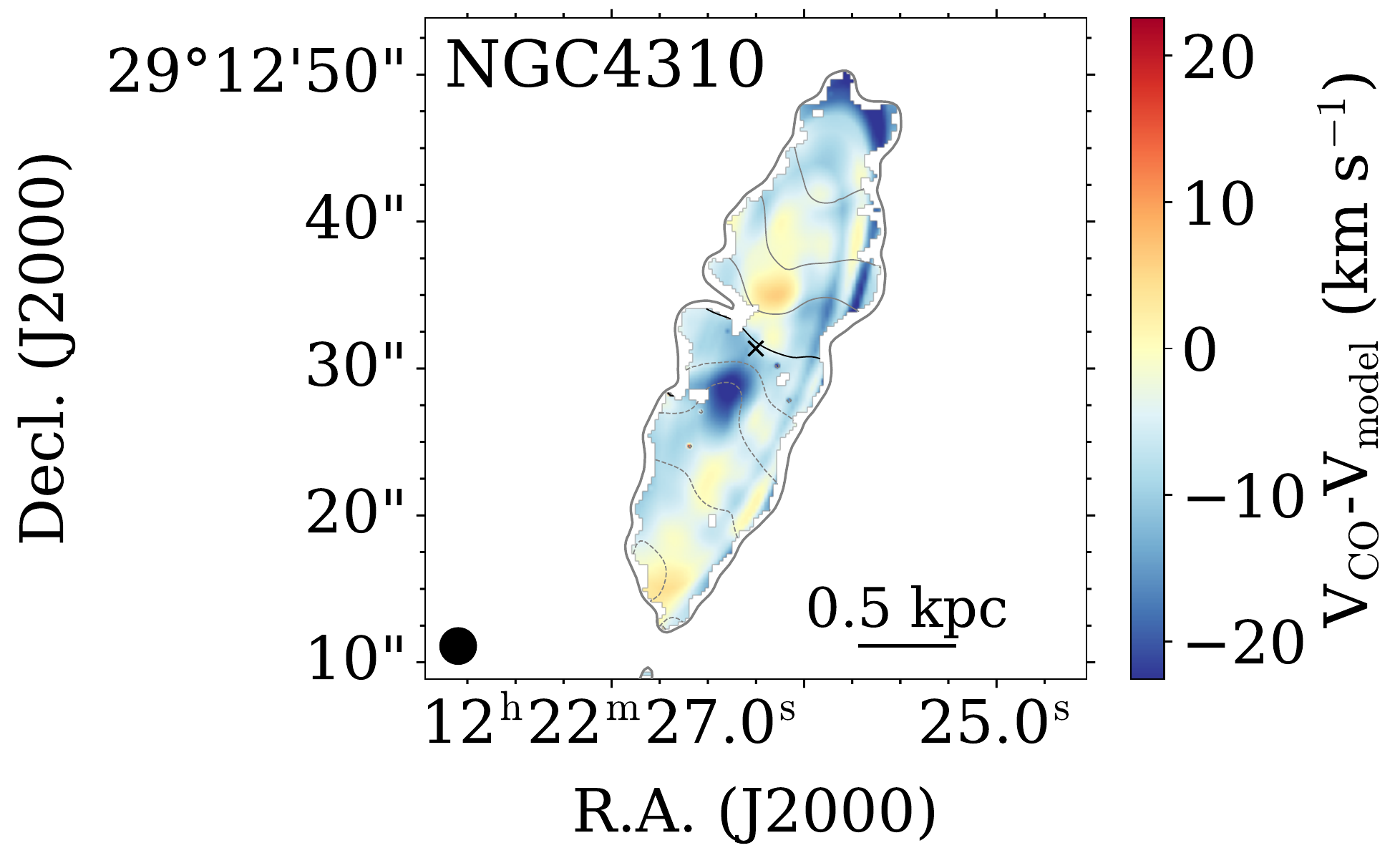}
        \includegraphics[width=0.33\textwidth]{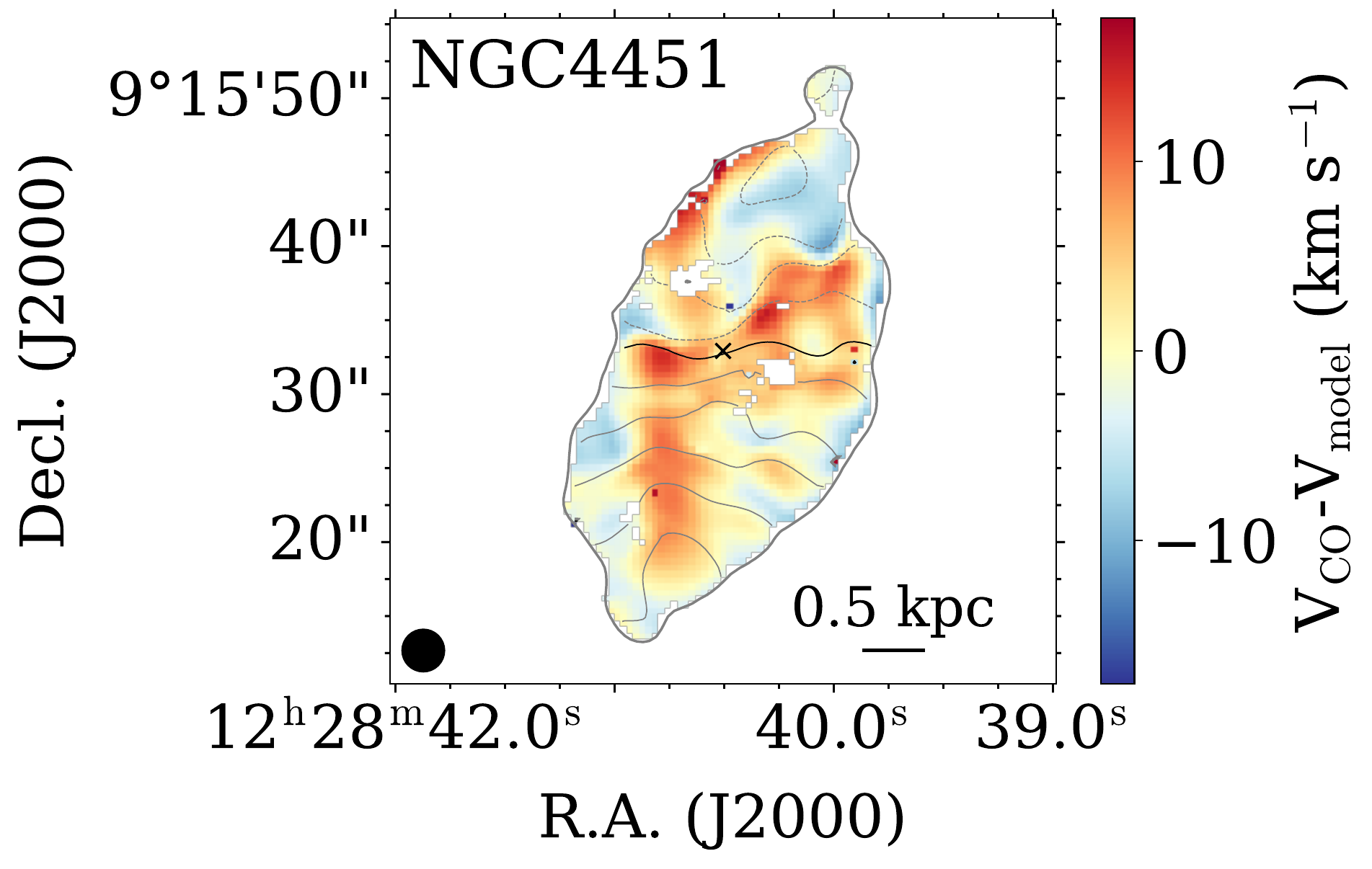}
        
        \includegraphics[width=0.33\textwidth]{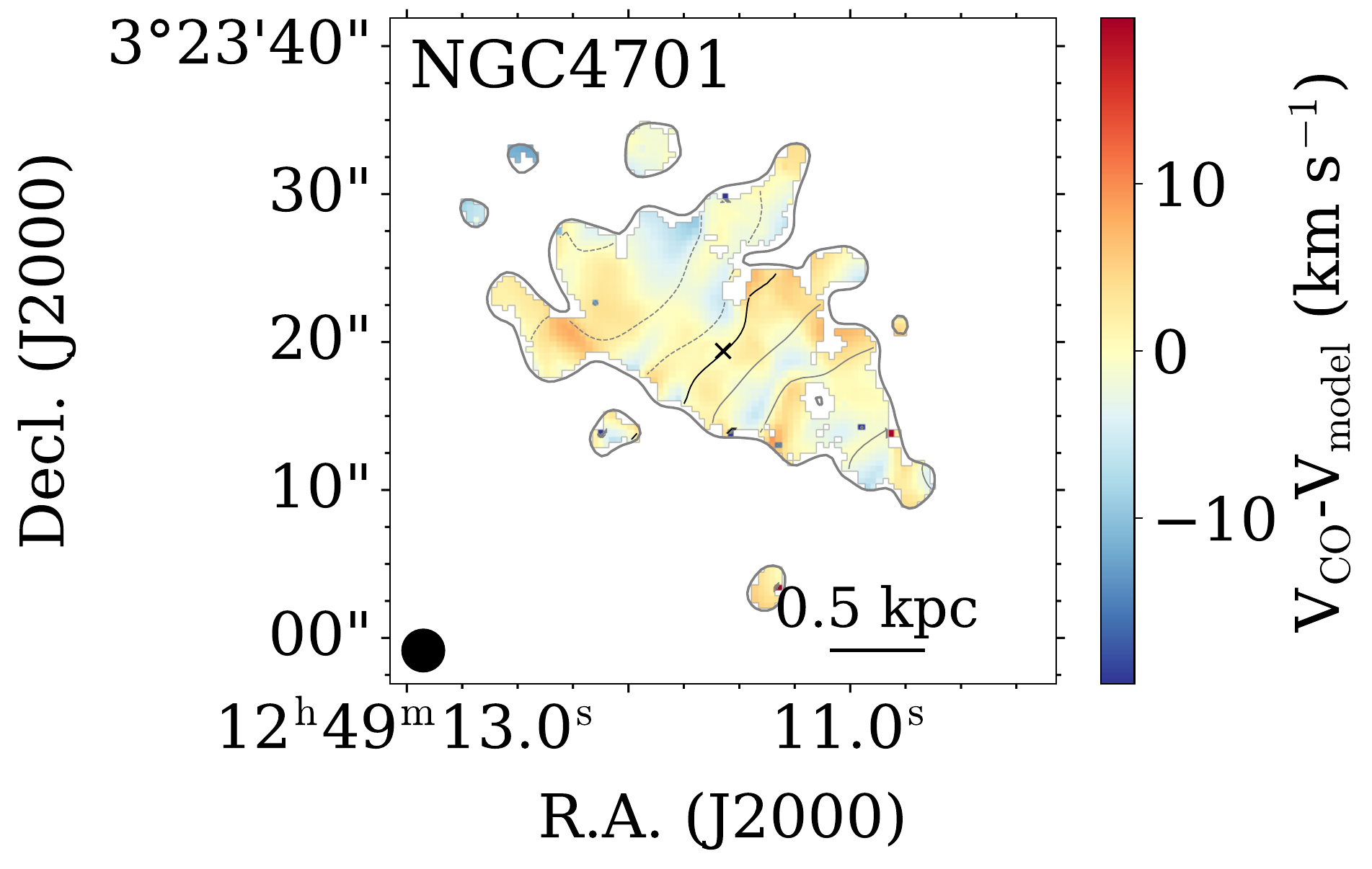}
        \includegraphics[width=0.33\textwidth]{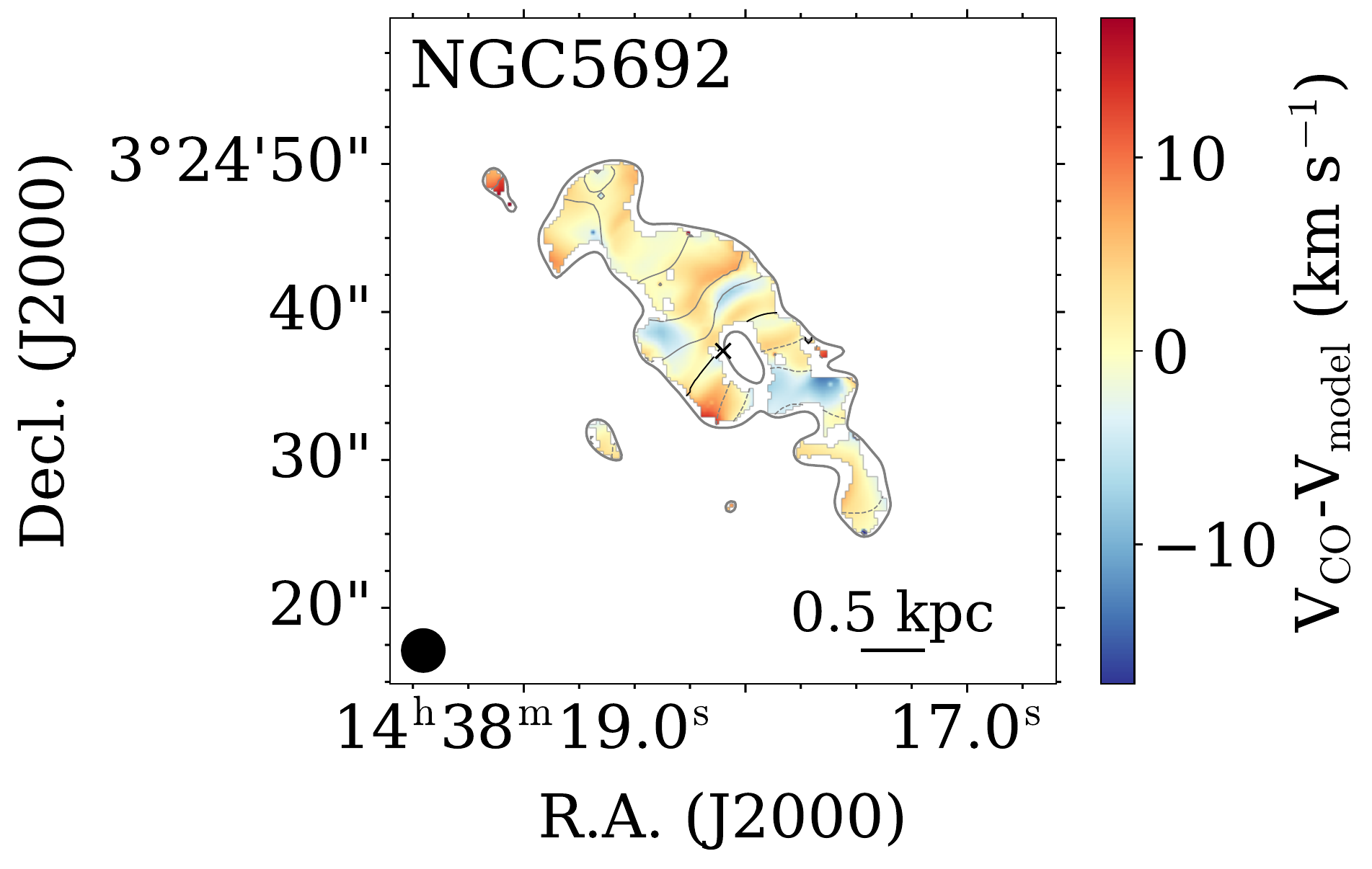}
        \includegraphics[width=0.33\textwidth]{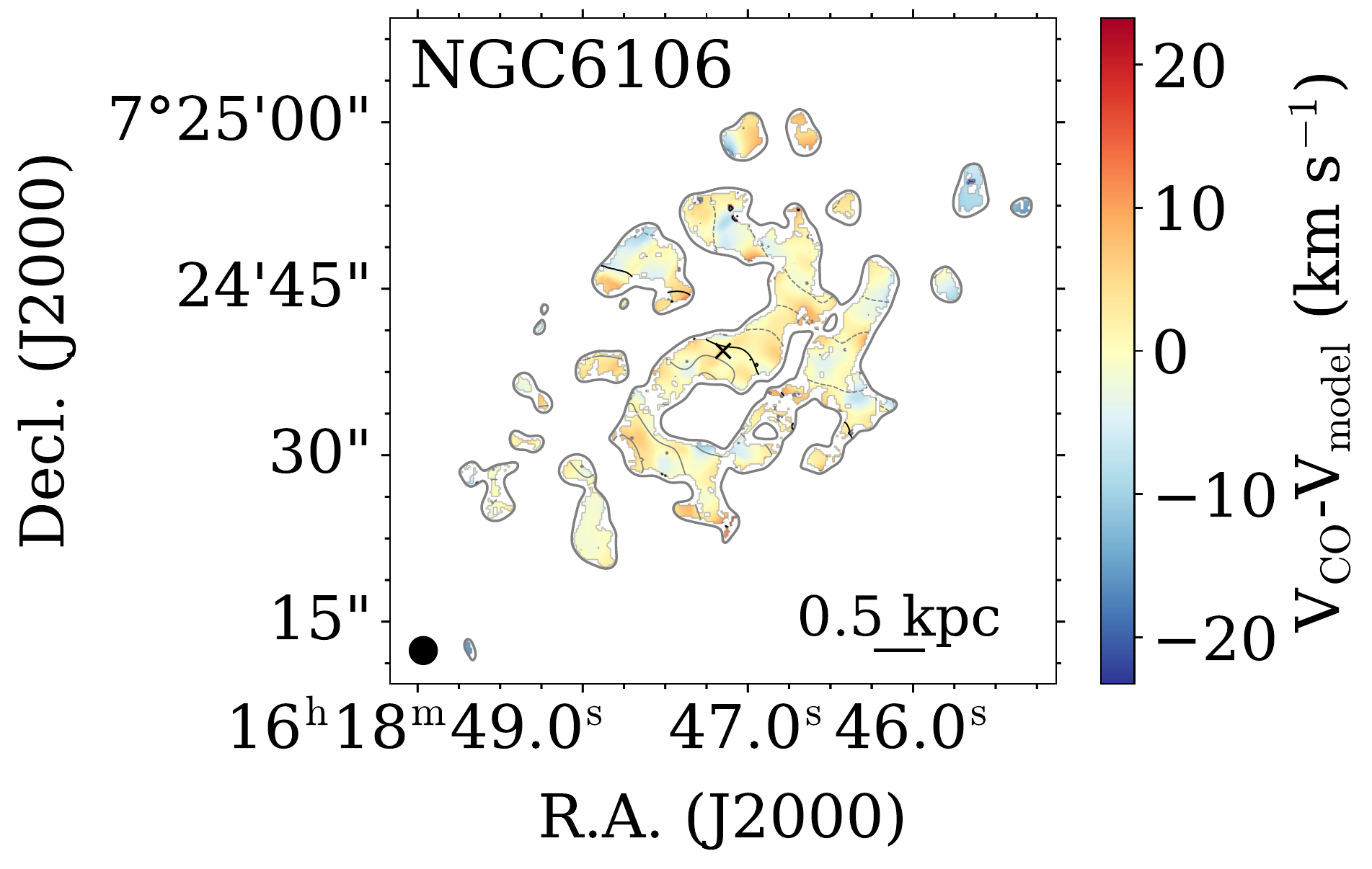}

    \caption{\change{Residual velocity fields, showing the best-fitting harmonic decompositions removed from the observed CO velocity fields. The crosses show the location of the kinematic \changes{centre} ({Table~\ref{tab:geomparams}}). The grey contours surrounding the residual velocity field show SNR = 3 based on the CO peak intensity. The CO velocity map is masked to this level. The black ellipse in the lower left corner shows the FWHM beam size. The grey contours over the residual velocity shows the CO velocity (as in {Figure~\ref{fig:maps}}) from -100~\kms\ to 100~\kms\ in 20~\kms\ intervals. Solid contours show positive velocities, dashed contours show negative velocities, and the black contour shows $V=0$~\kms. Note that, unlike {Figure~\ref{fig:maps}}, the panels are not all cropped to the same angular size.}}
    \label{fig:velresid}
\end{figure*}

\change{As a check on the rotation curve fitting, we compute the residual velocity field for each galaxy. We construct a model velocity field from the best-fitting harmonic decomposition ({Equation~\ref{eq:harmdecomp}}). The residual velocity fields are the difference between  observed CO velocity fields and the models ($\rm{V_{CO}-V_{model}}$) and are shown in {Figure~\ref{fig:velresid}}. In general, the residual velocities are small with no distinguishable patterns that may indicate improper geometric parameters \citep[e.g.,][]{vanderkruit78}. In NGC\,4310, there are small residuals near the centre. Because these residuals are small and near the centre (which is excluded from the analysis), they have a negligible effect. In NGC\,4451, the residual velocities are positive along the spiral arms. The magnitudes of these residual velocities are small, so they likely have only a minor effect on the robustness of the derived rotation curve. In principle, we could use a higher order harmonic decomposition that could better represent the spiral arms. However, given the relatively few angular bins in our CO rotation curves, a higher order fit would likely be poorly constrained. }

\section{Fitting the Dark Matter Profiles}
\label{app:mcmc}

\begin{figure*}
    \centering
        \includegraphics[width=0.4\textwidth]{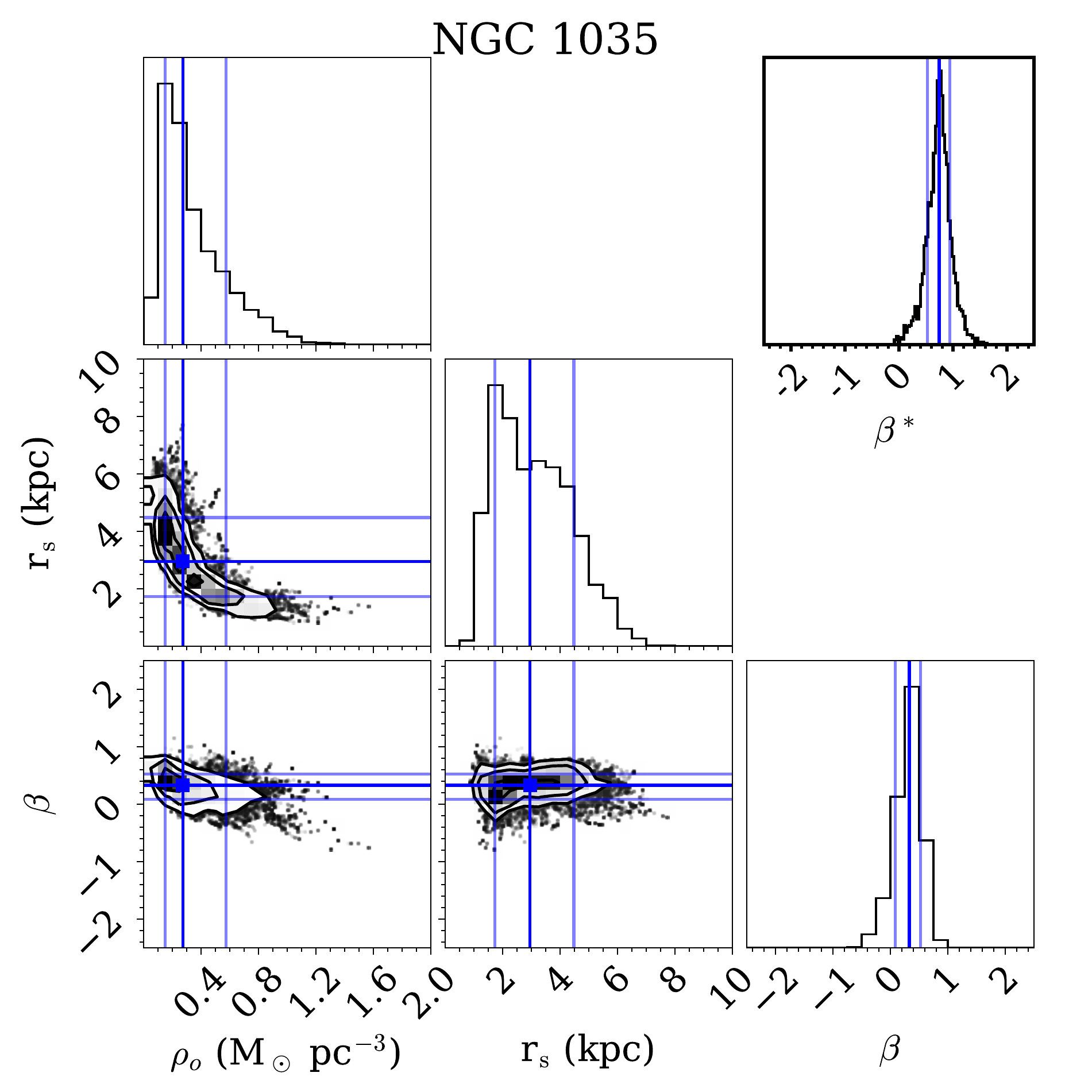}
        \includegraphics[width=0.4\textwidth]{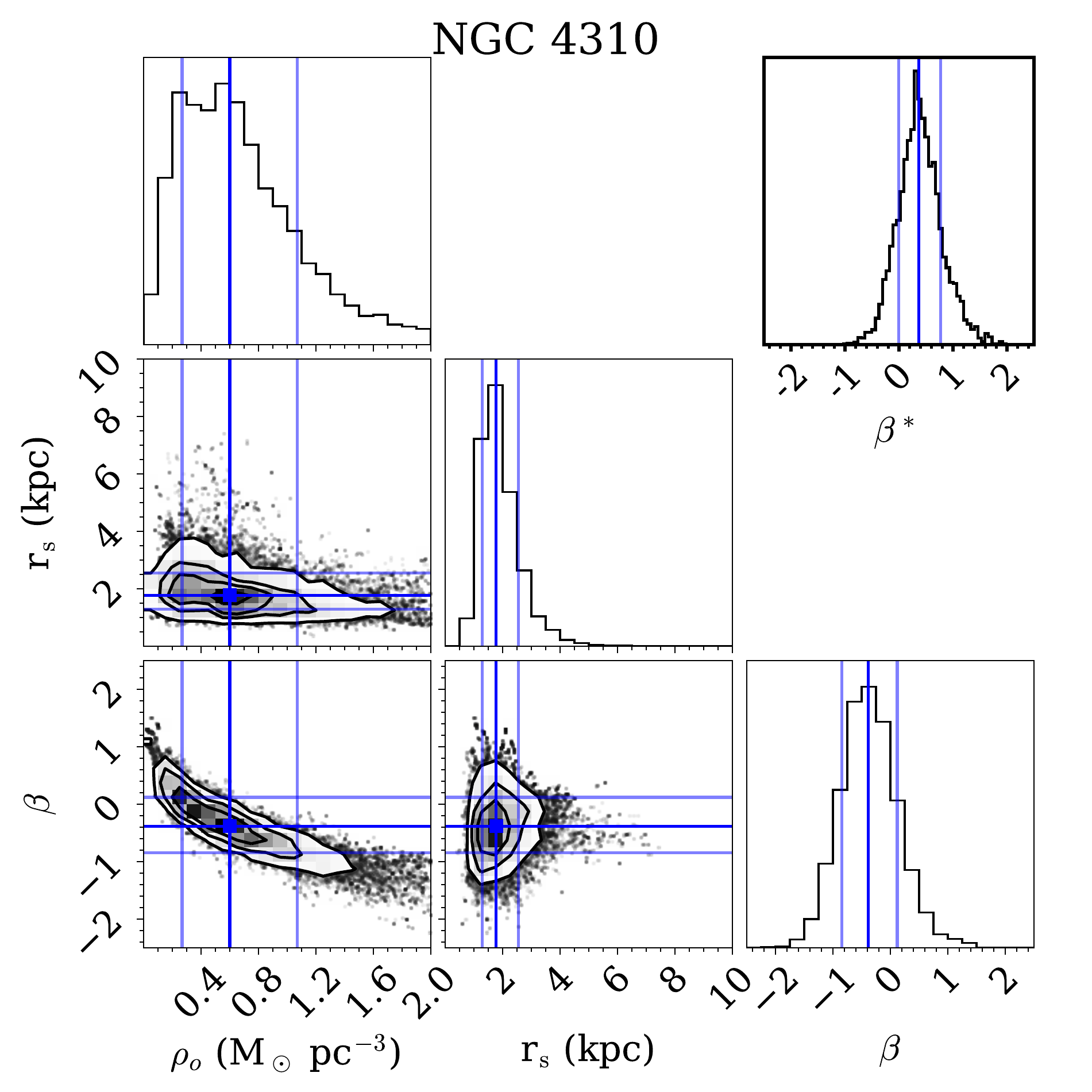}
       
        \includegraphics[width=0.4\textwidth]{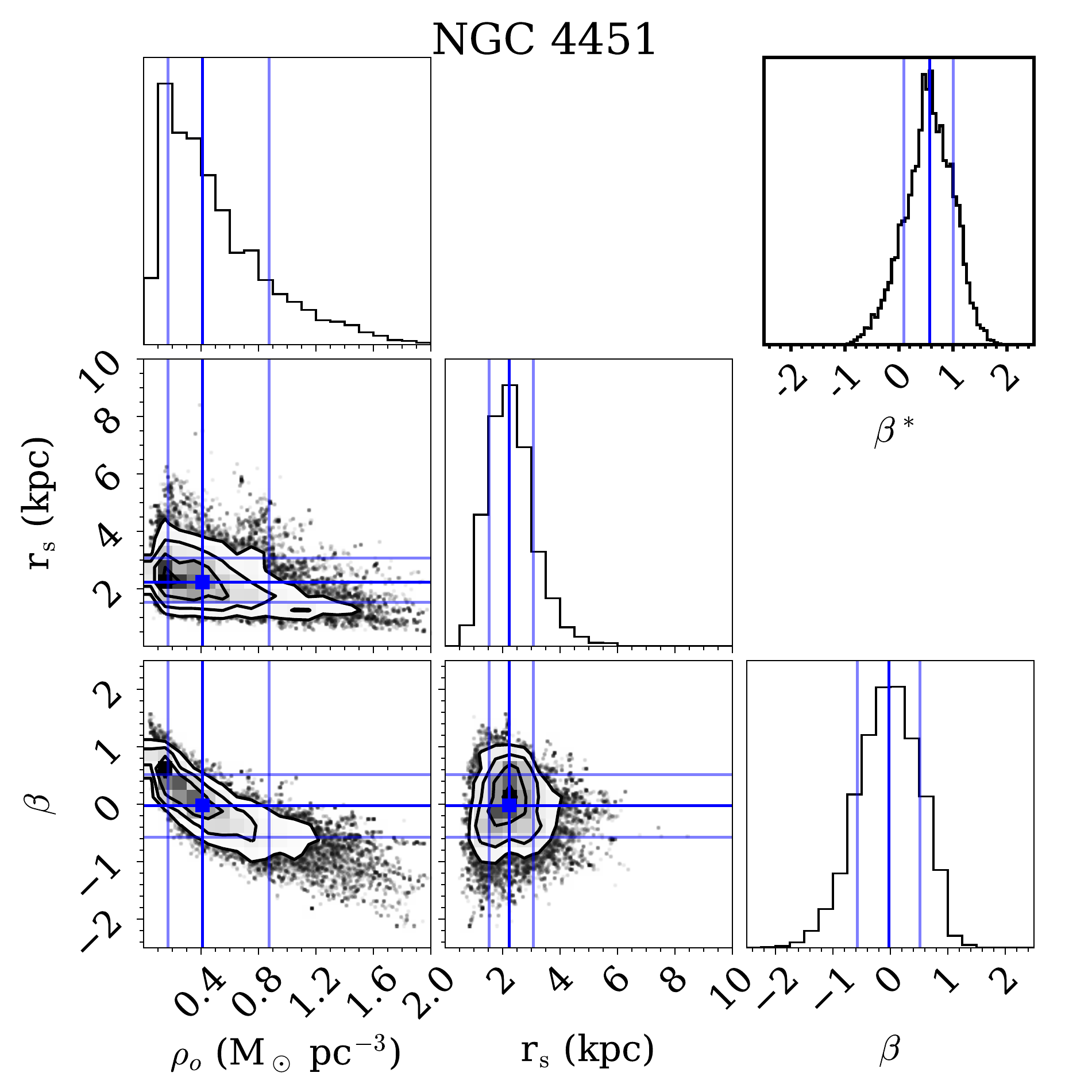}
        \includegraphics[width=0.4\textwidth]{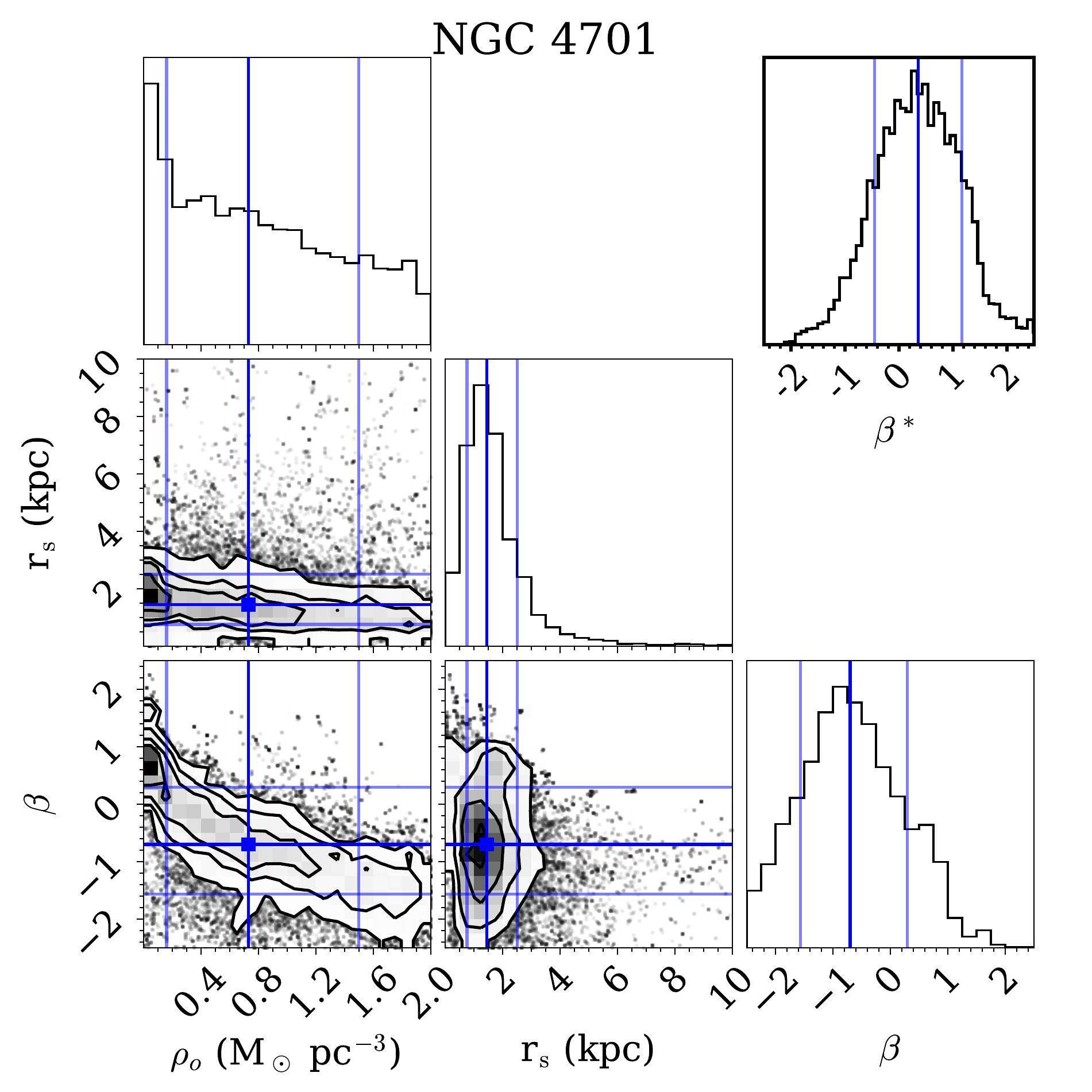}

        \includegraphics[width=0.4\textwidth]{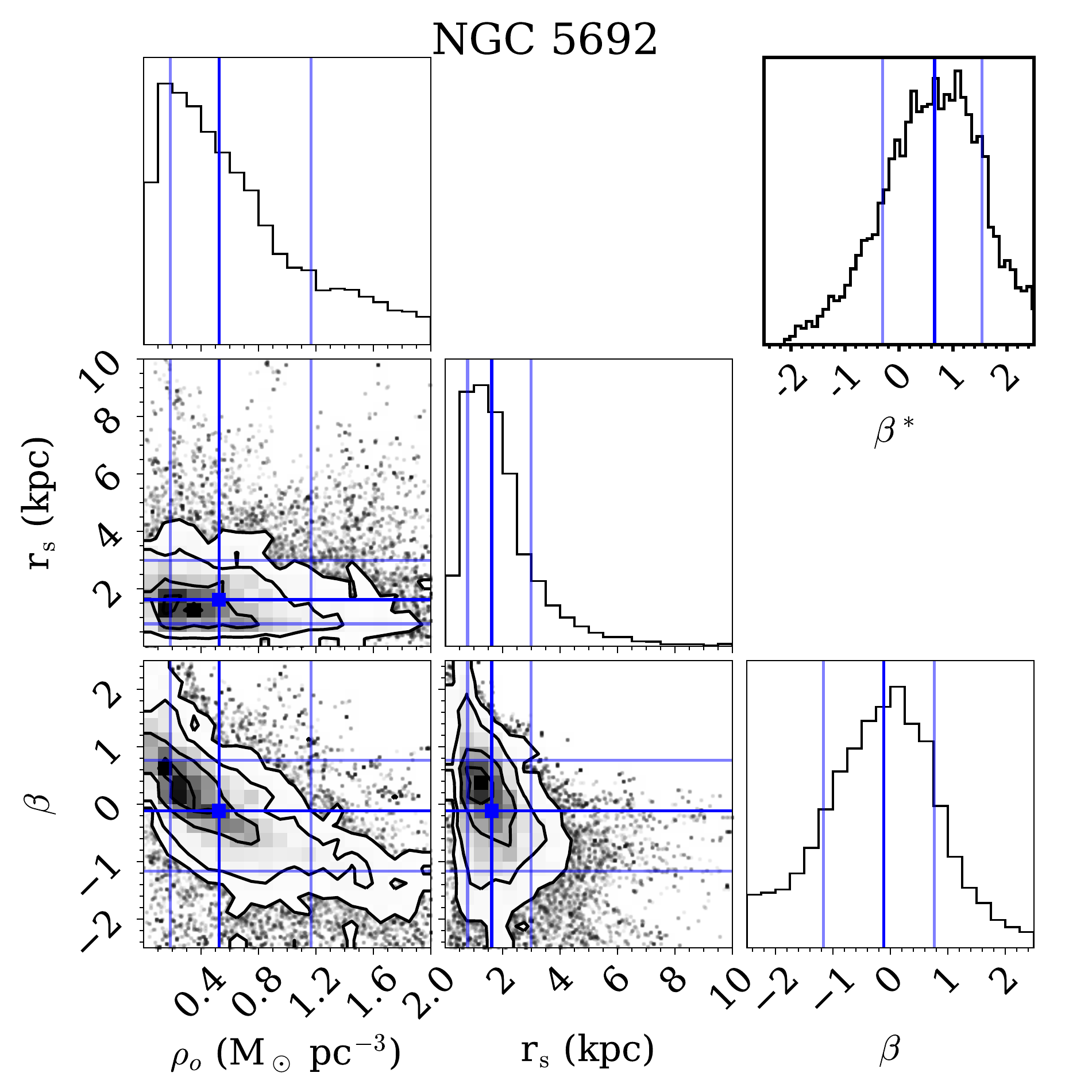}
        \includegraphics[width=0.4\textwidth]{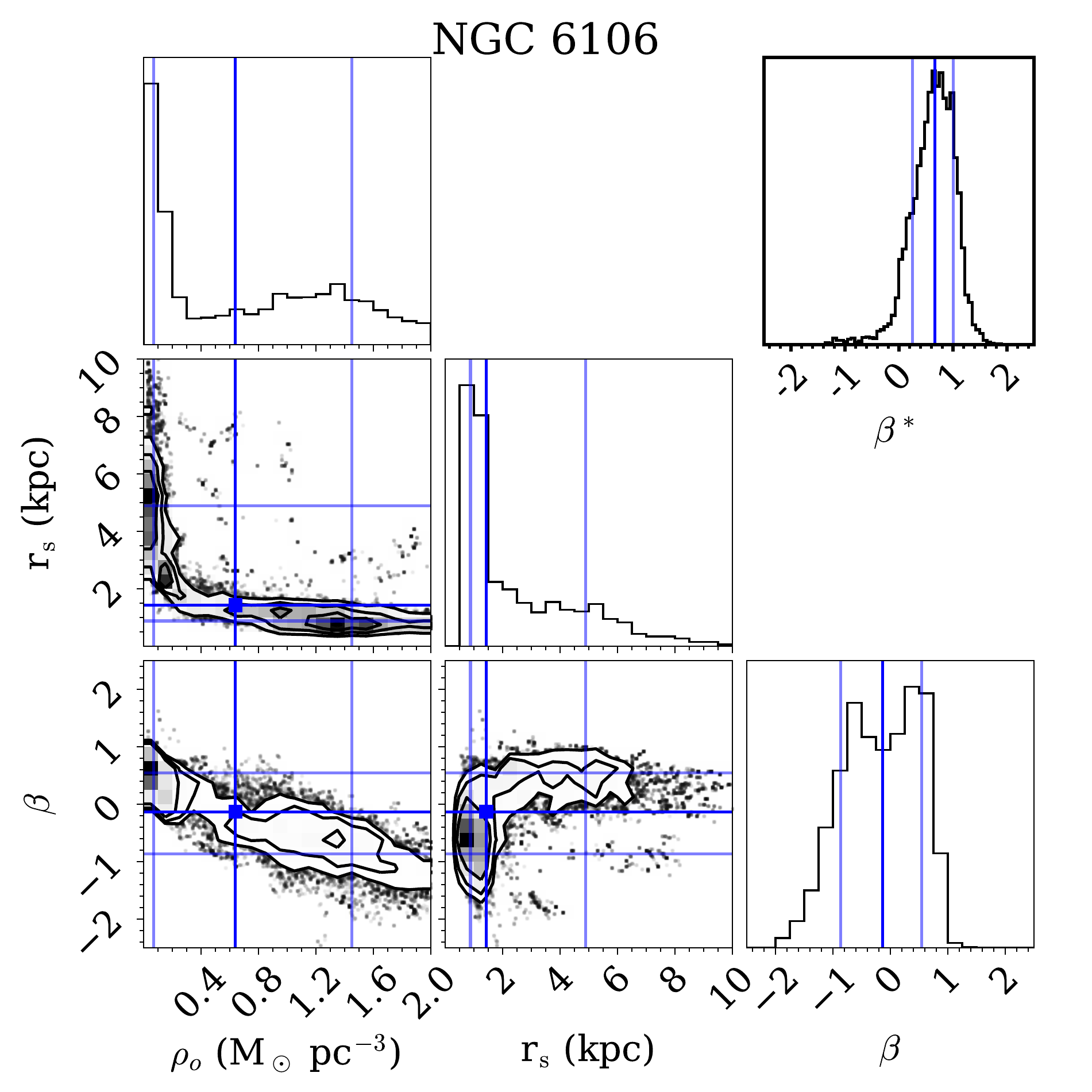}

    \caption{The posterior distributions for $\rho_o$, $r_s$, and $\beta$ in the gNFW dark matter profile given in {Equation~\ref{eq:NFW}} are shown in these corner plots. The plots in the upper right corner of each panel with thick outlines show the resulting distribution of \betastar. The dark blue lines show the median and the light blue lines show the 16$^{\rm th}$ and 84$^{\rm th}$ percentiles of the marginalized posterior likelihood distributions.}
    \label{fig:mcmcposteriors}
\end{figure*}

The MCMC method used to decompose the dark matter density profile is described in {Section~\ref{sec:DMdensity},} and the final fitted parameters and uncertainties are reported in {Table~\ref{tab:fittedparams}.} We show the posterior distributions of the fitted parameters for each galaxy in {Figure~\ref{fig:mcmcposteriors}.} While some of the posterior distributions are wide, the value of \betastar\ is relatively insensitive to the precise individual values of $\rho_o$, $r_s$, and $\beta$. Rather, it is the combination of these parameters yielding the dark matter density distribution that determines the value of \betastar.

\begin{figure*}
\label{fig:betastar_dmprofiles}
\centering
    \includegraphics[width=\textwidth]{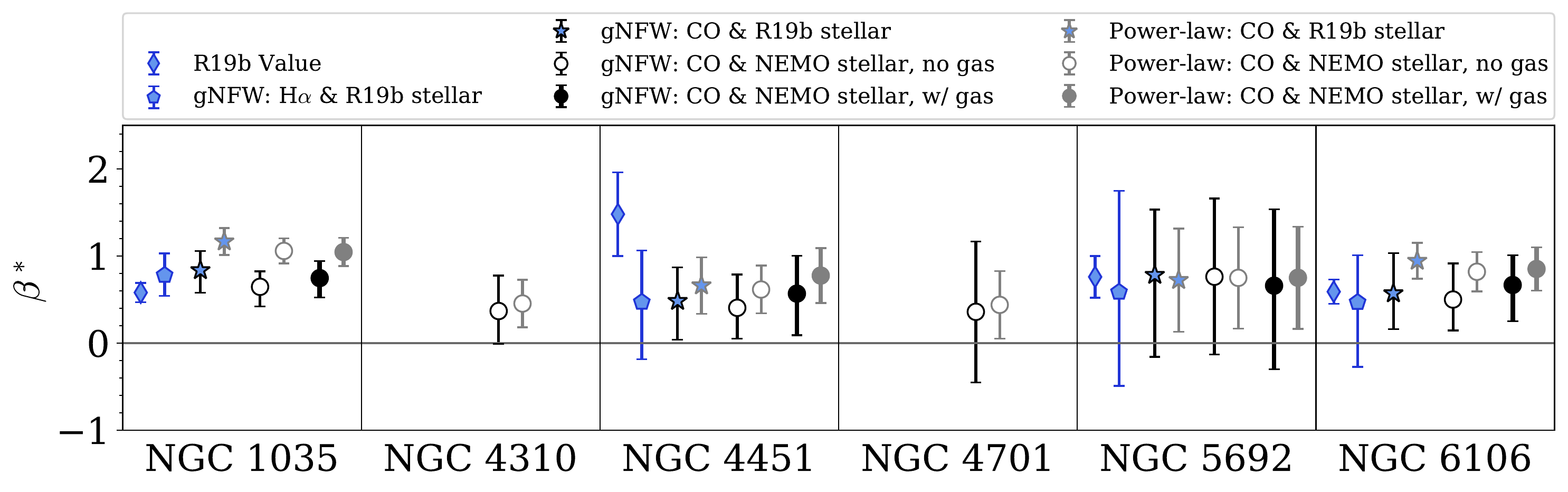}
\caption{\change{A comparison of the \betastar\ values derived for different forms of the dark matter density profiles. This is the same as {Figure~\ref{fig:betastarcomp}}, but we have added the symbols outlined in grey to this figure, which show the measurements of \betastar\ using a power-law parametrization of the dark matter density profile {(Equation~\ref{eq:dm_pl})}. Symbol shapes are the same as in {Figure~\ref{fig:betastarcomp}}.} In general, the \betastar\ values from the power-law fits agree with those from the gNFW within the uncertainties. Within each galaxy bin, points are artificially offset along the horizontal axis for clarity. } 
\end{figure*}

\begin{table*}
\begin{center}
\caption{Fitted Dark Matter Density Parameters}
\label{tab:fittedparams}
\begin{tabular}{ccccccc}
\hline
Name & $\rho_o$ & $r_s$ & $\beta$ & $\Upsilon_*$ & PA & $i$\\
 & (M$_\odot$ pc$^{-3}$) & (kpc) & & (M$_\odot$ L$_\odot^{-1}$) & (deg) & (deg)\\
\hline
NGC\,1035 & 0.27{\raisebox{0.5ex}{\tiny$^{+0.30}_{-0.12}$}} & 3.0{\raisebox{0.5ex}{\tiny$^{+1.5}_{-1.2}$}} & 0.3{\raisebox{0.5ex}{\tiny$^{+0.2}_{-0.2}$}} & 0.20{\raisebox{0.5ex}{\tiny$^{+0.02}_{-0.02}$}} & 145{\raisebox{0.5ex}{\tiny$^{+2}_{-2}$}} & 80{\raisebox{0.5ex}{\tiny$^{+2}_{-3}$}}\\
NGC\,4310 & 0.60{\raisebox{0.5ex}{\tiny$^{+0.47}_{-0.33}$}} & 1.8{\raisebox{0.5ex}{\tiny$^{+0.8}_{-0.5}$}} & -0.4{\raisebox{0.5ex}{\tiny$^{+0.5}_{-0.5}$}} & 0.22{\raisebox{0.5ex}{\tiny$^{+0.02}_{-0.03}$}} & 334{\raisebox{0.5ex}{\tiny$^{+5}_{-3}$}} & 71{\raisebox{0.5ex}{\tiny$^{+3}_{-2}$}}\\
NGC\,4451 & 0.41{\raisebox{0.5ex}{\tiny$^{+0.47}_{-0.24}$}} & 2.2{\raisebox{0.5ex}{\tiny$^{+0.8}_{-0.7}$}} & -0.0{\raisebox{0.5ex}{\tiny$^{+0.5}_{-0.6}$}} & 0.22{\raisebox{0.5ex}{\tiny$^{+0.02}_{-0.03}$}} & 177{\raisebox{0.5ex}{\tiny$^{+4}_{-3}$}} & 50{\raisebox{0.5ex}{\tiny$^{+3}_{-3}$}}\\
NGC\,4701 & 0.73{\raisebox{0.5ex}{\tiny$^{+0.77}_{-0.57}$}} & 1.4{\raisebox{0.5ex}{\tiny$^{+1.1}_{-0.7}$}} & -0.7{\raisebox{0.5ex}{\tiny$^{+1.0}_{-0.9}$}} & 0.21{\raisebox{0.5ex}{\tiny$^{+0.03}_{-0.02}$}} & 230{\raisebox{0.5ex}{\tiny$^{+3}_{-4}$}} & 49{\raisebox{0.5ex}{\tiny$^{+3}_{-3}$}}\\
NGC\,5692 & 0.53{\raisebox{0.5ex}{\tiny$^{+0.64}_{-0.34}$}} & 1.6{\raisebox{0.5ex}{\tiny$^{+1.4}_{-0.8}$}} & -0.1{\raisebox{0.5ex}{\tiny$^{+0.9}_{-1.1}$}} & 0.85{\raisebox{0.5ex}{\tiny$^{+0.10}_{-0.12}$}} & 35{\raisebox{0.5ex}{\tiny$^{+4}_{-3}$}} & 52{\raisebox{0.5ex}{\tiny$^{+3}_{-2}$}}\\
NGC\,6106 & 0.64{\raisebox{0.5ex}{\tiny$^{+0.81}_{-0.57}$}} & 1.4{\raisebox{0.5ex}{\tiny$^{+3.5}_{-0.6}$}} & -0.1{\raisebox{0.5ex}{\tiny$^{+0.7}_{-0.7}$}} & 0.21{\raisebox{0.5ex}{\tiny$^{+0.03}_{-0.03}$}} & 143{\raisebox{0.5ex}{\tiny$^{+3}_{-3}$}} & 59{\raisebox{0.5ex}{\tiny$^{+3}_{-2}$}}\\
\hline
\end{tabular}
\end{center}

\justifying\noindent{The density normalisation ($\rho_o$), radial scale length ($r_s$), and power-law index ($\beta$) are the three fitted parameters in the gNFW profile (Equation \ref{eq:NFW}). The fitted mass-to-light ratio ($\Upsilon_*$), position angle (PA), and inclination ($i$) agree with the fiducial values listed in Tables \ref{tab:geomparams} and \ref{tab:decompparams} within the uncertainties. The uncertainties are the 68 per cent confidence intervals of the marginalized posterior distributions (Figure \ref{fig:mcmcposteriors}). See Section \ref{sec:DMdensity} for details.}
\end{table*}

In addition to the gNFW profile in {Equation~\ref{eq:NFW},} we also tried a single power-law of the form
\begin{equation}
    \label{eq:dm_pl}
    \rho_{\rm DM}(r) = Ar^{-\beta}
\end{equation}
since our CO observations tend to probe small radii close to the centre and the \changes{three-parameter} fits shown in {Figure~\ref{fig:mcmcposteriors}} appear under-constrained. The posteriors of the power-law fits are good and, in general, better constrained than the \changes{three-parameter} gNFW fits. 
The resulting power-law model rotation curves, however, tend to continue rising whereas the CO rotation curves tracing the total potential flatten. This suggests that while the CO observations tend to probe close to the galaxy centres, they do reach the radial range where a second power-law slope may be important. This is most evident for NGC\,6106, which has the largest ${\rm R_{max}}$ in the sample. The power-law fit continues to rise, producing a steeper dark matter slope in the centre, resulting in the slightly larger values of \betastar\ compared to the gNFW, all else being equal. We directly compare the derived values of \betastar\ in {Figure~\ref{fig:betastar_dmprofiles}.} For the same assumptions about the gas content and stellar profiles used \change{(see Section~\ref{ssec:checks})}, the \betastar\ values from the power-law and gNFW dark matter density profiles are consistent within the uncertainties.

These tests show that while the precise values for the gNFW dark matter density profile parameters may be under-constrained, their combination and hence the derivation of \betastar\ is fairly robust. \cta{relatores19b} also found that \betastar\ itself is better constrained than the individual gNFW parameters. \change{We therefore adopt} the \betastar\ values from the more physically motivated gNFW profile in the discussion.

\bsp	
\label{lastpage}
\end{document}